\documentclass[3p,authoryear,times]{elsarticle}

\usepackage{ecrc}
\usepackage{aas_macros}

\volume{00}

\firstpage{1}

\journalname{Treatise on Geophysics, 2nd Edition}

\runauth{T. Guillot \& D. Gautier}

\jid{}

\jnltitlelogo{}

\CopyrightLine{2014}{}

\usepackage{amssymb}

\usepackage[figuresright]{rotating}

\def\beq{\begin{equation}}
\def\eeq{\end{equation}}

\def\eref#1{(\ref{eq:#1})}

\def\bar{\,\rm{bar}}
\def\K{{\,\mbox{K}}}

\def\gcc{\,\rm{g\,cm^{-3}}}

\def\ti#1{$^{{#1}}$}

\def\sec{\,\mbox{s}}
\def\m{\,\mbox{m}}
\def\bfnabla{\mathbf{\nabla}}

\def\req{R_\rm{eq}}

\def\teff{T_\rm{eff}}

\def\eref#1{(\ref{eq:#1})}
\def\rfig#1{fig.~\ref{fig:#1}}

\def\wig#1{\mathrel{\hbox{\hbox to 0pt{%
          \lower.6ex\hbox{$\sim$}\hss}\raise.4ex\hbox{$#1$}}}}

\def\eref#1{(\ref{eq:#1})}
\def\rfig#1{fig.~\ref{fig:#1}}

\def\beq{\begin{equation}}
\def\eeq{\end{equation}}
\def\comment#1{\noindent{\bf }}
\newcommand{\dpar}[2]{{\partial #1\over\partial #2}}

\def\bfnab{{\bf\nabla}}
\def\bfOm{{\bf\Omega}}
\def\bfr{{\bf r}}

\def\wig#1{\mathrel{\hbox{\hbox to 0pt{%
          \lower.6ex\hbox{$\sim$}\hss}\raise.4ex\hbox{$#1$}}}}

\def\req{R_{\rm eq}}
\def\rpol{R_{\rm pol}}
\def\rbar{{\overline{r}}}

\def\mea{{\rm\,M_\oplus}}

\def\mjup{{\rm\,M_J}}

\def\teff{T_{\rm eff}}

\def\linc{L_{\rm *p}}

\def\sec{\rm\,s}
\def\m{\rm\, m}
\def\g{\rm\,g}

\def\K{\rm\,K}
\def\bar{\rm\,bar}
\def\gcc{\rm\,g\,cm^{-3}}

\def\p#1{\times 10^{#1}}
\def\ti#1{$^{\rm #1}$}

\def\bfnabla{\mbox{\boldmath $\nabla$}}

\def\apjs{ApJS}
\def\apjl{ApJL}

\def\aap{A\&A}

\begin{document}

\begin{frontmatter}



\dochead{{\em To appear in \/}Treatise on Geophysics, 2nd Edition, Eds. T. Spohn \& G. Schubert}

\title{Giant Planets}
\author[nice]{Tristan Guillot\corref{cor1}}
\ead{tristan.guillot@oca.eu}
\author[paris]{Daniel Gautier}
\cortext[cor1]{Corresponding author}
\address[nice]{Laboratoire Lagrange, Universit\'e de Nice-Sophia Antipolis, Observatoire de la C\^ote d'Azur, CNRS, CP 34229, 06304 NICE Cedex 04, France} 
\address[paris]{LESIA, Observatoire de Paris, CNRS FRE 2461, 5 pl. J. Janssen, 92195 Meudon Cedex, France}

\begin{abstract}
We review the interior structure and evolution of Jupiter, Saturn,
Uranus and Neptune, and giant exoplanets with
particular emphasis on constraining their global composition. Compared to the first edition of this review, we provide a new discussion of the atmospheric compositions of the solar system giant planets, we discuss the discovery of oscillations of Jupiter and Saturn, the significant improvements in our understanding of the behavior of material at high pressures and the consequences for interior and evolution models. We place the giant planets in our Solar System in context with the trends seen for exoplanets. 
\end{abstract}


\end{frontmatter}

\newpage

\tableofcontents

\newpage

\section{Introduction}

\begin{figure}[htb]
\centerline{\resizebox{12cm}{!}{\includegraphics[angle=0]{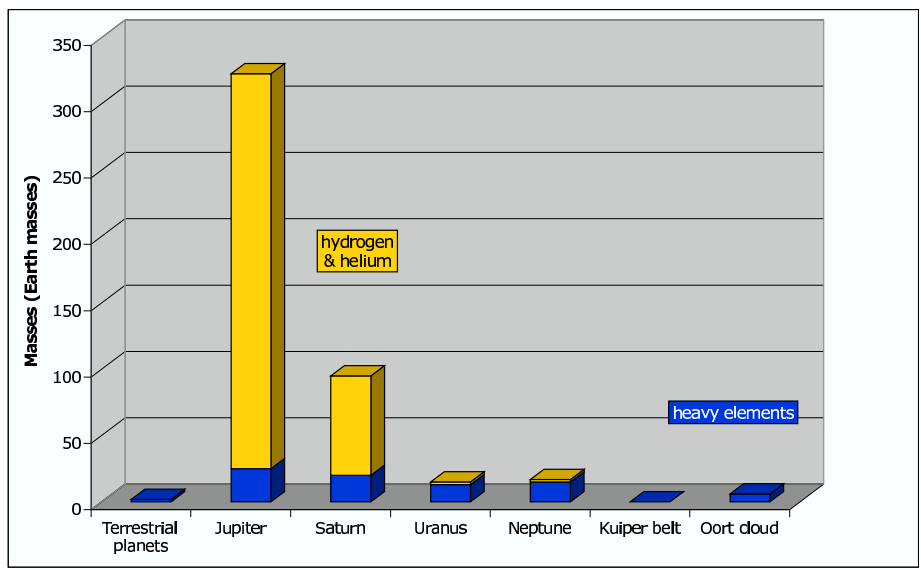}}}
\caption{An inventory of hydrogen and helium and all other elements
  (``heavy elements'') in the Solar System excluding the Sun (the Sun
  has a total mass of $332,960\mea$, including about $5000\mea$ in
  heavy elements, $1\mea$ being the mass of the Earth). The precise
  amount of heavy elements in Jupiter ($10-40\mea$) and Saturn
  ($20-30\mea$) is uncertain (see \S~\ref{sec:JupSat}).}
\label{fig:inventory}
\end{figure}

In our solar system, four planets stand out for their sheer mass and size. Jupiter, Saturn, Uranus, and Neptune indeed qualify as ``giant planets'' because they are larger than any terrestrial planet and much more massive than all other objects in the solar system, except the Sun, put together (Figure~\ref{fig:inventory}). Because of their gravitational might, they have played a key role in the formation of the solar system, tossing around many objects in the system, preventing the formation of a planet in what is now the asteroid belt, and directly leading to the formation of the Kuiper Belt and Oort Cloud. They also retain some of the gas (in particular hydrogen and helium) that was present when the Sun and its planets formed and are thus key witnesses in the search for our origins.

Because of a massive envelope mostly made of hydrogen helium, these planets are {\em fluid}, with no solid or liquid surface. In terms of structure and composition, they lie in between stars (gaseous and mostly made of hydrogen and helium) and smaller terrestrial planets (solid and liquid and mostly made of heavy elements), with Jupiter and Saturn being closer to the former and Uranus and Neptune to the latter (see fig.~\ref{fig:inventory}).

The discovery of many extrasolar planets of masses from a few thousands down to a few Earth masses and the possibility to characterize them by the measurement of their mass and size prompts a more general definition of giant planets. For this review, we will adopt the following: ``a giant planet is a planet mostly made of hydrogen and helium and too light to ignite deuterium fusion.'' This is purposely relatively vague -- depending on whether the inventory is performed by mass or by atom or molecule, Uranus and Neptune may be included or left out of the category. Note that Uranus and Neptune are indeed relatively different in structure than Jupiter and Saturn and are generally referred to as ``ice giants'', due to an interior structure that is consistent with the presence of mostly ``ices'' (a mixture formed from the condensation in the protoplanetary disk of low- refractivity materials such as H$_2$O, CH$_4$ and NH$_3$, and brought to the high-pressure conditions of planetary interiors -- see below).

Globally, this definition encompasses a class of objects that have similar properties (in particular, a low viscosity and a non-negligible compressibility) and inherited part of their material directly from the same reservoir as their parent star. These objects can thus be largely studied with the same tools, and their formation is linked to that of their parent star and the fate of the circumstellar gaseous disk present around the young star.

We will hereafter present some of the key data concerning giant planets in the solar system and outside. We will then present the theoretical basis for the study of their structure and evolution. On this basis, the constraints on their composition will be discussed and analyzed in terms of consequences for planet formation models.

\section{Observations and global properties}

\subsection{Visual appearances}

\begin{figure}[htb]
\centerline{\resizebox{16cm}{!}{\includegraphics[angle=0]{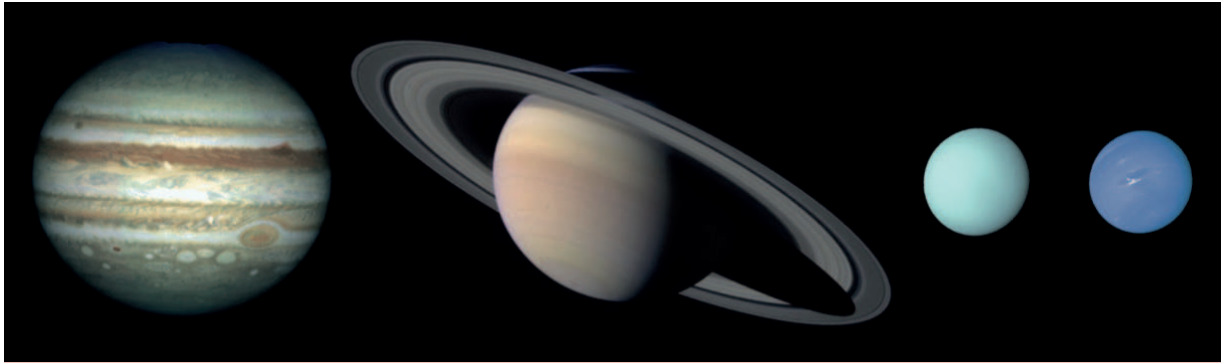}}}
\caption{Photomontage from images of Voyager 2 (Jupiter, Uranus, and Neptune) and Cassini (Saturn). The planets are shown to scale, with their respective axial inclinations.}
\label{fig:visual}
\end{figure}

In spite of its smallness, the sample of four giant planets in our solar system exhibits a large variety of appearances, shapes, colors, variability, etc. As shown in Figure~\ref{fig:visual}, all four giant planets are flattened by rotation and exhibit a more or less clear zonal wind pattern, but the color of their visible atmosphere is very different (this is due mostly to minor species in the high planetary atmosphere), their clouds have different compositions (ammonia for Jupiter and Saturn, methane for Uranus and Neptune) and depths, and their global meteorology (number of vortexes, long-lived anticyclones such as Jupiter's Great Red Spot, presence of planetary-scale storms, convective activity) is different from one planet to the next.

We can presently only wonder about what is in store for us with extrasolar giant planets since we cannot image and resolve them. But with orbital distances from as close as 0.01 AU to 100\,AU and more, a variety of masses, sizes, and parent stars, we should expect to be surprised!

\subsection{Gravity fields}
\label{sec:gravity}

The mass of our giant planets can be obtained with great accuracy from the observation of the motions of their natural satellites: 317.834,
95.161, 14.538 and 17.148 times the mass of the Earth ($1\mea =5.97369\times 10^{27}\g$)  for Jupiter, Saturn, Uranus and Neptune, respectively. More precise measurements of their gravity field can be obtained through the analysis of the trajectories of a spacecraft during flyby, especially when they come close to the planet and preferably in a near-polar orbit. The gravitational field thus measured departs from a purely spherical function due to the planets' rapid rotation. The measurements are generally expressed by expanding the components of the gravity field in Legendre polynomials $P_i$ of progressively higher orders:
\beq
V_{\rm ext}(r,\theta)=-\frac{GM}{r}\left\{
1-\sum_{i=1}^\infty \left(\frac{R_{\rm eq}}{r} \right)^i J_i P_i(\cos\theta)
\right\},
\eeq
where $V_{\rm ext}(r,\theta)$ is the gravity field evaluated outside
the planet at a distance $r$ and colatitude $\theta$, 
$R_{\rm eq}$ is the equatorial radius, and $J_i$ are the
gravitational moments. Because the giant planets are very close to
hydrostatic equilibrium the coefficients of even order are the only
ones that are not negligible. We will see how these gravitational
moments, as listed in table~\ref{tab:moments}, help us constrain the
planets' interior density profiles. 

\begin{table}[htb]
\begin{center}
\caption{Characteristics of the gravity fields and radii}
\label{tab:moments}
\small
\begin{tabular}{l r@{.}l r@{.}l r@{.}l r@{.}l} \hline\hline 
	&\multicolumn{2}{c}{\bf Jupiter} &\multicolumn{2}{c}{\bf
Saturn} &\multicolumn{2}{c}{\bf Uranus} &\multicolumn{2}{c}{\bf
Neptune} \\ \hline
$M\p{-26}$ [kg] & 18&986112(15)\ti{a} & 5&68463036(16)\ti{b} & 0&8683205(34)\ti{c} & 1&0243547861(15)\ti{d} \\
$R_{\rm eq}\p{-7}$ [m] &  7&1492(4)\ti{e} & 6&0268(4)\ti{f}  & 2&5559(4)\ti{g} & 2&4766(15)\ti{g} \\
$R_{\rm pol}\p{-7}$ [m] &  6&6854(10)\ti{e} & 5&4364(10)\ti{f}  & 2&4973(20)\ti{g} & 2&4342(30)\ti{g} \\
$\overline{R}\p{-7}$ [m] & 6&9894(6)\ti{h} & 5&8210(6)\ti{i} & 2&5364(10)\ti{i} &  2&4625(20)\ti{i} \\
$\overline{\rho}\p{-3}$ [$\rm kg\,m^{-3}$] & 1&3275(4) & 0&6880(2) & 1&2704(15) & 1&6377(40) \\
$R_{\rm ref}\p{-7}$ [m] & 7&1398\ti{a} & 6&0330\ti{b}  & 2&5559  & 2&5225\ti{d} \\
$J_2\p{2}$ & 1&4736(1)\ti{a} & 1&629071(27)\ti{b} & 0&35160(32)\ti{c} & 0&34084(45) \ti{d}  \\
$J_4\p{4}$ & $-5$&87(5)\ti{a} & $-9$&358(28)\ti{b} & $-0$&354(41)\ti{c} & $-0$&334(29)\ti{d} \\
$J_6\p{4}$ & 0&31(20)\ti{a} & 0&861(96)\ti{b} & \multicolumn{2}{c}{\dots} & \multicolumn{2}{c}{\dots} \\
$P_{\omega}\p{-4}$ [s] & 3&57297(41)\ti{j} & 3&83624(47)\ ?\ti{j,k} & 6&206(4)\ti{l} & 5&800(20)\ti{m} \\
$q$ & 0&08923(5) & 0&15491(10) & 0&02951(5) & 0&02609(23) \\
$C/M\req^2$ & 0&258 & 0&220 & 0&230 & 0&241 \\

\hline\hline
\multicolumn{9}{p{13.5cm}}{%
The numbers in parentheses are the uncertainty in the last digits 
of the given value. The value of the gravitational constant used to
calculate the masses of Jupiter and Saturn is $G=6.67259\p{-11}\,\rm
N\,m^2\,kg^{-2}$ \citep{1987RvMP...59.1121C}. The values of the radii and
density correspond to the one bar pressure 
level (1\,bar$=10^5$\,Pa). Gravitational moments are normalized at the reference radius $R_{\rm ref}$. Only values published in refereed journals are considered.}\\
\multicolumn{9}{l}{\ti{a} \citet{1985AJ.....90..364C}}\\
\multicolumn{9}{l}{\ti{b} \citet{2006AJ....132.2520J}}\\
\multicolumn{9}{l}{\ti{c} \citet{1987JGR....9214877A}}\\
\multicolumn{9}{l}{\ti{d}  \citet{Jacobson2009}}\\ 
\multicolumn{9}{l}{\ti{e} \citet{1981JGR....86.8721L}; \citet{HelledGuillot2013} derive slightly different values.}\\
\multicolumn{9}{l}{\ti{f} \citet{1985AJ.....90.1136L}}\\
\multicolumn{9}{l}{\ti{g} \citet{1992AJ....103..967L}; \citet{Helled+2010} derive slightly different values.}\\
\multicolumn{9}{l}{\ti{h} From 4th order figure theory}\\
\multicolumn{9}{l}{\ti{i} $(2\req+\rpol)/3$ (Clairaut's approximation)}\\
\multicolumn{9}{l}{\ti{j} \citet{1986CeMec..39..103D}}\\
\multicolumn{9}{p{13.5cm}}{\ti{k} This measurement from the {\it
    Voyager} era is now 
  in question and values down to $37955$\,s have been proposed (see
  \S~\ref{sec:magnetic fields})}\\ 
\multicolumn{9}{l}{\ti{l} \citet{1986Sci...233..102W}. See however \citet{Helled+2010}.}\\
\multicolumn{9}{l}{\ti{m} \citet{1989Sci...246.1498W}. See however \citet{Helled+2010}.}\\
\end{tabular}
\normalsize
\end{center}
\end{table}

Table~\ref{tab:moments} also indicates the radii obtained with the
greatest accuracy by radio-occultation experiments. An important
consequence obtained is the fact that these planets have low
densities, from $0.688\gcc$ for Saturn to $1.64\gcc$ for Neptune, to
be compared with densities of 3.9 to 5.5$\gcc$ for the terrestrial
planets in the solar system. Considering the compression that strongly increases with
mass, one is led naturally to the conclusion that these planets
contain an important proportion of light materials including hydrogen
and helium. It also implies that Uranus and Neptune which are less
massive must contain a relatively larger proportion of heavy elements
than Jupiter and Saturn. This may lead to a sub-classification between
the hydrogen-helium giant planets Jupiter and Saturn, and the ``ice
giants'' or ``sub giants'' Uranus and Neptune.

The planets are also relatively fast rotators, with periods of $\sim
10$ hours for Jupiter and Saturn, and $\sim 17$ hours for Uranus and
Neptune. The fact that this fast rotation visibly affects the figure
(shape) of these planets is seen by the significant difference between
the polar and equatorial radii. It also leads to gravitational moments
that differ significantly from a null value. However, it is important
to stress that there is no unique rotation frame for these fluid
planets: atmospheric zonal winds imply that different latitudes rotate
at different velocities (see \S~\ref{sec:dynamics}), and the magnetic
field provides another rotation period. Because the latter is tied to
the deeper levels of the planet, it is believed to be more relevant
when interpreting the gravitational moments. The rotation periods
listed in Table~\ref{tab:moments} hence correspond to that of the
magnetic field. The case of Saturn is complex and to be
discussed in the next section.

\subsection{Magnetic fields}
\label{sec:magnetic fields}

As the Earth, the Sun and Mercury, our four giant planets possess
their own magnetic fields. 
These magnetic fields $\bf B$ may be expressed in form of a
development in spherical harmonics of the scalar potential $W$, such
that ${\bf B}=-\bfnabla W$:
\begin{equation}
W=a\sum_{n=1}^{\infty} \left(\frac{a}{r} \right)^{n+1}
\sum_{m=0}^n \left\{g_n^m \cos(m\phi)+h_n^m \sin(m\phi)\right\}
P_n^m(\cos\theta).
\label{eq:W}
\end{equation}
$r$ is the distance to the planet's center, $a$ its radius, $\theta$
the colatitude, $\phi$ the longitude and $P_n^m$ the associated
Legendre polynomials. The coefficients $g_n^m$ and $h_n^m$ are the
magnetic moments that characterize the field. They are expressed in
magnetic field units.

One can show that the first coefficients of relation~\eref{W} (for
$n=0$ and $n=1$) correspond to the potential of a magnetic dipole such
that $W={\bf M\cdot r}/r^3$ of moment:
\beq
M=a^3 \left\{\left(g_1^0\right)^2 + \left(g_1^1\right)^2 +
\left(h_1^1\right)^2\right\}^{1/2}.
\eeq

As shown by the Voyager~2 measurements, Jupiter and Saturn have magnetic fields of essentially dipolar nature,
of axis close to the rotation axis ($g_1^0$ is much larger than the
other harmonics); Uranus and Neptune have magnetic fields that are
intrinsically much more complex. To provide an idea of the intensity
of the magnetic fields, the value of the dipolar moments for the four
planets are $4.27\,\rm Gauss\,R_J^3$, $0.21\rm\,Gauss\,R_S^3$,
$0.23\rm\,Gauss\,R_U^3$, $0.133\rm\,Gauss\,R_N^3$, respectively
\citep[][see also chapter by Connerney]{1982Natur.298...44C,1983JGR....88.8771A,1986Sci...233...85N,1989Sci...246.1473N}.

A true surprise from {\it Voyager} that has been confirmed by the {\it
Cassini-Huygens} mission is that Saturn's magnetic field is
axisymetric {\it to the limit of the measurement accuracy}: Saturn's
magnetic and rotation axes are perfectly aligned \citep[e.g.,][]{RussellDougherty2010}. Voyager measurements
indicated nevertheless a clear signature in the radio signal at $\rm 10^h 39^m 22.4^s$ believed to be a consequence of the rotation of the magnetic
field. New measurements of a slower spin period of $\rm 10^h 47^m 6^s$ by Cassini \citep{2005Sci...307.1255G,2006Natur.441...62G} have shown that the kilometric radiation was not directly tied to the period of the magnetic field but resulted from a complex interplay between the spin of the planetary magnetic field and the solar wind \citep[e.g.][]{2005JGRA..11012203C}. New periods have been proposed: $\rm 10^h 32^m 35^s$ based on a minimization of the zonal differential rotation \citep{AndersonSchubert2007} and $\rm 10^h 33^m 13^s$ based on the latitudinal distribution of potential vorticity \citep{Read+2009}. The problem still stands out. 

These magnetic fields must be generated in the conductive parts of the interiors, i.e., in metallic hydrogen at radii which are about 80\% and 60\% of the planetary radius for Jupiter and Saturn respectively (see sections~\ref{sec:EOS}, \ref{sec:JupSat} and e.g., \citet{StanleyGlatzmaier2010}). Recent models that consistently join the slowly convecting metallic interior with the non-conducting outer molecular envelope dominated by zonal flows result in a mainly dipolar magnetic field similar to the observations and further show that Jupiter's stronger magnetic field and Saturn's broader equatorial jet (see next section) can be interpreted as resulting from the deeper location of the transition region in Saturn \citep{HeimpelGomezPerez2011}. These explanations remain largely qualitative rather than quantitative and are further complicated by a necessary overforcing of the simulations \citep[see][and next section]{Showman+2011}. The question of why Saturn's magnetic field is much more axisymmetric than Jupiter's remains. Dipolar fields are obtained relatively naturally through the forcing of zonal jets extending down to the conducting region \citep{Guervilly+2012} but why Jupiter differs is unexplained. 
A possibility is that both fields are non-axisymmetric at deep levels but that Saturn's is filtered by a more extended helium sedimentation region \citep{1983RPPh...46..555S}, but in practice, a realistic solution yielding Saturn's measured field has not been found \citep[][and references therein]{StanleyGlatzmaier2010}. 

Within Uranus and Neptune, the magnetic field is believed to be generated within a layer in which water is in an ionic phase, below about 80\% of their total radius \citep[see][and section \ref{sec:UraNep} hereafter]{Redmer+2011}. Their complex, multipolar magnetic fields has been thought to be a consequence of a strong stratification and of a dynamo generated in a thin shell \citep{StanleyBloxham2004}. However, this point of view is now challenged by new simulations that generate both planets' magnetic fields and zonal wind structures through a thick shell dynamo \citep{Soderlund+2013}. Further work however must involve realistic variations of the interior density and conductivity.

\subsection{Atmospheric dynamics: winds and weather}
\label{sec:dynamics}

The atmospheres of all giant planets are evidently complex and
turbulent in nature. This can, for example, be seen from the mean zonal
winds (inferred from cloud tracking), which are very rapidly varying
functions of the latitude \citep[see e.g.,][]{Ingersoll+95}: while some
of the regions rotate at the same speed as the interior magnetic field
(in the so-called ``system III'' reference frame), most of the
atmospheres do not. Jupiter and Saturn both have superrotating
equators ($+100$ and $+400\m\sec^{-1}$ in system III, for Jupiter and
Saturn, respectively), Uranus and Neptune have subrotating equators,
and superrotating high latitude jets. Neptune, which receives the
smallest amount of energy from the Sun has the largest peak-to-peak
latitudinal variations in wind velocity: about $600\m\sec^{-1}$.  It
can be noted that, contrary to the case of the strongly irradiated
planets to be discussed later, the winds of Jupiter, Saturn, Uranus
and Neptune, are significantly slower than the planet itself under its
own spin (from 12.2\,km$\sec^{-1}$ for Jupiter to 2.6\,km$\sec^{-1}$
for Neptune, at the equator).

It is not yet clear whether the observed winds are driven from the bottom or from the top. 
The first possibility is that surface winds are related to motions in
the planets' interiors, which, according to the Taylor-Proudman
theorem, should be confined by the rapid rotation to the plane
perpendicular to the axis of rotation
\citep{Busse78}. This is now backed by simulations in the anelastic limit (i.e., accounting for compressibility) which show that the outcome strongly depends on the density stratification in the interior. A small density contrast (as expected in Jupiter and Saturn) leads to equatorial superrotation whereas for a large one (as expected for Uranus and Neptune), the equatorial jet tends to subrotate \citep{Glatzmaier+2009, Gastine+2013}. However, the application of these numerical results to the true conditions prevailing in the giant planets requires an extrapolation over at least 6 orders of magnitude \citep{Showman+2011}. The second possibility (not exclusive) is that winds are driven from the top by the injection of turbulence at the cloud level, which can also lead to the correct winds for the four giant planets \citep{LianShowman2010,LiuSchneider2011}. 

Information on the gravity field of the planets can be used to constrain the interior rotation profile, as in the case of Uranus and Neptune whose observed jets appear to only extend to the outer 0.15\% and 0.20\% of the mass (corresponding to pressures of 2 and 4\,kbar) for Uranus and Neptune, respectively \citep{Kaspi+2013}. The method is promising with the perspective of the Juno measurements at Jupiter \citep{Liu+2013}. 

Our giant planets also exhibit planetary-scale to small-scale storms
with very different temporal variations. For example, Jupiter's great
red spot is a 12000\,km-diameter anticyclone found to have lasted for
at least 300 years \citep[e.g.][]{SimonMiller+02}. Storms developing
over the entire planet have even been observed on Saturn
\citep{SanchezLavega+96}. Uranus and Neptune's storm system has been
shown to have been significantly altered since the Voyager era
\citep{Rages+02,Hammel+05,dePater+2011}. On Jupiter, small-scale storms related to
cumulus-type cloud systems have been observed
\citep[e.g.,][]{Gierasch+00,HSG02}, and lightning strikes have been
monitored by Galileo \citep[e.g.,][]{Little+99}. These represent only a
small arbitrary subset of the work concerning the complex atmospheres
of these planets.

It is tempting to extrapolate these observations to the objects
outside our Solar System as well. However, two features governing the weather in these are not necessarily present for exoplanets \citep[e.g.,][]{Guillot99b}: their rapid rotation, and the presence of abundant condensing species and in particular one, water, whose latent heat can fuel powerful storms. But as we will see briefly in section~\ref{sec:exoplanets} theoretical models for exoplanets are now complemented by measurements of wind speeds and of global temperature contrasts, offering the perspective of a global approach to planetary weather and atmospheric dynamics.

\subsection{Energy balance and atmospheric temperature profiles}

Jupiter, Saturn and Neptune are observed to emit more
energy than they receive from the Sun (see Table~\ref{tab:flux}). The
case of Uranus is less clear. Its intrinsic heat flux $F_{\rm int}$ is
significantly smaller than that of the other giant planets. 
With this caveat, all four giant planets can be said to emit more energy
than they receive from the Sun.  \citet{Hubbard68} showed in the case of
Jupiter that this can be explained simply by the progressive
contraction and cooling of the planets.

\begin{table}[htb]
\begin{center}
\caption{Energy balance as determined from Voyager IRIS data\ti{a}.}
\label{tab:flux}
\small
\begin{tabular}{l r@{$\pm$}l r@{$\pm$}l r@{$\pm$}l r@{$\pm$}l} \hline \hline
	& \multicolumn{2}{c}{\bf Jupiter} &
\multicolumn{2}{c}{\bf Saturn} &\multicolumn{2}{c}{\bf Uranus} 
&\multicolumn{2}{c}{\bf Neptune}  \\
\hline 
Absorbed power [$10^{16}$\, J\,s$^{-1}$] & 50.14&2.48 &
11.14&0.50 & 0.526&0.037 & 0.204&0.019 \\
Emitted power [$10^{16}$\, J\,s$^{-1}$] & 83.65&0.84 &
19.77&0.32 & 0.560&0.011 & 0.534&0.029 \\
Intrinsic power [$10^{16}$\, J\,s$^{-1}$]&
33.5& 2.6 & 8.63&0.60 & 
\multicolumn{2}{c}{0.034\,\parbox{2em}{\tiny $+0.038$ $-0.034$}} 
& 0.330& 0.035 \\
Intrinsic flux [J\,s$^{-1}$\,m$^{-2}$] & 5.44&
0.43 & 2.01& 0.14 & \multicolumn{2}{c}{0.042\,\parbox{1.2em}{\tiny
$+0.047$ $-0.042$}} & 0.433& 0.046 \\
Bond albedo [] & 0.343&0.032 & 0.342&0.030 & 0.300&0.049 & 0.290&0.067 \\
Effective temperature [K] & 124.4& 0.3 & 95.0&
0.4 & 59.1& 0.3 & 59.3& 0.8 \\ 
1-bar temperature\ti{b} [K] & 165&5 & 135&5 & 76&2 & 72&2 \\
\hline\hline 
\multicolumn{9}{l}{\ti{a} After \citet{PC91}} \\
\multicolumn{9}{l}{\ti{b} \citet{Lindal92}} 
\end{tabular}
\normalsize
\end{center}
\end{table}

A crucial consequence of the presence of an intrinsic heat flux is
that it requires high internal temperatures ($\sim 10,000\K$ or more),
and that consequently the giant planets are {\it fluid} (not solid)
(\cite{Hubbard68}; see also \cite{Hubbard+95}). Another consequence is
that they are essentially
convective, and that their interior temperature profile are close to
{\it adiabats}. We will come back to this in more detail. 

\begin{figure}[htb]
\resizebox{12cm}{!}{\includegraphics{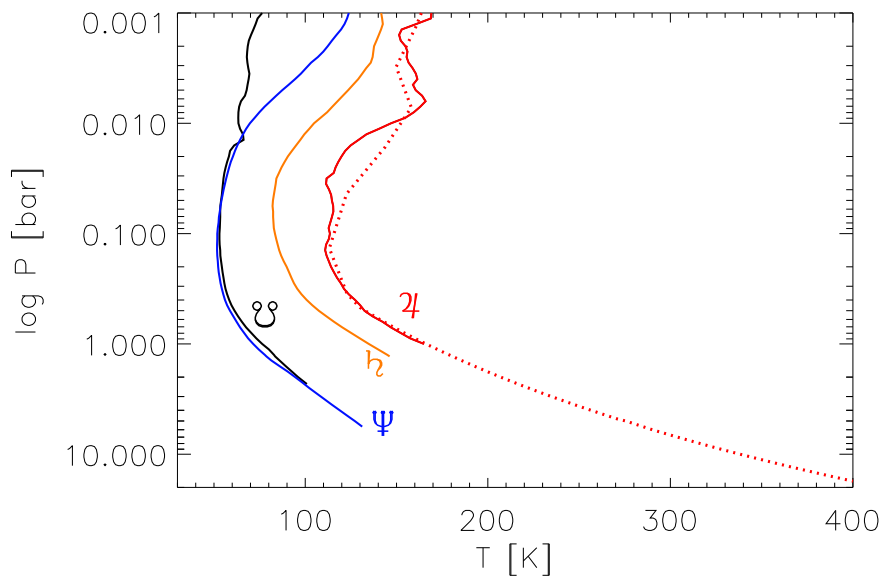}}
\caption{Atmospheric temperatures as a function of pressure for
Jupiter, Saturn, Uranus and Neptune, as obtained from Voyager
radio-occultation experiments \citep[see][]{Lindal92}. The dotted line
corresponds to the temperature profile retrieved by the Galileo probe,
down to 22\bar\ and a temperature of $428\K$ \citep{Seiff+98}.}
\label{fig:atm_temp}
\end{figure}

The deep atmospheres (more accurately tropospheres) of the four giant
planets are indeed observed to be close to adiabats, a result first
obtained by spectroscopic models \citep{Trafton67}, then verified by
radio-occultation experiments by the Voyager spacecrafts, and by the
{\it in situ} measurement from the Galileo probe
(fig.~\ref{fig:atm_temp}). The temperature profiles show a temperature
minimum, in a region near 0.2\bar\, called the tropopause. At higher
altitudes, in the stratosphere, the temperature gradient is negative
(increasing with decreasing pressure). In the regions that we will be
mostly concerned with, in the troposphere and in the deeper interior,
the temperature always increases with depth. It can be noticed that
the slope of the temperature profile in fig.~\ref{fig:atm_temp} becomes
almost constant when the atmosphere becomes convective, at pressures
of a fraction of a bar, in the four giant planets.

It should be noted that the 1 bar temperatures listed in
table~\ref{tab:flux} and the profiles shown in fig.~\ref{fig:atm_temp}
are retrieved from radio-occultation measurements using a helium to
hydrogen ratio which, at least in the case of Jupiter and Saturn, was
shown to be incorrect. The new values of $Y$ are found to lead to
increased temperatures by $\sim 5\K$ in Jupiter and $\sim 10\K$ in
Saturn \citep[see][]{Guillot99a}. However, the Galileo probe found a 1 bar
temperature of $166\K$ \citep{Seiff+98}, and generally a good
agreement with the Voyager radio-occultation profile with the wrong
He/H$_2$ value.

When studied at low spatial resolution, it is found that all four
giant planets, in spite of their inhomogeneous appearances, have a
rather uniform brightness temperature, with pole-to-equator
latitudinal variations limited to a few kelvins
\citep[e.g.,][]{Ingersoll+95}. However, in the case of Jupiter, some
small regions are known to be very different from the average of the
planet. This is the case of hot spots, which cover about 1\% of the
surface of the planet at any given time, but contribute to most of the
emitted flux at 5 microns, due to their dryness (absence of water
vapor) and their temperature brightness which can, at this wavelength,
peak to 260\K.

\subsection{Atmospheric compositions}\label{sec:compositions}

In fluid planets, the distinction between the atmosphere and the
interior is not obvious. We name ``atmosphere'' the part of the planet
which can directly exchange radiation with the exterior
environment. This is also the part which is accessible by remote
sensing. It is important to note that the continuity between the
atmosphere and the interior does not guarantee that compositions
measured in the atmosphere can be extrapolated to the deep interior,
even in a fully convective environment: Processes such as phase
separations \citep[e.g.,][]{Salpeter73, SS77a, FH03}, phase transitions
\citep[e.g.,][]{1989oeps.book..539H}, chemical reactions
\citep[e.g.,][]{FL94} and cloud formation \citep[e.g.][]{Rossow1978} can occur and decouple the surface
and interior compositions. Furthermore, imperfect mixing may also
occur, depending on the initial conditions
\citep[e.g.,][]{1985Icar...62....4S}.

The conventional wisdom is however that these processes are limited to
certain species (e.g. helium) or that they have a relatively small
impact on the global abundances, so that the hydrogen-helium envelopes
may be considered relatively uniform, from the perspective of the
global abundance in heavy elements. An important caveat is that measurements must probe deeper than the condensation altitude for any volatile (e.g. ammonia, water, etc.).
 We first discuss measurements 
made in the atmosphere before inferring interior compositions from
interior and evolution models. 

\subsubsection{Hydrogen and helium}\label{sec:hhe}

The most important components of the atmospheres of our giant planets
are also among the most difficult to detect: H$_2$ and He have a zero
dipolar moment and hence absorb very inefficiently visible and
infrared light. Absorption in the infrared becomes important only at
high pressures as a result of collision-induced absorption
\citep[e.g.,][]{1997AA...324..185B}. On the other hand, lines due to electronic transitions
correspond to very high altitudes in the atmosphere, and bear little
information on the structure of the deeper levels.  The only robust
result concerning the abundance of helium in a giant planet is by {\it
in situ} measurement by the Galileo probe in the atmosphere of Jupiter
\citep{1998JGR...10322815V}. The helium mole fraction (i.e., number of
helium atoms over the total number of species in a given volume) is
$q_{\rm He}=0.1359\pm 0.0027$. The helium mass mixing ratio $Y$ (i.e.,
mass of helium atoms over total mass) is constrained by its ratio over
hydrogen, $X$: $Y/(X+Y)=0.238\pm 0.05$. This ratio is by coincidence
that found in the Sun's atmosphere, but because of helium
sedimentation in the Sun's radiative zone, it was larger in the
protosolar nebula: $Y_{\rm proto}=0.275\pm 0.01$ and $(X+Y)_{\rm
proto}\approx 0.98$ \citep[e.g.,][]{1995RvMP...67..781B}. 
Less helium is therefore found in the atmosphere
of Jupiter than inferred to be present when the planet formed. We will
discuss the consequences of this measurement later: let us mention
that the explanation invokes helium settling due to a phase separation
in the interiors of massive and cold giant planets.

Helium is also found to be depleted compared to the protosolar value
in Saturn's atmosphere. However, in this case the analysis is
complicated by the fact that Voyager radio occultations combined with the far-IR sounding (to separate effects of helium from that of temperature) led
to a wrong value for Jupiter when compared to the Galileo probe data and hence are suspect for the other planets. The current adopted value from IR data only is now $Y=0.18-0.25$
\citep{ConrathGautier2000}, in agreement with values predicted by
interior and evolution models \citep{Guillot99b,Hubbard+99}. 

Finally, as shown in table~\ref{tab:comp} hereafter, Uranus and Neptune are found to have near-protosolar helium
mixing ratios, but with considerable uncertainty.

\subsubsection{Heavy elements}\label{sec:heavies}

\begin{table}[htbp]
\begin{center}
\parbox{12cm}{
\caption{Elemental abundances measured in the tropospheres 
of giant planets}\label{tab:comp}
}
\small
\begin{tabular}{lllllll} \hline\hline\vspace{.8ex}
 & {\bf Element} & {\bf Carrier} & {\bf Abundance ratio/H}$^\dagger$ & {\bf Protosun}$^a$ & $\frac{\mbox{\bf Planet}}{\mbox{\bf Protosun}}$ & {\bf Method}
\vspace{.3ex}\\ \hline
\multicolumn{7}{l}{\bf Jupiter}\\
&  He/H & He & $(7.85\pm 0.18)\times 10^{-2}$ & $9.69\times 10^{-2}$ & $0.810\pm 0.019$ & Galileo/GPMS $^b$\\
& C/H & CH$_4$ & $(1.185\pm 0.019)\times 10^{-3}$ & $2.75\times 10^{-4}$ & $4.31\pm 0.07$ & Galileo/GPMS$^c$ \\
& N/H & NH$_3$ & $(3.3\pm 1.3)\times 10^{-4}$ & $8.19\times 10^{-5}$ & $4.05\pm 1.55$ &Galileo/GPMS$^c$\\
& O/H & H$_2$O$^\star$ & $(1.49^{+0.98}_{-0.68})\times 10^{-4}$ & $6.06\times 10^{-4}$ & $0.25^{+0.16}_{-0.11}$ & Galileo/GPMS@19 bar$^c$ \\
& S/H & H$_2$S & $(4.5\pm 1.1)\times 10^{-5}$ & $1.55\times 10^{-5}$ & $2.88\pm 0.68$ & Galileo/GPMS$^c$\\
& Ne/H & Ne & $(1.20\pm 0.12)\times 10^{-5}$ & $1.18\times 10^{-4}$ & $0.10\pm 0.01$ & Galileo/GPMS$^d$\\
& Ar/H & Ar & $(9.10\pm 1.80)\times 10^{-6}$ & $3.58\times 10^{-6}$ & $2.54\pm 0.50$ & Galileo/GPMS$^d$\\
& Kr/H & Kr & $(4.65\pm 0.85)\times 10^{-9}$ & $2.15\times 10^{-9}$ & $2.16\pm 0.40$ & Galileo/GPMS$^d$\\
& Xe/H & Xe & $(4.45\pm 0.85)\times 10^{-10}$ & $2.11\times 10^{-10}$ & $2.11\pm 0.40$ & Galileo/GPMS$^d$\\
& P/H & PH$_3$$^\star$ & $(1.11\pm 0.06)\times 10^{-6}$ & $3.20\times 10^{-7}$ & $3.45\pm 0.18$ & Cassini/CIRS$^e$ \\
& Ge/H & GeH$_4$$^\star$ & $(4.1\pm 1.2)\times 10^{-10}$ & $4.44\times 10^{-9}$ & $0.09\pm 0.03$ & Voyager/IRIS$^f$ \\
& As/H & AsH$_3$$^\star$ & $(1.3\pm 0.6)\times 10^{-10}$ & $2.36\times 10^{-10}$ & $0.54\pm 0.27$ & Ground/IR$^g$ \\
\smallskip\\
\multicolumn{7}{l}{\bf Saturn}\\
 & He/H & He & $(6.75\pm 1.25)\times 10^{-2}$ & $9.69\times 10^{-2}$ & $0.70\pm 0.13$ & Voyager/IRIS$^h$\\
& C/H & CH$_4$ & $(2.67\pm 0.11)\times 10^{-3}$ & $2.75\times 10^{-4}$ & $9.72\pm 0.41$ & Cassini/CIRS$^i$ \\
& N/H & NH$_3$$^\star$ & $(2.27\pm 0.57)\times 10^{-4}$ & $8.19\times 10^{-5}$ & $2.77\pm 0.69$ & Cassini/VIMS$^j$ \\
& S/H & H$_2$S & $(1.25\pm 0.17)\times 10^{-4}$ & $1.55\times 10^{-5}$ & $8.08\pm 1.10$ & Ground/radio$^k$ \\
& P/H & PH$_3$$^\star$ & $(4.65\pm 0.32)\times 10^{-6}$ & $3.20\times 10^{-7}$ & $14.5\pm 1.0$ & Cassini/CIRS$^e$\\
& & & $(1.76\pm 0.17)\times 10^{-6}$ & $3.20\times 10^{-7}$ & $5.49\pm 0.53$ & Cassini/VIMS$^i$\\
& & & $(4.0^{+1.7}_{-1.1})\times 10^{-6}$ & $3.20\times 10^{-7}$ & $12.4^{+5.3}_{-3.5}$ &Ground/IR$^l$\\
& Ge/H & GeH$_4$$^\star$ & $(2.3\pm 2.3)\times 10^{-10}$ & $4.44\times 10^{-9}$ & $0.05\pm 0.05$ & Ground/IR$^l$ \\
& As/H & AsH$_3$$^\star$ & $(1.25\pm 0.17)\times 10^{-9}$ & $2.36\times 10^{-10}$ & $5.33\pm 0.73$ &  Cassini/VIMS$^i$\\
& & & $(1.71\pm 0.57)\times 10^{-9}$ & $2.36\times 10^{-10}$ & $7.3\pm 2.4$ & Ground/IR$^l$
\smallskip\\
\multicolumn{7}{l}{\bf Uranus}\\
 & He/H & He & $(9.0\pm 2.0)\times 10^{-2}$ & $9.69\times 10^{-2}$ & $0.93\pm 0.20$ & Voyager/IRIS+occult$^m$\\
& C/H & CH$_4$$^\star$  & $(2.36\pm 0.30)\times 10^{-2}$ & $2.75\times 10^{-4}$ & $85.9\pm 10.7$ & Hubble/STIS$^n$\\
& S/H & H$_2$S$^\star$  & $(3.2\pm 1.6)\times 10^{-4}$ & $1.55\times 10^{-5}$ & $21.0\pm 10.5$ & Ground/radio$^o$\\
\smallskip\\
\multicolumn{7}{l}{\bf Neptune}\\
& He/H & He & $(1.17\pm 0.20)\times 10^{-1}$ & $9.69\times 10^{-2}$ & $1.21\pm 0.20$ & Voyager/IRIS+occult$^p$\\
& C/H & CH$_4$$^\star$  & $(1.85\pm 0.43)\times 10^{-2}$ & $2.75\times 10^{-4}$ & $67.5\pm 15.8$ & Ground/IR$^q$\\
&         &                          & $(2.47\pm 0.62)\times 10^{-2}$ & $2.75\times 10^{-4}$ & $89.9\pm 22.5$ & Hubble/STIS$^r$\\
& S/H & H$_2$S$^\star$  & $(3.2\pm 1.6)\times 10^{-4}$ & $1.55\times 10^{-5}$ & $21.0\pm 10.5$ & Ground/radio$^o$ \\
\hline\hline
\end{tabular}
\end{center}
\noindent
$^\star$: Species which condense or are in chemical disequilibrium, i.e., with vertical/horizontal variations of their concentration. The global elemental abundances are estimated from the maximum measured mixing ratio, but like in the case of H$_2$O in Jupiter (believed to correspond to the measurement in a dry downdraft), they may only be lower limits to the bulk abundance. \\
$^\dagger$: Abundance ratios $r$ are measured with respect to atomic hydrogen. In these atmospheres dominated by molecular hydrogen and helium, mole fractions $f$ are found by $f=2r/(1+r_{\rm He})$ where $r_{\rm He}$ is the He/H abundance ratio. \\ 
$^a$: protosolar abundances from \cite{Lodders+2009}; $^b$: \cite{VonZahn+1998}; $^c$: \cite{Wong+04}; $^d$: \cite{Atreya+2003}; $^e$: \cite{Fletcher+2009b}; $^f$: \cite{Kunde+1982}; $^g$: \cite{Noll+1990}; $^h$:  \cite{ConrathGautier2000}; $^i$: \cite{Fletcher+2009a}; $^j$: \cite{Fletcher+2011}; $^k$: \cite{BS89}; $^l$: \cite{NollLarson1991}; $^m$: \cite{Conrath+1987}; $^n$:  \cite{Sromovsky+2011}; $^o$:  \cite{dePater+1991}; $^p$:  \cite{Conrath+1991}; $^q$:  \cite{BainesSmith1990}; $^r$:  \cite{KarkoschkaTomasko2011}.
\end{table}

The abundance of elements other than hydrogen and helium (that we will
call hereafter ``heavy elements'') bears crucial information for the
understanding of the processes that led to the formation of these
planets. Table~\ref{tab:comp} summarizes the present situation after {\em in situ} measurements in Jupiter by the Galileo probe, as well as spectroscopic measurements from spacecraft and from the ground for the other planets. 

The elemental abundances in the giant planets' atmospheres are most usefully compared to those in the Sun since they all originated from the protosolar disk. The solar abundances have seen very significant revisions in the past decade because it has been realized that convective motions in the Sun's atmosphere affect spectral lines more extensively than was previously thought. It is not yet clear at this date whether the solar abundances have converged. Furthermore, as discussed for helium, heavy elements gradually settle towards the Sun's interior so that a proper reference for the giant planets is not the solar atmosphere today, but its value 4.5 billion years ago which is model-dependent. Table~\ref{tab:comp} provides the values obtained for the protosun by \cite{Lodders+2009}. These are used as reference without accounting for their uncertainties. 

The most abundant heavy elements in the envelopes of our four giant
planets are O (presumably) and C, N and S. It is possible to model the chemistry of gases
in the tropospheres from the top of the convective zone down to the
2000 K temperature level \citep{FL94}. Models conclude
that, whatever the initial composition in these elements of
planetesimals which collapsed with hydrogen onto Jupiter and Saturn
cores during the last phase of the planetary formation, C in the upper
tropospheres of giant planets is mainly in the form of gaseous CH$_4$,
N in the form of NH$_3$, S in the form of H$_2$S, and O in the form of
H$_2$O. All these gases but methane in Jupiter and Saturn condense
in the upper troposphere and vaporize at deeper levels when the
temperature increases. Noble gases do not
condense even at the tropopauses of Uranus and Neptune, the coldest regions in these atmospheres. 

Jupiter is the planet which has been best characterized thanks to the measurements of the Galileo atmospheric probe which precisely measured the abundances of He, Ne, Ar, Kr, Xe, CH$_4$, NH$_3$, H$_2$S, and H$_2$O down to pressures around 22\,bars. As helium, neon was found to be depleted compared to the protosolar value, in line with theoretical predictions that this atom would fall in with the helium droplets \citep[][and section~\ref{sec:others}]{RS95,WilsonMilitzer2010}. C, N, and S were found to be
supersolar by a factor 2.5 to 4.5 \citep{Wong+04}, which was not
unexpected because condensation of nebula gases results in enriching
icy grains and planetesimals. The surprise came from Ar, Kr, Xe, which
were expected to be solar because they are difficult to condense, but
turned out to be supersolar by a factor $\sim 2$ \citep{Owen+99,Wong+04}.

H$_2$O is difficult to measure in all four giant planets because of
its condensation relatively deep. It was hoped that the Galileo probe
would provide a measurement of its deep abundance, but the probe fell
into one of Jupiter's 5-micron hot spots, now believed to be a
dry region mostly governed by downwelling motions
\citep[e.g.,][]{SI98}. As a result, and although the probe provided
measurements down to 22 bars, well below water's canonical 5 bar cloud
base, it is believed that this measurement of a water abundance equal
to a fraction of the solar value is only a lower limit. An indirect determination comes from the measurement of the disequilibrium species CO which has to be transported fast from the deep levels where H$_2$O and CH$_4$ tend to form more CO (and H$_2$). This predicts a mostly solar to slightly supersolar (by a factor 2) abundance of O in Jupiter \citep{VisscherMoses2011} and much larger enrichments in Neptune \citep{LF94}. This however depends crucially on the reaction network and somewhat on assumptions on mixing, both of which are not well known. The abundance of oxygen, the most abundant element in the Universe after hydrogen and helium and a crucial planetary building block is essentially unknown for what concerns our four giant planets. 

Three other species can help us probe the bulk elemental abundance inside Jupiter, although with larger difficulties perhaps because they are not necessarily in chemical equilibrium at the levels where they are detected and their measured abundances are thus not necessarily representative of their bulk abundance: these are PH$_3$, GeH$_4$ and AsH$_3$. All three where detected remotely rather than in situ. The first one is clearly supersolar in Jupiter, with an enrichment in between that measured for C and S. GeH$_4$ is clearly subsolar, but this is not surprising because of condensation into solid Ge and GeS \citep{FL94}. The same chemical models would predict that the measured abundance of AsH$_3$ should be close to its bulk abundance. The measured abundance therefore could be interpreted as a subsolar bulk abundance of As, but with considerable uncertainty. 

Table~\ref{tab:comp} shows that Saturn's atmosphere is more enriched in heavy elements than Jupiter. Unfortunately, unlike Jupiter, no in situ measurement has been performed in this planet and we can only rely on remote sensing. But we can confidently assess that the abundances of C, S, P and As are significantly higher than in Jupiter. Saturn has about twice more C (as CH$_4$) and S (as H$_2$S) than Jupiter, for a given mass of atmosphere. The situation for PH$_3$ is unclear, both because it is highly variable both vertically and latitudinally: the enrichment could be only slightly more than in Jupiter to more than 4 times that value \citep{Fletcher+2009b, Fletcher+2011}. Note that in the presence of horizontal variability (as for this molecule) table~\ref{tab:comp} indicates the maximum abundances measured - which should be closer to the bulk abundance, except if there exist mechanisms to preferentially trap certain species. The enrichment in N (as NH$_3$) is smaller than in Jupiter. This may be due to its condensation deeper as NH$_4$SH \citep{GJO78}, although it does not explain why Jupiter and Saturn would be that different in that respect. The enrichment in As (as AsH$_4$) is considerably larger than in Jupiter, which is also a mystery \citep{Fletcher+2009a}. At least, GeH$_4$ appears to be of equally low abundance in both planets, but this is probably more related to its condensation than to its bulk abundance. 

Finally, Uranus and Neptune provide all signs of a significant enrichment in heavy elements, even though very few elements have been detected. Methane is the most important one, although the fact that it condenses in these planets complicates the interpretation of the spectroscopic measurements. Large-scale variations with latitude are observed, in particular in Uranus \citep{Sromovsky+2011}, but less so in Neptune \citep{KarkoschkaTomasko2011}. However, the deep abundances provided in table~\ref{tab:comp} are very similar for both planets, with a $\sim 90$ times solar enrichment. This is much higher than the $\sim 30$ times solar enrichments discussed in past reviews \citep[e.g.,][]{Gautier+95} for two reasons: one is a decrease of the protosolar abundance itself. The other is the fact that it relied on spectroscopic measurements probing higher atmospheric levels affected by methane condensation \citep[see][]{KarkoschkaTomasko2011}. 

The other key species detected in Uranus and Neptune thanks to ground-based radio observations is H$_2$S, which points to a $10$ to $30$ enrichment in sulfur \citep{dePater+1991}. Because this element condenses at even greater depths than methane, the bulk abundance of S in these planets may be larger if the global circulation is indeed important down to the deep levels probed by the radio waves. The measurement is however a difficult one, with other potential absorbers affecting the smooth microwave spectra yielding degenerate solutions. 

Overall, the global picture that can be drawn is that of an increase of the abundance of heavy elements compared to the solar value with increasing distance to the Sun, from Jupiter which shows a $\sim2$ to 4 enrichment, Saturn a $3$ to $10$ one, and Uranus and Neptune which are enriched by a factor $\sim 90$ in carbon and by at least 10 to 30 in sulfur. In spite of their different atmospheric dynamics, and with the present accuracy of the measurements, the two ice giants have very similar abundances.

\subsection{Isotopic ratios}

\begin{table}[htbp]
\begin{center}
\parbox{12cm}{
\caption{Isotopic ratios measured in the tropospheres 
of giant planets}\label{tab:isotopes}
}
\small
\begin{tabular}{llllll} \hline\hline\vspace{.3ex}
 & {\bf Isotope} & {\bf Isotopic ratio} & {\bf Protosun}$^a$ & {\bf Planet/Protosun} & {\bf Comments}
\vspace{.3ex}\\ \hline
\multicolumn{6}{l}{\bf Jupiter}\\
& D/H & $(2.25\pm 0.35)\times 10^{-5}$&$1.94\times 10^{-5}$&$1.16\pm0.18$& ISO/SWS$^q$ \\
&$^3$He/$^4$He&$(1.66\pm 0.06)\times 10^{-4}$&$1.66\times 10^{-4}$&$1.00\pm0.03$ &Galileo/GPMS$^d$\\
&$^{13}$C/$^{12}$C&$(1.08\pm 0.05)\times 10^{-2}$&$1.12\times 10^{-2}$&$1.04\pm 0.05$&Galileo/GPMS$^d$\\
&$^{15}$N/$^{14}$N&$(2.30\pm 0.30)\times 10^{-3}$&$2.27\times 10^{-3}$&$0.99\pm0.13$&Galileo/GPMS$^d$\\
&$^{22}$Ne/$^{20}$Ne&$(7.7\pm 1.2)\times 10^{-2}$&$7.35\times 10^{-2}$&$0.96\pm 0.15$ &Galileo/GPMS$^d$\\
&$^{38}$Ar/$^{36}$Ar&$(1.79\pm 0.08)\times 10^{-1}$&$1.82\times 10^{-1}$&$1.02\pm 0.05$&Galileo/GPMS$^d$\\
&$^{128}$Xe/Xe&$(1.80\pm 0.20)\times 10^{-2}$&$2.23\times 10^{-2}$&$1.24\pm 0.14$&Galileo/GPMS$^d$\\
&$^{129}$Xe/Xe&$(2.85\pm 0.21)\times 10^{-1}$&$2.75\times 10^{-1}$&$0.96\pm 0.07$&Galileo/GPMS$^d$\\
&$^{130}$Xe/Xe&$(3.80\pm 0.50)\times 10^{-2}$&$4.38\times 10^{-2}$&$1.15\pm 0.15$&Galileo/GPMS$^d$\\
&$^{131}$Xe/Xe&$(2.03\pm 0.18)\times 10^{-1}$&$2.18\times 10^{-1}$&$1.07\pm 0.10$&Galileo/GPMS$^d$\\
&$^{132}$Xe/Xe&$(2.90\pm 0.20)\times 10^{-1}$&$2.64\times 10^{-1}$&$0.91\pm 0.06$&Galileo/GPMS$^d$\\
&$^{134}$Xe/Xe&$(9.10\pm 0.70)\times 10^{-2}$&$9.66\times 10^{-2}$&$1.06\pm 0.08$&Galileo/GPMS$^d$
\smallskip\\
\multicolumn{6}{l}{\bf Saturn}\\
 &D/H&$(1.60\pm 0.20)\times 10^{-5}$&$1.94\times 10^{-5}$&$0.83\pm 0.11$&Cassini/CIRS$^h$ \\
 & &$(1.70^{+0.75}_{-0.45})\times 10^{-5}$&$1.94\times 10^{-5}$&$0.88^{+0.39}_{-0.23}$& ISO/SWS$^q$ \\
&$^{13}$C/$^{12}$C&$(1.09\pm 0.10)\times 10^{-2}$&$1.12\times 10^{-2}$&$1.03\pm 0.09$&Cassini/CIRS$^h$
\smallskip\\
\multicolumn{6}{l}{\bf Uranus}\\
&D/H&$(4.40\pm 0.40)\times 10^{-5}$&$1.94\times 10^{-5}$&$2.27\pm 0.21$&Herschel/PACS$^r$
\smallskip\\
\multicolumn{6}{l}{\bf Neptune}\\
&D/H&$(4.10\pm 0.40)\times 10^{-5}$&$1.94\times 10^{-5}$&$2.11\pm 0.21$&Herschel/PACS$^r$  \\
\hline\hline
\end{tabular}
\end{center}
\noindent
$^a$: protosolar abundances from \cite{Lodders+2009}, except $^{15}$N/$^{14}$N which is corrected by \cite{Marty+2011}; $^d$: \cite{Atreya+2003}; $^h$: \cite{Fletcher+2009a}; $^q$: \cite{Lellouch+2001}; $^r$: \cite{Feuchtgruber+2013}. 
\end{table}

The measurement of isotopic ratios in planetary atmospheres is a powerful tool to understand their origin. Table~\ref{tab:isotopes} provides the ensemble of isotopic ratios measured in our giant planets, and a comparison to their values in the Sun. 

Of course, the Galileo probe and its onboard mass spectrometer have provided us a strikingly clear picture of Jupiter's atmosphere: it is directly formed from the same material as our Sun, with isotopic ratios which are, to the accuracy of the measurements, indistinguishable (i.e., within 2 sigma) from the solar values and for elements as diverse as D, He, C, N, Ne, Ar and Xe with as many as 6 isotopes measured. This was expected because indeed Jupiter's composition is globally similar to that of the Sun, but given the fact that the abundances of elements are far from being Sun-like, it is perhaps surprising to find such a good match! By extension, this applies to Saturn although only the deuterium to hydrogen and $^{13}$C/$^{12}$C isotopic ratios could be measured by remote spectroscopic observations. This confirms that the atmospheres and envelopes of Jupiter and Saturn originated from the same material that formed the Sun and that mechanisms leading to isotopic fractionation (e.g., atmospheric evaporation) were of limited importance. 

In the case of Uranus and Neptune, only the deuterium to hydrogen ratio was measured, from the ground in the infrared \citep{Irwin+2014} and most precisely by recent far infrared spectroscopy from Herschel \citep{Feuchtgruber+2013}. Interestingly, it is about twice larger than the protosolar value, and a factor 2 to 6 times smaller than the D/H value in comets. Given our present knowledge of the interiors of Uranus and Neptune, \citet{Feuchtgruber+2013} conclude that either these planets contain much more rocks than expected or that the ices in their interior have not been fully mixed.

\subsection{Moons and rings}

A discussion of our giant planets motivated by the opportunity to
extrapolate the results to objects outside our solar system would be
incomplete without mentioning the moons and rings that these planets
all possess (see chapters by Breuer \& Moore, by Peale \& Canup and by Hussmann
et al.). First, the satellites/moons can be distinguished from their
orbital characteristics as regular or irregular. The first ones have
generally circular, prograde orbits. The latter tend to have
eccentric, extended, and/or retrograde orbits. 

These satellites are numerous: After the Voyager era, Jupiter was
known to possess 16 satellites, Saturn to have 18, Uranus 20 and
Neptune 8.  Recent extensive observation programs have seen the number
of satellites increase considerably, with a growing list of satellites
presently reaching 62, 56, 27 and 13 for Jupiter, Saturn, Uranus and
Neptune, respectively. All of the new satellites discovered since
Voyager are classified as irregular.

The presence of regular and irregular satellites is due in part to the 
history of planet formation. It is believed that the regular
satellites have mostly been formed in the protoplanetary
subnebulae that surrounded the giant planets (at least Jupiter and
Saturn) at the time when they accreted their envelopes. On the other
hand, the irregular satellites are thought to have been captured by
the planet. This is, for example, believed to be the case of Neptune's
largest moon, Triton, which has a retrograde orbit. 

A few satellites stand out by having relatively large masses: it is
the case of Jupiter's Io, Europa, Ganymede and Callisto, of Saturn's
Titan, and of Neptune's Triton. Ganymede is the most massive of them,
being about twice the mass of our Moon. However, compared to the mass
of the central planet, these moons and satellites have very small
weights: $10^{-4}$ and less for Jupiter, $1/4000$ for Saturn,
$1/25000$ for Uranus and $1/4500$ for Neptune. All these satellites
orbit relatively closely to their planets. The farthest one,
Callisto revolves around Jupiter in about 16 Earth days. 

The four giant planets also have rings, whose material is probably 
constantly resupplied from their satellites. The ring of Saturn stands
out as the only one directly visible with binoculars. In this
particular case, its enormous area allows it to reflect a sizable
fraction of the stellar flux arriving at Saturn, and makes this
particular ring as bright as the planet itself. The occurrence of such
rings would make the detection of extrasolar planets slightly easier,
but it is yet unclear how frequent they can be, and how close to the
stars rings can survive both the increased radiation and tidal
forces.

 \subsection{Seismology}

The best way to directly probe planetary (or stellar) interiors is through seismology, i.e., by measuring the spectrum of waves propagating through the interior. Our knowledge of the interior structure of the Earth, the Sun and even other stars is largely due to the ability to detect the oscillations of these objects. Because Jupiter (and by extent the other giant planets in our Solar System) are similar in composition and density to the Sun, the possibility to detect pressure waves in their atmosphere has been proposed in the 1970's. In particular, \cite{VZL76} showed that waves with periods smaller than about 10 minutes would be trapped and reflected downwards in Jupiter's and Saturn's atmospheres, creating the possibility for resonant waves similar to those observed in the Sun to exist. On the theoretical side, the possibility of how these waves may be excited has however remained problematic \citep[e.g.,][]{BS87}. 

After two decades of promising but slow progress, the case for the existence of detectable free oscillations of giant planets has recently taken a new turn. First, using ground-based Doppler imaging of Jupiter, \cite{Gaulme+11} detected an oscillation pattern with frequencies between 0.8 and 2 mHz and a characteristic spacing of the peak of $155.3\pm 2.2\,\mu$Hz, in agreement with theoretical models. Separately, \cite{HedmanNicholson2013} confirmed that waves observed by the Cassini spacecraft in Saturn's rings cannot be caused by satellites and therefore must result from oscillations in Saturn, as had been proposed by \cite{MarleyPorco1993}. 

Further observations are required to better characterize the oscillations of these planets and start using them as probes of the planetary interiors \cite[e.g.,][]{Jackiewicz+2012}. This is very promising however and should lead to a revolution in our understanding of the giant planets. 


\subsection{Exoplanets}\label{sec:exoplanets}

\begin{figure}[htbp]
\resizebox{16cm}{!}{\includegraphics{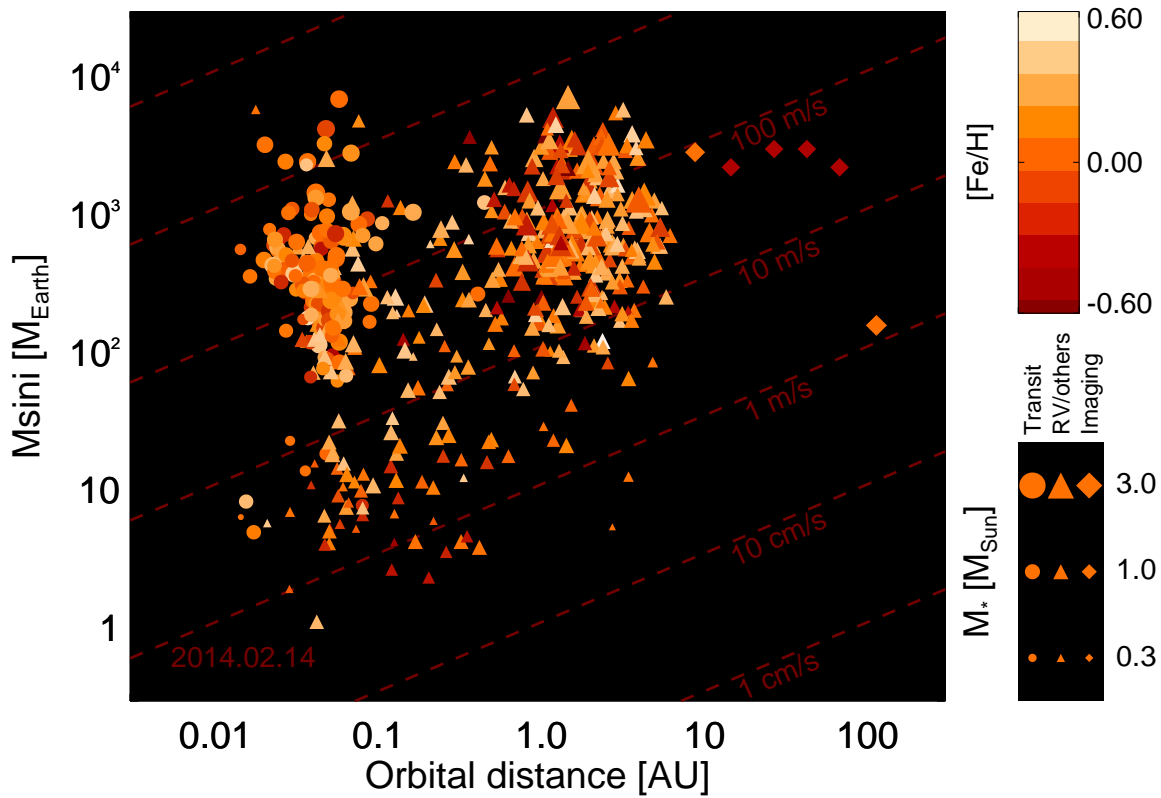}}
\caption{Masses and orbital distances of the extrasolar planets discovered by early 2014. Planets detected in transit are shown with circles, those detected by direct imaging with diamonds. All other systems (mostly detected by radial velocimetry) are shown as triangles. In the case of planets discovered by radial velocimetry, the displayed masses correspond to masses multiplied by the sine of the unknown inclination.  The size of the symbols is proportional to the mass of the parent star (from 0.1 to 3.1 stellar masses). The color (from reddish to white) is proportional to the stellar metallicity (from ${\rm [Fe/H]}=-0.8$ to $0.6$). The radial velocimetry thresholds from 1 cm/s to 10 km/s are  indicated as dashed lines.}
\label{fig:exo}
\end{figure}

Huge progress has been made in the field of extrasolar planets
since the detection of the first giant planet orbiting a solar-type
star by \citet{MQ95}. As shown in figure~\ref{fig:exo}, more than a thousand planets are known at the
time of this review, and importantly, more than four hundred planets that transit their
star at each orbital revolution have been identified \citep{Wright+11,Schneider+11}. These transiting planets are especially
interesting because of the possibility to measure both their mass and
size and thus obtain constraints on their global composition.

In spite of their particular location just a few stellar radii away
from their stars, the transiting giant planets that have been discovered bear
some resemblance with their Solar System cousins in the sense that
they are also mostly made of hydrogen and helium
\citep[e.g.,][]{Burrows+00, Guillot05, Baraffe+05}. They are, however,
much hotter due to the intense irradiation that they receive. 

Although obtaining direct information on these planets represents a
great observational challenge, several key steps have been
accomplished: Atomic sodium, predicted to be detectable \citep{SS00},
has indeed been detected by transit spectroscopy\citep{CBNG02} early on, and a tentative abundance measured in planet HD209458\,b: According to \citet{Sing+2009}, it appears to be oversolar by a factor $\sim 2$ at pressures deeper than about $\sim 3\,$mbar and undersolar above that level \citep[see also][]{Vidal-Madjar+2011}. Hydrodynamically escaping
species (including hydrogen, oxygen, carbon, nitrogen and heavier ions) have also been detected around the brightest hot Jupiters \cite[e.g.,][]{VidalMadjar+03,Fossati+10,Bourrier+13}. A theoretical study of the atmospheric dynamics of hot Jupiters \citep{SG02} predicted strong day-night temperature variations (up to 100's K), fast km/s zonal jets and a displacement of the hottest point west of the substellar point. These have been confirmed by observations of transiting and non-transiting planets in the infrared \citep{Harrington+06,Knutson+07} and by doppler-imaging of planetary CO lines \citep{Snellen+10}. 

Unfortunately, the list of chemical species thought to have been detected \citep[see][for a review]{SD10} has dwindled in recent years due to the realization that instrumental effects could mimic spectral signatures \citep[e.g.,][]{Desert+2009,Gibson+2011,Crouzet+2012}, due to new observations with a better instrument \citep[e.g.,][]{Deming+2013} and generally because of the unexpected prevalence of hazes in these close-in exoplanets \citep[e.g.,][]{Sing+2009,Pont+13}. After examination, claims of a high C/O ratio in some of these atmospheres also appear to be highly uncertain \citep{Crossfield+2012}. 

In any case, in spite of the hiccups, there is obviously a big potential for growth in this young field,
and the comparison between fine observations made for giant planets in
our Solar System and the more crude, but also more statistically
significant data obtained for planets around other stars promise to
be extremely fruitful to better understand these objects.

\section{The calculation of interior and evolution models}

\subsection{Basic equations}\label{sec:basic}

The structure and evolution of a
giant planet is governed by the following hydrostatic, thermodynamic,
mass conservation and energy conservation equations:
\begin{eqnarray}
\dpar{P}{r}&=&-\rho g \label{eq:dpdr}\\
{\partial T\over\partial r}&=&{\partial P\over \partial r}{T\over
P}\nabla_T. \label{eq:dtdr}\\
{\partial m\over\partial r}&=&4\pi r^2\rho. \label{eq:dmdr}\\
{\partial L\over\partial r}&=&4\pi r^2\rho \left(\dot{\epsilon}-
T{\partial S\over \partial t}\right),\label{eq:dldr}
\end{eqnarray}
where $P$ is the pressure, $\rho$ the density, and $g=Gm/r^2$ the
gravity ($m$ is the mass, $r$ the radius and $G$ the gravitational
constant). The temperature gradient $\nabla_T\equiv(d\ln T/d\ln P)$
depends on the process by which the internal heat is transported.
$L$ is the intrinsic luminosity, $t$ the time, $S$ the
specific entropy (per unit mass), and $\dot{\epsilon}$ accounts for
the sources of energy due e.g., to radioactivity or more importantly
nuclear reactions. Generally it is a good approximation to assume
$\dot{\epsilon}\sim 0$ for objects less massive than $\sim 13\mjup$,
i.e., too cold to even burn deuterium (but we will see that in certain
conditions this term may be useful, even for low mass planets). 

The boundary condition at the center is trivial: $r=0$; ($m=0$,
$L=0$). The external boundary condition is more difficult to obtain
because it depends on how energy is transported in the atmosphere. One
possibility is to use the Eddington approximation, and to write
\citep[e.g.,][]{Chandrasekhar39}: $r=R$; ($T_0=\teff$,
$P_0=2/3\,g/\kappa$), where $\teff$ is the effective temperature
(defined by $L=4\pi R\sigma\teff^4$, with $\sigma$ being the
Stephan-Boltzmann constant), and $\kappa$ is a mean opacity. Note for
example that in the case of Jupiter $\teff=124$\,K,
$g=26\rm\,m\,s^{-2}$ and $\kappa\approx 5\times 10^{-3}
(P/1\rm\,bar)\,m^2\,kg^{-1}$. This implies $P_0\approx 0.2$\,bar
(20,000\,Pa), which, given the simplicity of the calculation, is surprisingly close to the location of Jupiter's real tropopause where
$T\approx 110$\,K. Actually, the properties of the opacities of important absorbing chemical species like water and their pressure dependence imply that photospheres around 0.1 bar should be common \citep{RobinsonCatling2014}. 

However, the Eddington boundary condition should not be used in the case of irradiated atmospheres because it does not properly account for both the incoming flux (mostly at visible wavelengths for planets around solar-type stars) and the intrinsic flux (in the infrared). The fact that opacities differ at these wavelengths yields the possibility of thermal inversions (higher visible than infrared opacities) or a greenhouse effect (lower visible than infrared opacities) and thus a hotter interior, something that cannot be captured without accounting for the different fluxes. Analytical solutions of the radiative transfer problem exist in the semi-grey case (two opacities for the visible and infrared, respectively) \citep{Hansen2008, Guillot2010}, and can even be extended to include non-grey effects \citep{ParmentierGuillot2014}. Numerical solutions in the non-irradiated, solar-composition case are provided by \citet{Saumon+96}, and for the irradiation levels and compositions relevant for the solar system giant planets by \citet{Fortney+2011}. In that case, a grid is used to relate the atmospheric 
temperature and pressure at a given level to the radius $R$, intrinsic
luminosity $L$ and incoming stellar luminosity $\linc$: $r=R$;
($T_0=T_0(R,L,\linc)$, $P_0=P_0(R,L,\linc)$). $P_0$ is chosen to
satisfy the condition that the corresponding optical depth at that
level should be much larger than unity. 


\subsection{High pressure physics \& equations of state}\label{sec:EOS}

\subsubsection{Hydrogen}

\begin{figure}[htbp]
\resizebox{14cm}{!}{\includegraphics[angle=0]{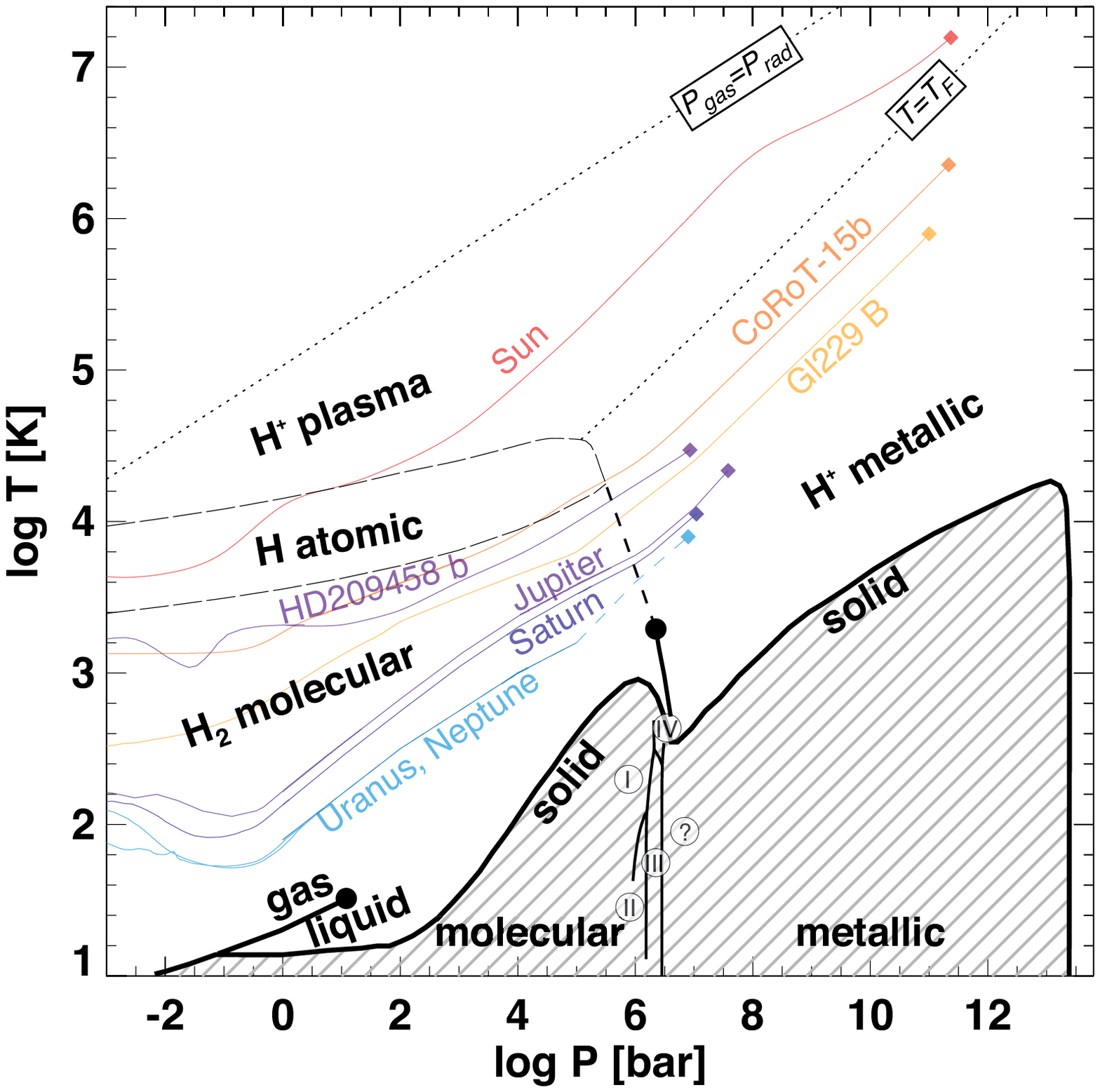}}
\caption{Phase diagram for hydrogen in the pressure-temperature plane, with pressure in bars ($\rm 1\,bar=10^5\,Pa=10^6\,dyn\,cm^{-2}$).  The thick lines indicate the first order (discontinuous) phase transitions, the black circles the critical points \citep{McMahon+2012, Morales+2013a}. Phase transitions in solid hydrogen with the different known phases are labelled I, II, III, IV \citep{McMahon+2012}. The region where 50\% of all hydrogen is in atomic form \citep[from][]{SCvH95} is shown by a thin contour. The approximate location of the molecular to metallic hydrogen (continuous) transition \citep{Loubeyre+2012, Morales+2013a} is indicated by a dashed line. The $T=T_{\rm F}$ line separate the non-degenerate region (at low pressures and high temperatures) from the region in which electrons are degenerate (see text). The $P_{\rm gas}=P_{\rm rad}$ line shows the region which is dominated by radiation pressure (at high temperatures). Colored lines show profiles for a selection of noteworthy substellar objects: giant planets from Uranus to Jupiter \citep{Guillot05}, the hot Jupiter HD~209458~b \citep{GS02, Burrows+2014}, the brown dwarfs Gl~229~B \citep{Marley+1996}, CoRoT-15~b \citep{Bouchy+2011} and our Sun \citep{Christensen-Dalsgaard+1996}. The diamonds correspond to the conditions either at the core/envelope interface for Jupiter and Saturn or at the center for the other objects.}
\label{fig:phase_diag}
\end{figure}

In terms of pressures and temperatures, the interiors of giant planets
lie in a region for which accurate equations of state (EOS) are
extremely difficult to calculate. This is because both molecules,
atoms, and ions can all coexist, in a fluid that is partially degenerate
(free electrons have energies that are determined both by quantum and
thermal effects) and partially coupled (Coulomb interactions
between ions are not dominant but must be taken into account). The
presence of many elements and their possible interactions further
complicate matters. For lack of space, this section will mostly focus
on hydrogen whose EOS has seen the most important developments in
recent years. A phase diagram of hydrogen (fig.~\ref{fig:phase_diag})
illustrates some of the important phenomena that occur in giant
planets.

The photospheres of giant planets are generally relatively cold (50 to
3000\,K) and at low pressure (0.1 to 10\,bar, or $10^4$ to
$10^6$\,Pa), so that hydrogen is in molecular form and the perfect gas
conditions apply.  As one goes deeper into the interior hydrogen and
helium progressively become fluid. (The perfect gas relation tends to
underestimate the pressure by 10\% or more when the density becomes
larger than about $0.02\gcc$ ($P\wig{>} 1\,$kbar in the case of
Jupiter)).

Characteristic interior pressures are considerably larger however: as
implied by Eqs.~\ref{eq:dpdr} and \ref{eq:dmdr}, $P_{\rm c}\approx
GM^2/R^4$, of the order of 10-100\,Mbar for Jupiter and Saturn. As shown in fig.~\ref{fig:phase_diag}, all the central pressures and temperatures of giant planets and brown dwarfs (from Uranus to CoRoT-15b) lie in a regime of high pressures and temperatures lower than the corresponding Fermi temperature $T_{\rm F}$, implying that electrons are
degenerate: their pressure is mostly a function of the density of the material. In Jupiter and Saturn, the degeneracy parameter
$\theta=T/T_{\rm F}$ is always close to $0.03$. Even for the warmer CoRoT-15b (a $\sim 60\rm\,M_{Jup}$ brown dwarf discovered in transit in front of its parent star -- see \citet{Bouchy+2011}), $\theta\approx 0.2$. This implies that for these objects, the thermal component is small so that the energy of
electrons in the interior is expected to be only slightly larger than
their non-relativistic, fully degenerate limit: $u_{\rm e}\ge
3/5\,kT_{\rm F} =15.6\left(\rho/\mu_{\rm e}\right)^{2/3}\ \rm eV$,
where $k$ is Boltzmann's constant, $\mu_{\rm e}$ is the number of
nucleons per electron and $\rho$ is the density in $\rm
g\,cm^{-3}$. For pure hydrogen, when the density reaches $\sim
0.8\gcc$, the average energy of electrons becomes larger than
hydrogen's ionization potential, even at zero temperature: hydrogen
pressure-ionizes and becomes metallic. This molecular to metallic
transition occurs near Mbar pressures, but exactly how this happens
is a result of the complex interplay of thermal,
Coulomb and degeneracy effects.  

Recent laboratory measurements on fluid deuterium have been able to
reach extremely high pressures up to 20\,Mbar \citep{Mochalov+2012}. Beyond that experimental feat, most of the progress of the decade in the domain has been the improvement in our understanding of the hydrogen metallization region at pressures of a fraction to a few Mbars, both from an experimental and numerical point of view \citep[see the very complete review by][]{McMahon+2012}. Already in the 1990's, gas-gun experiments had been able to measure 
a rise in the conductivity of molecular hydrogen up to $T\sim 3000\K$, $P\sim 1.4\,$Mbar, a sign
that metallization had been reached \citep{WMN96}.  A very sharp transition, probably discontinuous, was then later measured by isentropic convergent explosive shock experiments \citep{Fortov+2007} at pressures between 1.5 and 2.5\,Mbar but uncertain temperatures below 4000\,K\citep[see][]{McMahon+2012}. New experiments at higher temperatures using laser compression, directly \citep{Sano+2011} and from precompressed targets \citep{Loubeyre+2012}, confirmed that this transition from a weakly conducting molecular fluid to a metal-like hydrogen fluid occurs around pressures near 1\,Mbar and temperatures as high as 15,000\,K, but that it is continuous at these temperatures. In parallel, {\it ab initio\/} calculations of the behavior of fluid hydrogen in this thermodynamical regime predicted the existence of a first order liquid-liquid phase transition (the so-called PPT for {\it Plasma Phase Transition}) with a critical point near $T\sim 1500-2000\,$K and $P\sim 2.2\,$Mbar \citep{Morales+2010, Morales+2013a}. This discontinuous transition at low temperatures merges into a continuous transition at temperatures above the critical point, in good agreement with the experimental data.  The PPT is indicated by a thick almost vertical line in fig.~\ref{fig:phase_diag}. It is prolonged by a dashed line indicating the location of the continuous transition from molecular to metallic hydrogen that extends up to the region of thermal dissociation and ionization of hydrogen. 

The controversy that had arisen between laser-induced shock compression \citep{daSilva+97,Collins+98} and pulsed-power shock compression \citep{Knudson+04} regarding the maximum compression of deuterium along the principal shock Hugoniot has now been resolved in favor of the latter thanks to new experiments \citep{Boriskov+2005,Hicks+2009} and the realization that the equation of state of quartz used to calibrate the laser-induced shock experiments was incorrect \citep{KnudsonDesjarlais2009}. Similarly, the existence of a PPT of hydrogen at high temperatures in a regime crossing the adiabats of Jupiter and Saturn \citep{SCvH95} have now been shown to be a spurious effect resulting from the different treatment of molecules, atom and ions within the so-called chemical picture \citep{Chabrier+2007}.  Both laboratory experiments and independent models based on first-principles \citep{Militzer+01,Desjarlais03,BMG04,Vorberger+2007,French+2012} now agree and show that the transition from molecular to metallic hydrogen should occur continuously in all known giant planets.


Progress has also been made on the issue of the solidification of hydrogen \citep[see][and references therein]{McMahon+2012}. This has led to the confirmation that the interiors of the hydrogen-helium giant planets and brown dwarfs
are {\it fluid} whatever their age, a result expected since the pioneering study by \citet{Hubbard68}. Of course, because of their initial gravitational energy, these objects are warm enough to avoid the
critical point for the liquid gas transition in hydrogen and helium,
at very low temperatures, but they also lie comfortably above the
solidification lines for hydrogen and helium. (An {\it isolated}
Jupiter should begin partial solidification only after at least $\sim
10^3\,$Ga of evolution.) They are considered to be fluid because at
the high pressures and relatively modest temperatures in their
interiors, Coulomb interactions between ions play an important role in
the EOS and yield a behavior that is more reminiscent of that of a
liquid than that of a gas, contrary to what is the case in
e.g., solar-like stars.  For Uranus and Neptune, the situation is
actually more complex because at large pressures they are not expected
to contain a significant amount of hydrogen (see next section).

As fig.~\ref{fig:phase_diag} highlights, while some highly irradiated planets and brown dwarfs like CoRoT-15b have temperature profiles that get close to the hydrogen thermal dissociation line, most of them are well within the molecular hydrogen regime at low-pressures and in the metallic, degenerate regime at high pressures. Stars like our Sun lie in a higher temperature regime for which the EOS is dominated by thermal effects and electrons are essentially non-degenerate.

\subsubsection{Other elements and mixtures}\label{sec:others}

Hydrogen is of course a key element, but it is not sufficient to describe the structure of all giant planets. A description of the high-pressure behavior of other elements would go beyond the scope of the present review. We only sketch a few important results here. 

In order to obtain tractable equations of state in the entire domain of pressure and
temperature spanned by the planets during their evolution, one has to consider simplifications, among which the first one is to consider that an element (e.g., hydrogen) dominates, and that others can be considered as a perturbation. This is done for the hydrogen-helium mixture for the now classical EOS by \citet{SCvH95}, and now with more up-to-date EOSs \citep{Caillabet+2011,Nettelmann+2012,MilitzerHubbard2013}. The addition of other elements can be done through the so-called additive volume rule which is generally a good approximation given other sources of uncertainty (\citet{Vorberger+2007}; see also \citet{ChabrierAshcroft1990}). 

Equations of state for elements other than hydrogen and helium in the parameter range relevant for giant planetary interiors have traditionally been difficult to obtain, and are often not easily shared. Beyond an extrapolation from the classical ANEOS and Sesame tables \citep[see][and references therein]{SG04}, new results have become available. In particular, ab-initio simulations revealed that when compressing water or ammonia along an isentrope from conditions relevant to the atmospheres of Uranus and Neptune, they transition from a molecular to a ionic fluid, then to a superionic fluid and finally to a plasma \citep{Cavazzoni+1999}. The superionic fluid corresponds to a state in which protons move relatively freely among a lattice of oxygen atoms. An equation of state for water has been calculated from first-principles by \citet{French+2009} and is found to be in good agreement with experiments \citep{Knudson+2012}. A similar equation of state for ammonia is presented by \citet{Bethkenhagen+2013}. In models of Uranus and Neptune, for water, the ionic transition occurs near 0.1\,Mbar and the superionic transition near 1\,Mbar \citep{Redmer+2011}. 

\begin{figure}[htbp]
\resizebox{12cm}{!}{\includegraphics[angle=0]{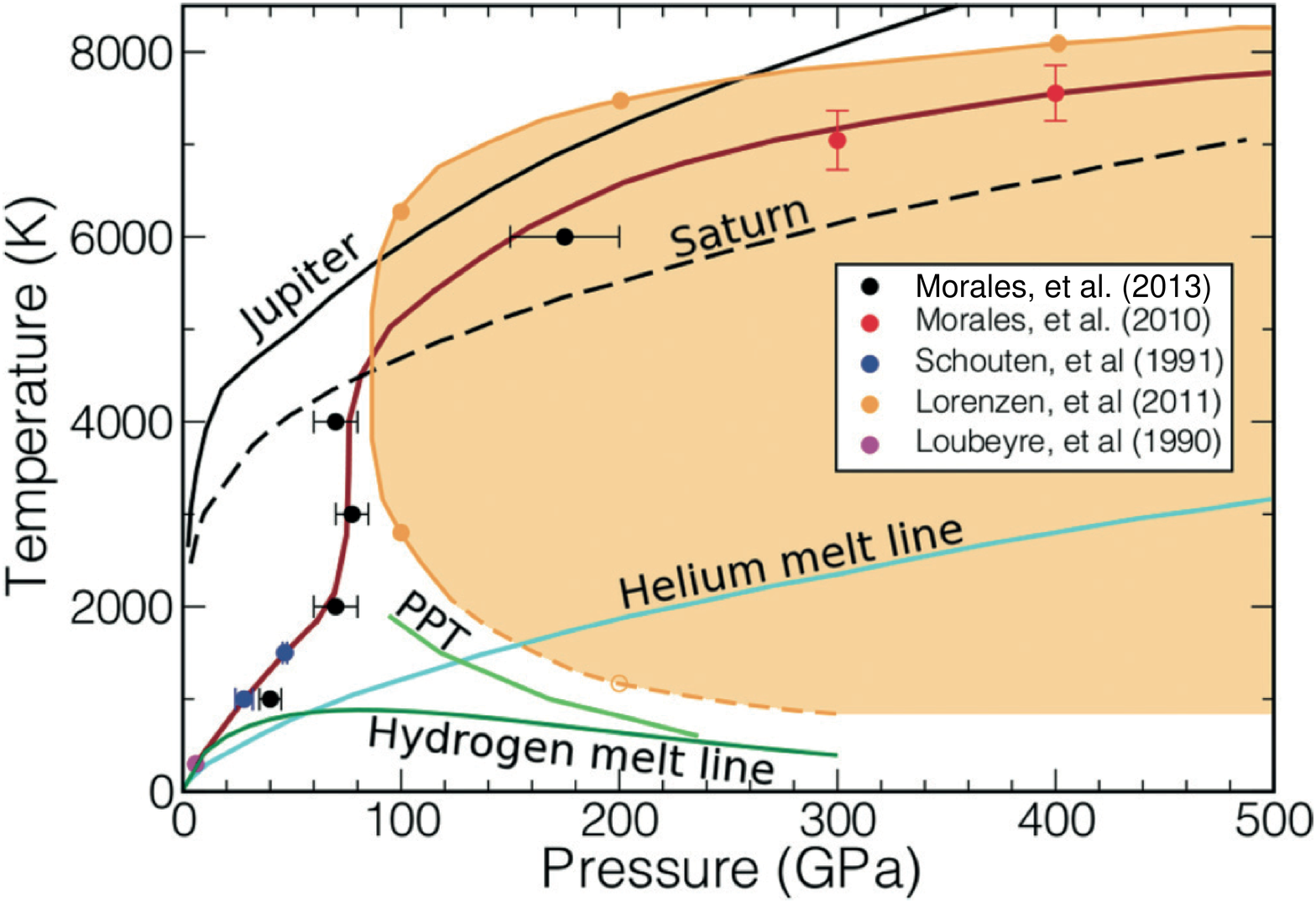}}
\caption{Phase diagram for the hydrogen-helium mixture for a helium mole concentration of 8\%. The orange region shows where the two elements are expected to separate from each other according to the calculations of \citet{Lorenzen+2011}. The red curve correspond to the critical temperature for that separation according to \cite{Morales+2013b}. Numerical results by \cite{Schouten+1991} and experimental determinations by \cite{Loubeyre+1991} are also shown. The back curves show the isentropes of Jupiter (plain) and Saturn (dashed) respectively. [From \cite{Morales+2013b}].}
\label{fig:hhe}
\end{figure}

In some cases however, generally at low enough temperatures for a given pressure range, mixtures cannot remain homogeneous. This further complicates the calculation of equations of state and has important physical consequences for the planetary structure: the two components having different molecular weights, they tend to be separated by gravity so that the heavier component settles down under the lighter one. This is the case of the hydrogen and helium mixture for which it was proposed that such a separation would occur in Saturn already in the 1970s \citep{Salpeter73, SS77b}, but for which realistic calculations have only become possible in the past decade or so \citep[see][]{Morales+2009,Lorenzen+2011,Morales+2013b}. Figure~\ref{fig:hhe} shows the comparison between two of these calculations based on first-principles simulations. According to these calculations, Saturn's interior is in the phase separation region below 1\,Mbar but whether Jupiter is too depends on which calculation is considered. 

The separation of other mixtures have also been calculated and is relevant to understand the initial formation and subsequent possible erosion of the cores of giant planets \citep[see][]{GSHS04}. First-principles calculations of the water in metallic hydrogen (at pressures above 10\,Mbar) predict a critical phase separation temperature of less than 4000\,K \citep{WilsonMilitzer2010} implying that water is completely soluble in hydrogen in the interiors of Jupiter, Saturn and generally gas giants. This is also the case of iron, with a critical temperature of around 2000\,K \citep{Wahl+2013}.  Finally, mixing rocks (specifically MgO) and metallic hydrogen is also relatively easy, even though the critical temperature for the same pressure range is higher, of order 10,000\,K or less \citep{WilsonMilitzer2012}.


\subsection{Heat transport}

Giant planets possess hot interiors, implying that a
relatively large amount of energy has to be transported from the deep
regions of the planets to their surface. This can either be done by
radiation, conduction, or, if these processes are not sufficient, by
convection. Convection is generally ensured by the rapid rise of the
opacity with increasing pressure and temperature.  At pressures of a
bar or more and relatively low temperatures (less than 1000\,K), the
three dominant sources of opacities are water, methane and
collision-induced absorption by hydrogen molecules.

However, in the intermediate temperature range between $\sim 1200$ and
$1500\K$, the Rosseland opacity due to the hydrogen and helium
absorption behaves differently: the absorption at any given wavelength
increases with density, but because the temperature also rises, the
photons are emitted at shorter wavelengths, where the monochromatic
absorption is smaller. As a consequence, the opacity can
decrease. This was shown by Guillot et al. (1994) to potentially lead
to the presence of a deep radiative zone in the interiors of Jupiter,
Saturn and Uranus.

This problem must however be reanalyzed in the light of 
observations and analyses of brown dwarfs. Their spectra show
unexpectedly wide sodium and potassium absorption lines (see Burrows,
Marley \& Sharp 2000), in spectral regions where hydrogen, helium,
water, methane and ammonia are relatively transparent. It thus appears
that the added contribution of these elements (if they are indeed
present) would wipe out any radiative region at these
levels \citep{GSHS04}. 

At temperatures above $1500\sim 2000\K$ two important sources of
opacity appear: (i) the rising number of electrons greatly enhances
the absorption of H$_2^-$ and H$^-$; (ii) TiO, a very strong absorber
at visible wavelengths is freed by the vaporization of
CaTiO$_3$. Again, the opacity rises rapidly which ensures a convective
transport of the heat. Still deeper, conduction by free electrons
becomes more efficient, but the densities are found not to be high
enough for this process to be significant, except perhaps near the
central core \citep[see][]{Hubbard68, SS77a}.

While our giant planets seem to possess globally convective interiors,
strongly irradiated extrasolar planets must develop a radiative zone
just beneath the levels where most of the stellar irradiation is
absorbed. Depending on the irradiation and characteristics of the
planet, this zone may extend down to kbar levels, the deeper levels
being convective. In this case, a careful determination of the
opacities is necessary (but generally not possible) as these control
the cooling and contraction of the deeper interior \citep[see][for a discussion of opacities and tables for substellar
atmospheres and interiors]{Freedman+2008}.

\subsection{The contraction and cooling histories of giant planets}\label{sec:virial}

The interiors of giant planets are expected to evolve with time from a
high entropy, high $\theta$ value, hot initial state to a low entropy,
low $\theta$, cold degenerate state. The essential underlying physics can
be derived from the well-known virial theorem and the energy
conservation which link the planet's internal energy $E_{\rm i}$,
gravitational energy $E_{\rm g}$ and luminosity through:
\begin{eqnarray}
\xi E_{\rm i} + E_{\rm g} &=&0,\label{eq:virial_E}\\
L &=& -{\xi-1\over \xi}{dE_{\rm g}\over dt},\label{eq:virial_L}
\end{eqnarray}
where $\xi=\int_0^M 3(P/\rho)dm / \int_0^M u dm\approx <\!3P/\rho
u\!>$, the brackets indicating averaging, and $u$ is the specific
internal energy. For a diatomic perfect gas, $\xi=3.2$; for
fully-degenerate non-relativistic electrons, $\xi=2$.

Thus, for a giant planet or brown dwarf beginning its life mostly as a
perfect H$_2$ gas, two third of the energy gained by contraction is
radiated away, one third being used to increase $E_{\rm i}$. The
internal energy being proportional to the temperature, the effect is
to heat up the planet. This represents the slightly counter-intuitive
but well known effect that a star or giant planet initially heats up
while radiating a significant luminosity \citep[e.g.,][]{KW94}.

Let us now move further in the evolution, when the contraction has
proceeded to a point where the electrons have become degenerate.  For
simplicity, we will ignore Coulomb interactions and exchange terms,
and assume that the internal energy can be written as $E_{\rm
i}=E_{\rm el}+E_{\rm ion}$, and that furthermore $E_{\rm el}\gg E_{\rm
ion}$ ($\theta$ is small). Because $\xi\approx 2$, we know that half of
the gravitational potential energy is radiated away and half of it
goes into internal energy.  The problem is to decide how this energy
is split into an electronic and an ionic part.  The gravitational
energy changes with some average value of the interior density as
$E_{\rm g}\propto 1/R \propto \rho^{1/3}$. The energy of the
degenerate electrons is essentially the Fermi energy: $E_{\rm
el}\propto \rho^{2/3}$. Therefore, $\dot{E}_{\rm g}\approx 2(E_{\rm
g}/ E_{\rm el})\dot{E}_{\rm el}$. Using the virial theorem and specifically eq.~(\ref{eq:virial_E}) we get that $\dot{E}_{\rm g}\approx -\dot{E}_{\rm el}$. The luminosity is by definition $L=-(\dot{E}_{\rm g}+\dot{E}_{\rm i})$ and therefore 
\begin{equation}
L \approx -\dot{E}_{\rm ion} \propto -\dot{T}.  
\end{equation}
In this limit, the gravitational energy lost is entirely absorbed by the increase in pressure of the degenerate
electrons and the observed luminosity is due to the thermal cooling
of the ions \citep{Guillot05}. 

Several simplifications limit the applicability of this result (that
would be valid in the white dwarf regime). In particular, the
Coulomb and exchange terms in the EOS introduce negative
contributions that cannot be neglected. However, the approach is
useful to grasp how the evolution proceeds: in its
very early stages, the planet is very compressible. It follows a
standard Kelvin-Helmholtz contraction. When degeneracy sets in, the
compressibility becomes much smaller ($\alpha T\sim 0.1$, where
$\alpha$ is the coefficient of thermal expansion), and the planet
gets its luminosity mostly from the thermal cooling of the ions. The
luminosity can be written in terms of a modified Kelvin-Helmholtz
formula: 
\begin{equation}
L\approx \eta {GM^2\over R\tau},
\label{eq:lapprox}
\end{equation}
where $\tau$ is the age, and $\eta$ is a factor that hides most of the
complex physics. In the approximation that Coulomb and exchange
terms can be neglected, $\eta\approx\theta/(\theta +1)$. The poor
compressibility of giant planets in their mature evolution stages
imply that $\eta\ll 1$ ($\eta\sim 0.03$ for Jupiter): the luminosity
is not obtained from the entire gravitational potential, but from the
much more limited reservoir constituted by the thermal internal
energy. Equation~(\ref{eq:lapprox}) shows that to first order, $\log
L\propto -\log\tau$: very little time is spent at high luminosity
values. In other words, the problem is (in most cases) weakly
sensitive to initial conditions. However, it is to be noticed
that with progress in our capability to detect very young objects,
i.e., planets and brown dwarfs of only a few million years of age, the
problem of the initial conditions does become important \citep{Marley+06}. Interestingly, at these early stages, their luminosity appears to strongly depend on their core mass \citep{Mordasini2013}.

\begin{figure}[tbp]
  \centerline{\resizebox{12cm}{!}{\includegraphics[angle=-90]{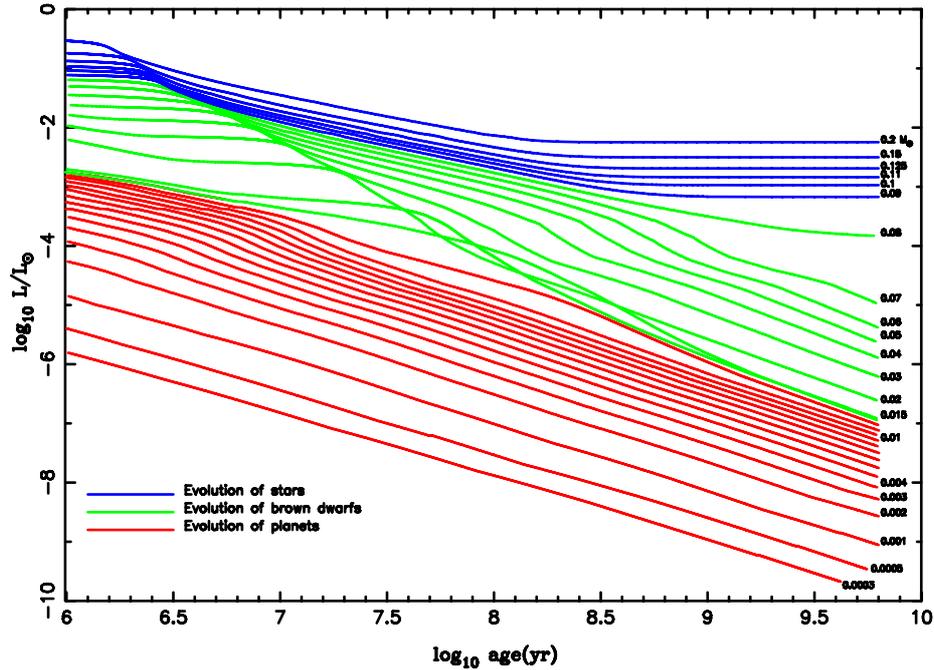}}}
\caption{Evolution of the luminosity (in L$_\odot$) of
solar-metallicity M dwarfs and substellar objects vs. time (in yr)
after formation. In this figure, "brown dwarfs" are arbitrarily
designated as those objects that burn deuterium, while those that do
not are tentatively labelled "planets". Stars are objects massive
enough to halt their contraction due to hydrogen fusion.  Each curve
is labelled by its corresponding mass in M$_\odot$, with the lowest three
corresponding to the mass of Saturn, half the mass of Jupiter, and the
mass of Jupiter.  [From \citet{Burrows+97}].}
\label{fig:L vs t}
\end{figure}

Figure~\ref{fig:L vs t} shows more generally how giant planets, but
also brown dwarfs and small stars see their luminosities evolve as a
function of time. The $1/\tau$ slope is globally conserved, with some
variations for brown dwarfs during the transient epoch of deuterium
burning, and of course for stars, when they begin burning efficiently
their hydrogen and settle on the main sequence: in that case, the
tendency of the star to contract under the action of gravity is
exactly balanced by thermonuclear hydrogen fusion.

\subsection{Mass-radius relation}

\begin{figure}[htbp]
\resizebox{15cm}{!}{\includegraphics[angle=0]{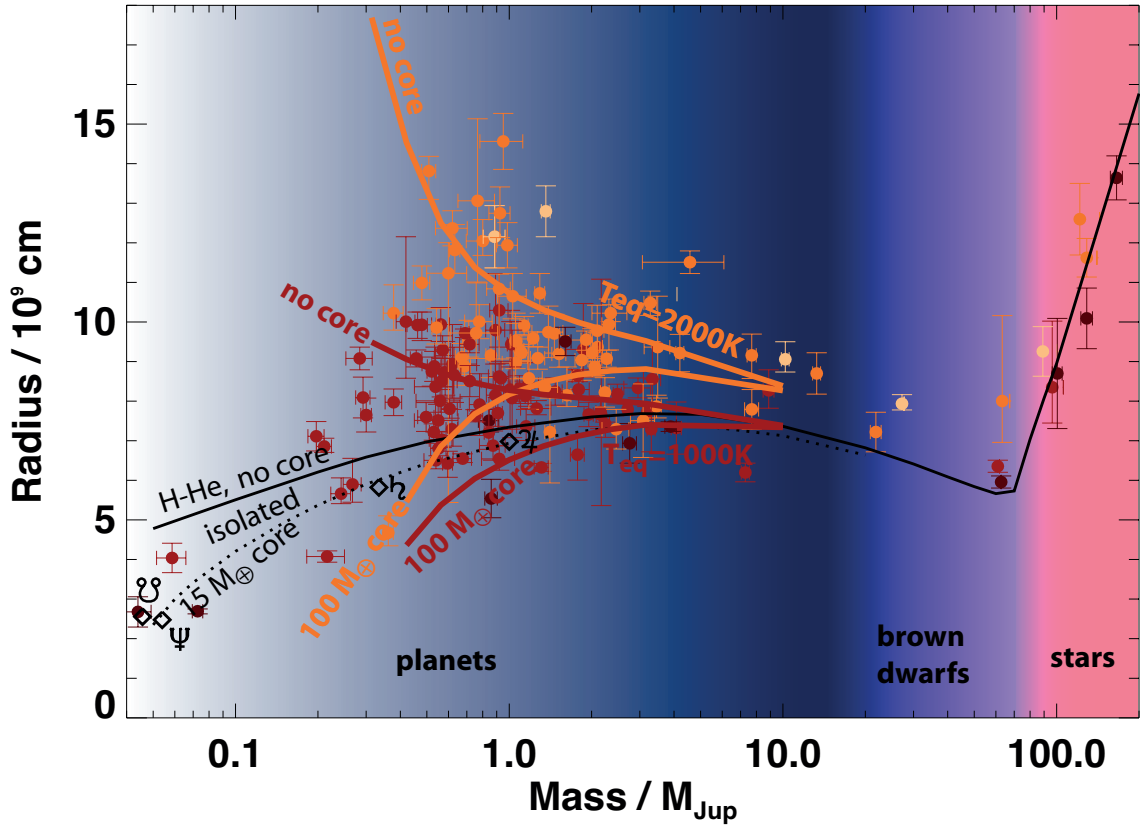}}
\caption{Theoretical and observed mass-radius relations. The black
  line is applicable to the evolution of solar composition planets,
  brown dwarfs and stars, when isolated or nearly isolated (as
  Jupiter, Saturn, Uranus and Neptune, defined by diamonds and their
  respective symbols), after 5 Ga of evolution. The dotted line shows
  the effect of a $15\mea$ core on the mass-radius relation. Orange
  and yellow curves represent the mass-radius relations for heavily
  irradiated planets with equilibrium temperatures of 1000 and
  2000\,K, respectively, and assuming that 0.5\% of the incoming
  stellar luminosity is dissipated at the center (see
  section~\ref{sec:irradiated}). For each irradiation level, two cases
  are considered: a solar-composition planet with no core (top curve),
  and one with a $100\mea$ central core (bottom curve). Circles with error bars correspond to known planets, brown dwarfs and low-mass stars, color-coded as a function of their equilibrium temperature(below 750, 1500, 2250\,K and above 2250\,K, respectively, from darkest to lightest).}
\label{fig:mass_rad}
\end{figure}

The relation between mass and radius has very fundamental astrophysical
applications. Most importantly it allows one to infer the gross
composition of an object from a measurement of its mass and
radius. This is especially relevant in the context of the discovery 
of extrasolar planets with both radial velocimetry and the transit
method, as the two techniques yield relatively accurate determination
of $M$ and $R$, these determinations being often limited by the uncertainty on the stellar parameters themselves. 

Figure~\ref{fig:mass_rad} shows mass-radius relations for compact
degenerate objects from giant planets to brown dwarfs and low-mass
stars. The right-hand side of the diagram shows a rapid increase of
the radius with mass in the stellar regime which is directly due to
the onset of stable thermonuclear reactions. In this regime,
observations and theoretical models agree \citep[see however][for a
more detailed discussion]{Ribas06}. The left-hand side of the diagram
is obviously more complex, and this can be understood by the fact that
planets have much larger variations in composition than stars, and
because external factors such as the amount of irradiation they
receive do affect their contraction in a significant manner.

Let us first concentrate on isolated or nearly-isolated gaseous
planets. The black curves have a local maximum near $4\mjup$: at
small masses, the compression is small so that the radius
increases with mass. At large masses, degeneracy sets in and the
radius decreases with mass.

This can be understood on the basis of polytropic models based on the
assumption that $P=K\rho^{1+1/n}$, where $K$ and $n$ are
constants. Because of degeneracy, a planet of large mass will tend to
have $n\rightarrow 1.5$, while a planet of smaller mass will be less
compressible ($n\rightarrow 0$). Indeed, it can be shown that in their
inner 70 to 80\% in radius isolated solar composition planets of 10, 1
and $0.1\mjup$ have $n=1.3$, 1.0 and 0.6, respectively. From
polytropic equations \citep[e.g.,][]{Chandrasekhar39}:
\begin{equation}
R\propto K^{n\over 3-n} M^{1-n\over 3-n}.
\label{eq:m-r-k}
\end{equation}
Assuming that $K$ is independant of mass, one gets $R\propto
M^{0.16}$, $M^{0}$, and $M^{-0.18}$ for $M=10$,
1 and $0.1\mjup$, respectively, in relatively good agreement with
\rfig{mass_rad} (the small discrepancies are due to the fact that the
intrinsic luminosity and hence $K$ depend on the mass considered).

Figure~\ref{fig:mass_rad} shows already that the planets in our Solar
System are not made of pure hydrogen and helium and require an
additional fraction of heavy elements in their interior, either in the
form of a core, or distributed in the envelope (dotted line). 

For extrasolar planets, the situation is complicated by the fact that
the intense irradiation that they receive plays a major role in their
evolution. The present sample is already quite diverse, with
equilibrium temperature (defined as the effective temperature
corresponding to the stellar flux received by the planet) ranging from
1000 to 2500\,K. Their compositions are also quite variable, with some
planets having large masses of heavy elements
\citep{Sato+05,Guillot+06}. The orange and yellow curves in
fig.~\ref{fig:mass_rad} show theoretical results for equilibrium
temperatures of 1000 and 2000\,K, respectively. Two extreme models
have been plotted: assuming a purely solar composition planet (top
curve), and assuming the presence of a $100\mea$ central core (bottom
curve). In each case, an additional energy source proportional to
0.5\% of the incoming luminosity was also assumed (see discussion in
\S~\ref{sec:irradiated} hereafter).

The increase in radius for decreasing planetary mass for irradiated,
solar-composition planets with little or no core can be understood
using the polytropic relation (eq.~\ref{eq:m-r-k}), but accounting for
variations of $K$ as defined by the atmospheric boundary
condition. Using the Eddington approximation, assuming $\kappa\propto
P$ and a perfect gas relation in the atmosphere, one can show that
$K\propto (M/R^2)^{-1/2n}$ and that therefore $R\propto M^{1/2-n\over
2-n}$. With $n=1$, one finds $R\propto M^{-1/2}$. Strongly irradiated
hydrogen-helium planets of small masses are hence expected to have the
largest radii which qualitatively explain the positions of the
extrasolar planets in \rfig{mass_rad}.  Note that this estimate implicitly
assumes that $n$ is constant throughout the planet. The real situation
is more complex because of the growth of a deep radiative region in
most irradiated planets, and because of structural changes between the
degenerate interior and the perfect gas atmosphere \citep{Guillot05}.

In the case of the presence of a fixed mass of heavy elements, the
trend is inverse because of the increase of mean molecular mass (or
equivalently core/envelope mass) with decreasing total mass. Thus,
small planets with a core are much more tightly bound and less
subject to evaporation than those that have no core.

\subsection{Rotation and the figures of planets}
\label{sec:rotation}

The mass and radius of a planet informs us on its global
composition. Because planets are also rotating, one is allowed to
obtain more information on their deep interior structure. 
The hydrostatic equation becomes more complex however:
\begin{equation}
{\bfnab P\over \rho}=\bfnab\left(G\int\!\!\!\int\!\!\!\int 
{\rho(\bfr')\over |\bfr - \bfr'|}d^3\bfr'\right) - \bfOm\times(\bfOm\times\bfr),
\label{eq:full_hydrostat}
\end{equation}
where $\bfOm$ is the rotation vector. 
The resolution of eq.~(\ref{eq:full_hydrostat}) is a complex
problem. It can however be somewhat simplified by assuming that
$|\bfOm|\equiv\omega$ is such that the centrifugal force can be
derived from a potential. The hydrostatic equilibrium then writes
$\nabla P = \rho \nabla U$, and the {\it figure} of the rotating
planet is then defined by the $U=constant$ level surface. 

One can show \citep[e.g.,][]{ZT78} that the hydrostatic
equation of a fluid planet can then be written in terms of the mean
radius $\rbar$ (the radius of a sphere containing the same volume as
that enclosed by the considered equipotential surface):
\begin{equation}
{1\over \rho}\dpar{P}{\rbar}=-{Gm\over \rbar^2}+{2\over 3}\omega^2
\rbar + {GM\over \overline{R}^3} \rbar\varphi_\omega,
\end{equation}
where $M$ and $\overline{R}$ are the total mass and mean radius of the
planet, and $\varphi_\omega$ is a slowly varying function of
$\rbar$. (In the case of Jupiter, $\varphi_\omega$ varies from about
$2\times 10^{-3}$ at the center to $4\times 10^{-3}$ at the surface.)
Equations~(\ref{eq:dtdr}-\ref{eq:dldr}) remain the same with the
hypothesis that the level surfaces for the pressure, temperature, and
luminosity are equipotentials.  The significance of rotation is
measured by the ratio of the centrifugal acceleration to the gravity:
\begin{equation}
q={\omega^2 \req^3\over GM}.
\end{equation}


As discussed in section~\ref{sec:gravity}, in some cases, the external
gravity field of a planet can be accurately measured in the form of
gravitational moments $J_{k}$ (with zero odd moments for a planet in
hydrostatic equilibrium) that measure the departure from spherical
symmetry. Together with the mass, this provides a constraint on the
interior density profile (see \cite{ZT74} -see also
chapters by Van Hoolst and Sohl \& Schubert):
\begin{eqnarray*}
M&=&\int\!\!\!\int\!\!\!\int \rho(r,\theta) d^3\tau, \\
J_{2i} &=& -{1\over M R_{\rm eq}^{2i}}\int\!\!\!\int\!\!\!\int \rho(r,\theta) r^{2i}
P_{2i}(\cos\theta) d^3\tau,
\end{eqnarray*}
where $d\tau$ is a volume element and the integrals are performed over
the entire volume of the planet.

\begin{figure}[htbp]
\resizebox{10cm}{!}{\includegraphics[angle=0]{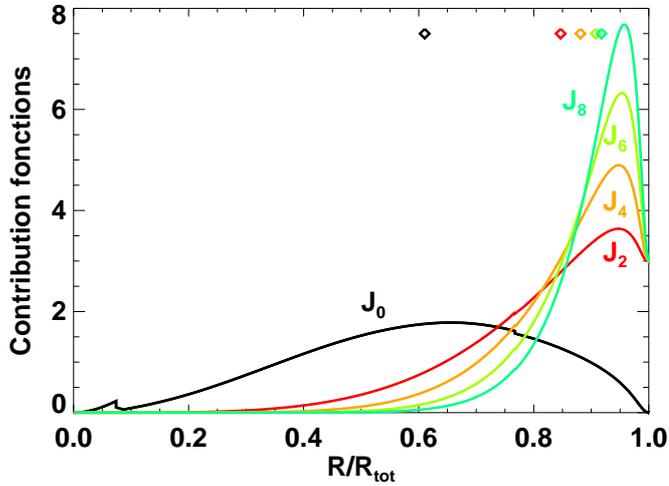}}
\caption{Contribution of the level radii to the gravitational moments
  of Jupiter. $J_0$ is equivalent to the planet's mass. The small
  discontinuities are caused by the following transitions, from left
  to right: core/envelope, helium rich/helium poor
  (metallic/molecular). Diamonds indicate the median radius for each
  moment. They correspond to pressures of 9.1, 1.3, 0.75, 0.45 and 0.35 Mbar, respectively, from left to right ($J_0$ to $J_8$).}
\label{fig:contrib}
\end{figure}

Figure~\ref{fig:contrib} shows how the different layers inside a
planet contribute to the mass and the gravitational moments. The
figure applies to Jupiter, but would remain relatively similar for
other planets. Note however that in the case of Uranus and Neptune,
the core is a sizable fraction of the total planet and contributes
both to $J_2$ and $J_4$. Measured gravitational moments thus provide
information on the external levels of a planet. It is only indirectly,
through the constraints on the outer envelope that the presence of a
central core can be infered. As a consequence, it is impossible to
determine this core's state (liquid or solid), structure
(differentiated, partially mixed with the envelope) and composition
(rock, ice, helium...) from the gravity field data.

The Juno \citep{Bolton2010} and Cassini Solstice \citep{Spilker2012} missions are expected to yield considerable improvements in our determination of the gravity fields of Jupiter and Saturn, respectively. Because the theory of figures is limited by its expansion in terms of the rotation parameter $q$ and because purely barotropic solutions are possible only in the limit of solid-body rotation and of pure rotation on cylinders, these high precision measurements will require new approaches to include rotation in planetary models \citep[see][]{Hubbard1999, Hubbard2013}. Separately, these measurements will enable new constraints such as the determination of the planets' angular momentum through the measurement of the relativistic Lense-Thirring effect \citep{Iorio2010} and the determination of their moment of inertia \citep{Helled2011,Helled+2011b}. But probably the most important prospects lie in the possibility to couple measurements on gravity field, magnetic fields and wind speeds with combined tri-dimensional magnetohydrodynamical models (see sections~\ref{sec:magnetic fields} and \ref{sec:dynamics}). 

For planets outside the solar system, although measuring their
gravitational potential is presently beyond reach, an indirect measurement of the planets' Love number $k_2$ may be possible in systems of planets in which one is locked into a so-called fixed-point eccentricity. So far, one such transiting system is known, HAT-P-13 \citep{Batygin+2009b,Mardling2010}, because it contains a transiting planet, HAT-P-13b and a companion, HAT-P-13c whose minimum mass $M\sin i$ and orbital eccentricity are known \citep{Bakos+2009}. The value of $k_2$ constrains the interior structure in a way that is very similar to $J_2$ and has been used to obtain first constraints on the interior structure of this planet \citep{Kramm+2012}.


\section{Interior structures and evolutions}
\subsection{Jupiter and Saturn}\label{sec:JupSat}
\begin{figure}[htbp]
  \centerline{\resizebox{12cm}{!}{\includegraphics[angle=90,bb=4.5cm
	7cm 15cm 24cm]{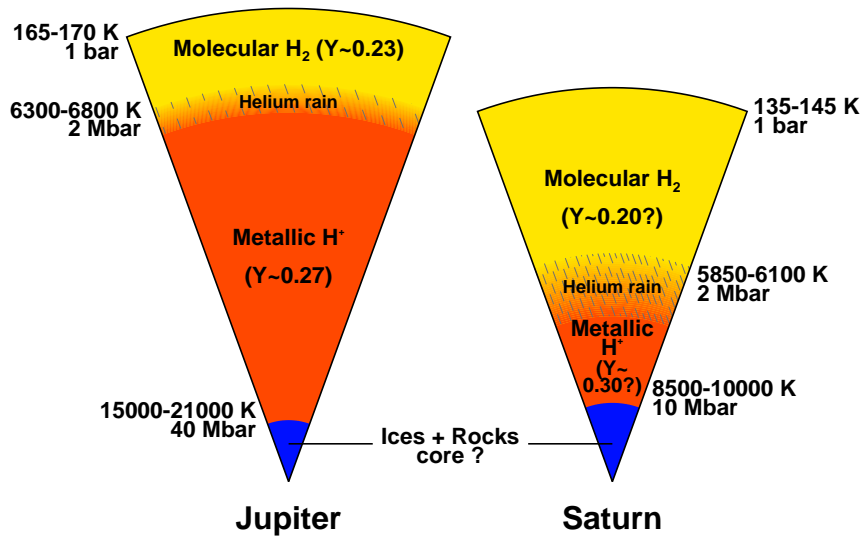}}}
\caption{Schematic representation of the interiors of Jupiter and
  Saturn. The range of temperatures is estimated using homogeneous
  models and including a possible radiative zone indicated by the hashed
  regions. Helium mass mixing ratios $Y$ are indicated. The size of the
  central rock and ice cores of Jupiter and Saturn is very uncertain
  (see text). In the case of Saturn, the inhomogeneous region may
  extend down all the way to the core which would imply the formation
  of a helium core. [Adapted from \citet{Guillot99b}].}
\label{fig:intjupsat}
\end{figure}

As illustrated by fig.~\ref{fig:intjupsat}, the simplest interior
models of Jupiter and Saturn matching all observational constraints
assume the presence of three main layers: (i) an outer hydrogen-helium
envelope, whose global composition is that of the deep atmosphere;
(ii) an inner hydrogen-helium envelope, enriched in helium because the
whole planet has to fit the H/He protosolar value; (iii) a central
dense core. Because the planets are believed to be mostly convective,
these regions are expected to be globally homogeneous. (Many
interesting thermochemical transformations take place in the deep
atmosphere, but they are of little concern to
us). 

The transition from a helium-poor upper envelope to a helium-rich
lower envelope is thought to take place through the formation of
helium-rich droplets that fall deeper into the planet due to their
larger density. These droplets form when the temperature-pressure profiles enter the separation region for the initial helium abundance, as shown in section~\ref{sec:others}.  Three-layer models implicitly make the hypothesis that
this region is adiabatic (this is justified only if convection is not inhibited by the formation of helium droplets, as discussed by \cite{SS77b}) and narrow. Figure~\ref{fig:hhe} shows that this zone may be extended, especially in present-day Saturn \citep[see also][]{FH03}. 

As discussed by \citet{SS77b,Stevenson82} the planets would start from an initially hot and homogeneous state and would start entering the phase separation region progressively, leading to a depletion of helium in the outer region and its increase in the deeper interior. According to the calculations by \cite{Lorenzen+2011} and \cite{Morales+2013b} the separation would first occur at a pressure between 1 and 2 Mbar and the inhomogeneous region would grow inward but not so much towards lower pressures because of the higher solubility of helium in molecular hydrogen. According to the simulations, although needed to explain the abundances in Jupiter's atmosphere (section~\ref{sec:compositions}), it is not yet clear that the process has begun in this planet. Fully consistent calculations that account both for the constraints on the planets' gravitational moments and their atmospheric composition should become possible, especially with the additional information brought by the Juno \citep{Bolton2010} and Cassini Solstice \citep{Spilker2012} space missions.  

In the absence of these calculations, adiabatic three-layer models can be
used as a useful guidance to a necessarily hypothetical ensemble of
allowed structures and compositions of Jupiter and Saturn. A
relatively extensive exploration of the parameter space has been
performed by several authors \citep{SG04,FortneyNettelmann2010,Nettelmann+2012,Nettelmann+2013b,HelledGuillot2013}. 
The calculations account for a transition from a helium-poor outer envelope to a helium-rich inner envelope. The abundance of heavy elements may or may not be held constant across this transition. Many sources of uncertainties are taken into
account however; among them, the most significant are on the equations
of state of hydrogen and helium, the uncertain values of $J_4$ and
$J_6$, the presence of differential rotation deep inside the planet,
the location of the helium-poor to helium-rich region, and the
uncertain helium to hydrogen protosolar ratio.

Their results indicate that Jupiter's core is smaller than $\sim
10\mea$, and that its global composition is pretty much unknown
(between 10 to 42$\mea$ of heavy elements in total). The models
indicate that Jupiter is enriched compared to the solar value by a
factor 1.5 to 8 times the solar value. This enrichment is compatible
with a global uniform enrichment of all species near the atmospheric
Galileo values, but allows many other possibilities as well. 

Other models of Jupiter based on an ab-initio equation of state by \citep{Militzer+2008} led to a solution with a large core mass and a very small enrichment in heavy elements in the envelope incompatible with either the Galileo probe measurements or the protosolar helium abundance. A significant update in the EOS is presented by \citet{MilitzerHubbard2013}. This updated EOS yields a warmer interior than the 2008 models and should therefore lead to a smaller core mass and larger amount of heavy elements in the envelope, in line with the other results. At the same time, the \citet{MilitzerHubbard2013} EOS is much more accurate than the range of EOSs used by \citet{SG04} and it differs slightly from the other ab-initio EOS used by \citet{Nettelmann+2012} and \citet{Nettelmann+2013b}. New constraints should therefore be derived on the basis of those new data. 



In the case of Saturn, the solutions depend less on the hydrogen EOS
because the Mbar pressure region is comparatively smaller. The total
amount of heavy elements present in the planet can therefore be
estimated with a better accuracy than for Jupiter, and is between $16$
and $30\mea$ \citep{Nettelmann+2013b,HelledGuillot2013}. The uncertainty on the core mass is found to be larger than for Jupiter because a heavy-element rich inner envelope can mimic the gravitational signature of the core. As a result only an upper limit on the core mass of $20\mea$ is derived.  

\begin{figure}[htb]
\centerline{\resizebox{8cm}{!}{\includegraphics[angle=0]{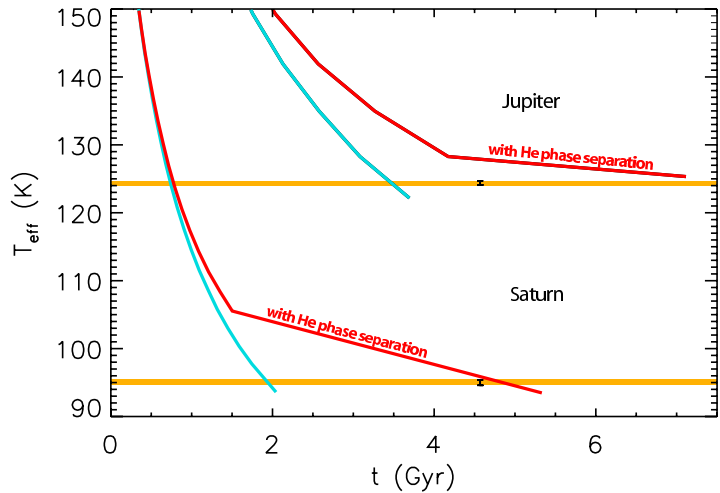}}}
\vspace*{1cm}
\caption{Final stages of evolution of Jupiter and Saturn. The present
  effective temperatures, reached after $\sim 4.55$\,Ga of
  evolution, are indicated as horizontal orange lines. For each planet
  two models represent attempts to bracket the ensemble of
  possibilities, with the faster evolution corresponding to that of an
  homogeneous planet, while the slowest evolution includes the effect
  of helium settling in the last evolution phase. [Adapted from
  \cite{Hubbard+99} and \cite{FH03}].}
\label{fig:jup-sat-evolution}
\end{figure}

Concerning the {\it evolutions} of Jupiter and Saturn, the three main
sources of uncertainty are, by order of importance: (1) the magnitude
of the helium separation; (2) the EOS; (3) the atmospheric boundary
conditions. Figure~\ref{fig:jup-sat-evolution} shows an ensemble of
possibilities that attempts to bracket the minimum and maximum
cooling. In all these quasi-adiabatic cases, helium sedimentation is needed to explain
Saturn's present luminosity \citep[see][]{Salpeter73, SS77b,
Hubbard77,Hubbard+99,FH03}. In the case of Jupiter, the sedimentation of
helium that appears to be necessary to explain the low atmospheric
helium abundance poses a problem for evolution models because it
appears to generally prolong its evolution beyond 4.55\,Ga, the age of
the Solar System \citep{Fortney+2011, Nettelmann+2012}. However, different solutions are possible, including
improvements of the EOS and atmospheric boundary conditions, or even
the possible progressive erosion of the central core that would yield
a lower luminosity of Jupiter at a given age \citep{GSHS04}.

A different set of solutions appears when one considers the possibility that the envelopes of Jupiter and Saturn are {\em not} homogeneous and adiabatic on large scales, but are instead of variable composition, with heavier elements at the bottom. In that case, convection can be strongly inhibited which requires the temperature gradient to be larger in order to transport the same intrinsic luminosity. In the case of giant planets, this process, known as semiconvection or diffusive convection leads to the formation of a non-static staircase structure with diffusive interfaces with abrupt variations of temperature and composition and a homogeneous adiabatic structure inbetween \citep{Stevenson1985, Rosenblum+2011}. By assuming that this structure is maintained over the entire envelopes of Jupiter and Saturn, \citet{LeconteChabrier2012} derive much warmer interior structures with also 30\% to 60\% more heavy elements than in conventional models. However, the assumption that this non-homogenous composition is maintained in the entire envelopes is ad hoc. In reality, one may expect semi-convection to be confined to a much smaller region and thus have a more limited effect. The problem is open however and requires further study. Its understanding is also critical for deciding whether Jupiter's core can erode into its envelope \citep[see][]{GSHS04,WilsonMilitzer2012}.

As discussed in sections~\ref{sec:magnetic fields} and \ref{sec:dynamics}, classical interior models of Jupiter and Saturn based on the assumption of a homogeneous structure in the molecular envelope and an increase of the conductivity mainly due to hydrogen metallization do yield solutions for their magnetic fields and atmospheric zonal winds that globally match the observations. However, important ``details'' such as why Saturn's magnetic field is axisymmetric and Jupiter is not remain unexplained. Coupling interior and dynamical models in order to fit both Jupiter and Saturn's gravity and magnetic fields as well as their observed zonal winds (see sections~\ref{sec:magnetic fields} and \ref{sec:dynamics}) should bring a more global understanding of the planetary structures.  

\subsection{Uranus and Neptune}\label{sec:UraNep}
\begin{figure}[htbp]
\centerline{\resizebox{11cm}{!}{\includegraphics[angle=90,bb=4.5cm 4.5cm 15cm 24cm]{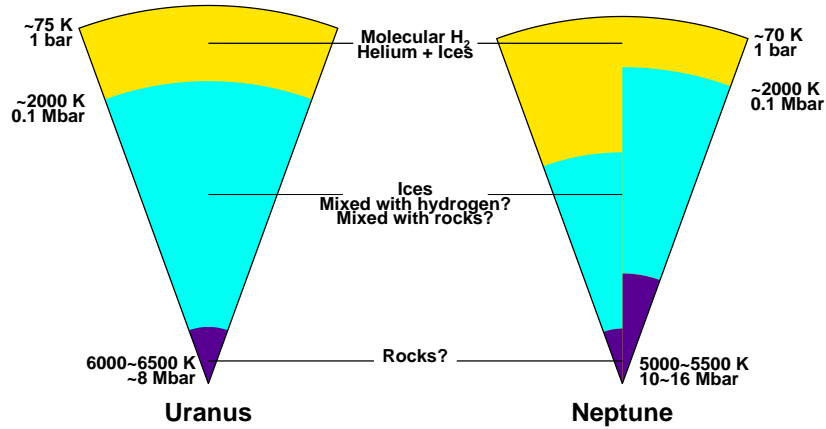}}}
\caption{Schematic representation of the interiors of Uranus and
  Neptune. The ensemble of possibilities for Neptune is larger. Two possible structures are shown. [Adapted from \cite{Guillot99b} using results from \cite{Nettelmann+2013a}].}
\label{fig:inturanep}
\end{figure}

Although the two planets are relatively similar, table~\ref{tab:comp} already
shows that Neptune's larger mean density compared to Uranus has to be
due to a slightly different composition: either more heavy elements
compared to hydrogen and helium, or a larger rock/ice ratio.  The
gravitational moments impose that the density profiles lie close to
that of ``ices'' (a mixture initially composed of e.g., H$_2$O, CH$_4$ and
NH$_3$, but which rapidly becomes a ionic fluid of uncertain chemical
composition in the planetary interior), except in the outermost
layers, which have a density closer to that of hydrogen and helium
\citep{MGP95,PPM00}. As illustrated in
\rfig{inturanep}, three-layer models of Uranus and Neptune consisting
of a central ``rocks'' core (magnesium-silicate and iron material), an
ice layer and a hydrogen-helium gas envelope have been calculated
\citep{PHS91, Hubbard+95, FortneyNettelmann2010, Helled+2011, Nettelmann+2013a}.

According to the models of \cite{Nettelmann+2013a}, Uranus contains a minimum of $1.8$ to $2.2\mea$ of hydrogen and helium and Neptune $2.7$ to $3.3\mea$. The global ice to rock ratio that is derived is very high (19 to 36) in Uranus, while Neptune has a wide range of solutions from $3.6$ to $14$. These values are much larger than the canonical ice to rock ratio of 2 to 3 for the protosun that accounts for the abundances of all elements condensing at low temperatures (``ices'') versus that of more refractory elements (``rocks''). The fact that either planet would have accreted much less rocks than ices is puzzling and unexplained by formation models. It is probably an artefact from assuming ices being confined to the envelope and rocks to the core.

The evolution of the two planets also remains a mystery. While Neptune's present luminosity may be explained by the adiabatic cooling of the planet over the age of the Solar System, this is not the case of Uranus's very low luminosity \citep{PWM95,Fortney+2011,Nettelmann+2013a}. This could be explained by the presence of a strongly inhibiting compositional gradient decoupling an inner region which would remain hot and an outer envelope which would cool progressively\citep{PWM95}. Such regions could also be present in Neptune but considerably deeper. Unfortunately, this qualitative explanation cannot be tied to the inferred interior structures. Apart from the latest models by \citep{Nettelmann+2013a}, the models of Uranus and Neptune are too similar (and so are their magnetic fields -- see section~\ref{sec:magnetic fields}) to explain why Uranus would have such a small intrinsic heat flux and not Neptune. 

In fact, it is likely that all present models of Uranus and Neptune are inadequate because of the assumption of an adiabatic temperature structure across interfaces with different compositions. Instead, diffusive-convection should occur and lead to strongly superadiabatic temperature gradient \citep[e.g.,][]{Rosenblum+2011}. As in the case of Jupiter and Saturn \citep[see][]{LeconteChabrier2012}, this would lead to higher temperatures in the interior and very different constraints on the interior composition. The amount of rocks required to fit the mean density and gravitational moments would certainly rise, potentially solving the ice to rock ratio problem. The evolution of the planets would be very different as the present-day luminosity would be mostly governed by the leak of heat from the hot interior by diffusion at the interfaces. 


\subsection{Irradiated giant planets}
\label{sec:irradiated}

\subsubsection{Interior structure and dynamics}

The physics that governs the calculation of interior structure and evolution models of giant planets described in the previous sections can be applied in principle to any gaseous exoplanet and brown dwarf. We focus the discussion on the ones that orbit extremely close to their star
because of the possibility to directly characterise them and measure
their mass, radius and in some cases even the properties of their atmosphere. Two
planets are proxies for this new class of objects: the first
extrasolar giant planet discovered, 51\,Peg\,b, with an orbital period
of $P=4.23$ days, and the first {\em transiting\/} extrasolar giant
planet, HD\,209458\,b, with $P=3.52$ days. Following widespread usage, we call these planets ``hot Jupiters'' (a.k.a ``Pegasids'' since these two archetypes have been discovered in the Pegasus constellation). 

With such a short orbital period, these planets are for most of them
subject to an irradiation from their central star that is so intense
that the absorbed stellar energy flux can be about $\sim 10^4$ times
larger than their intrinsic flux. The atmosphere is thus prevented
from cooling, with the consequence that a radiative zone develops and
governs the cooling and contraction of the interior
\citep{Guillot+96}. Typically, for a planet like HD\,209458\,b, this
radiative 
zone extends to kbar levels, $T\sim 4000\K$, and is located in the
outer 5\% in radius ($0.3\%$ in mass) \citep{GS02}.

\begin{figure}[htbp]
\begin{center}
  \centerline{\resizebox{10cm}{!}{\includegraphics[angle=0]{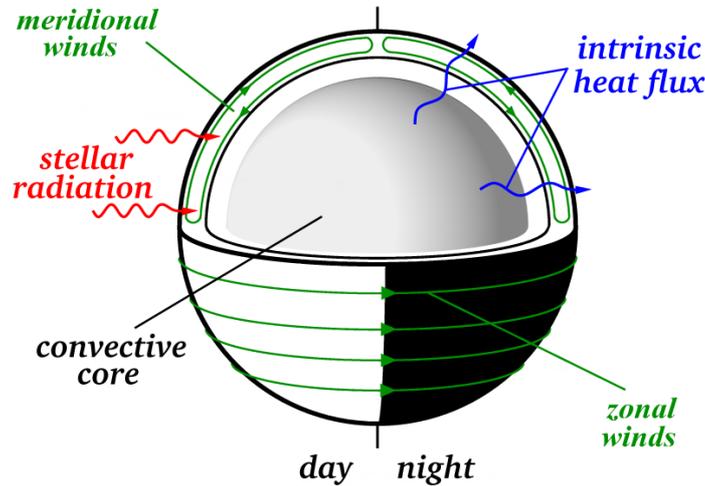}}}
\caption{Conjectured dynamical structure of hot Jupiters (strongly
  irradiated extrasolar giant planets): At pressures larger than
  $0.1--1$\,kbar, the intrinsic heat flux must be transported by
  convection. The convective core is at or near synchronous rotation
  with the star and has small latitudinal and longitudinal temperature
  variations. At lower pressures a radiative envelope is present. The
  top part of the atmosphere is penetrated by the stellar light on the
  day side. The spatial variation in insolation should drive winds
  that transport heat from the day side to the night side. [From
  \cite{SG02}].}
\label{fig:circ}
\end{center}
\end{figure}

Problems in modeling the evolution of hot Jupiters arise 
because of the uncertain outer boundary condition. The intense stellar
flux implies that the atmospheric temperature profile is extremely
dependent upon the opacity sources considered. Depending on the chosen
composition, the opacity data used, the assumed presence of clouds,
the geometry considered, resulting temperatures in the deep atmosphere
can differ by up to $\sim 600\K$ \citep{SS00, Goukenleuque+00, BHA01,
SBH03, IBG05, Fortney+06}. Furthermore, as illustrated by \rfig{circ},
the strong irradiation and expected synchronization of the planets' spin
implies that strong inhomogeneities should exist in the atmosphere
with in particular strong ($\sim 500$\,K) day-night and
equator-to-pole differences in effective temperatures \citep{SG02,
IBG05, CS05, BHA05}. 

\begin{figure}[htbp]
\begin{center}
\centerline{\resizebox{16cm}{!}{\includegraphics[angle=0]{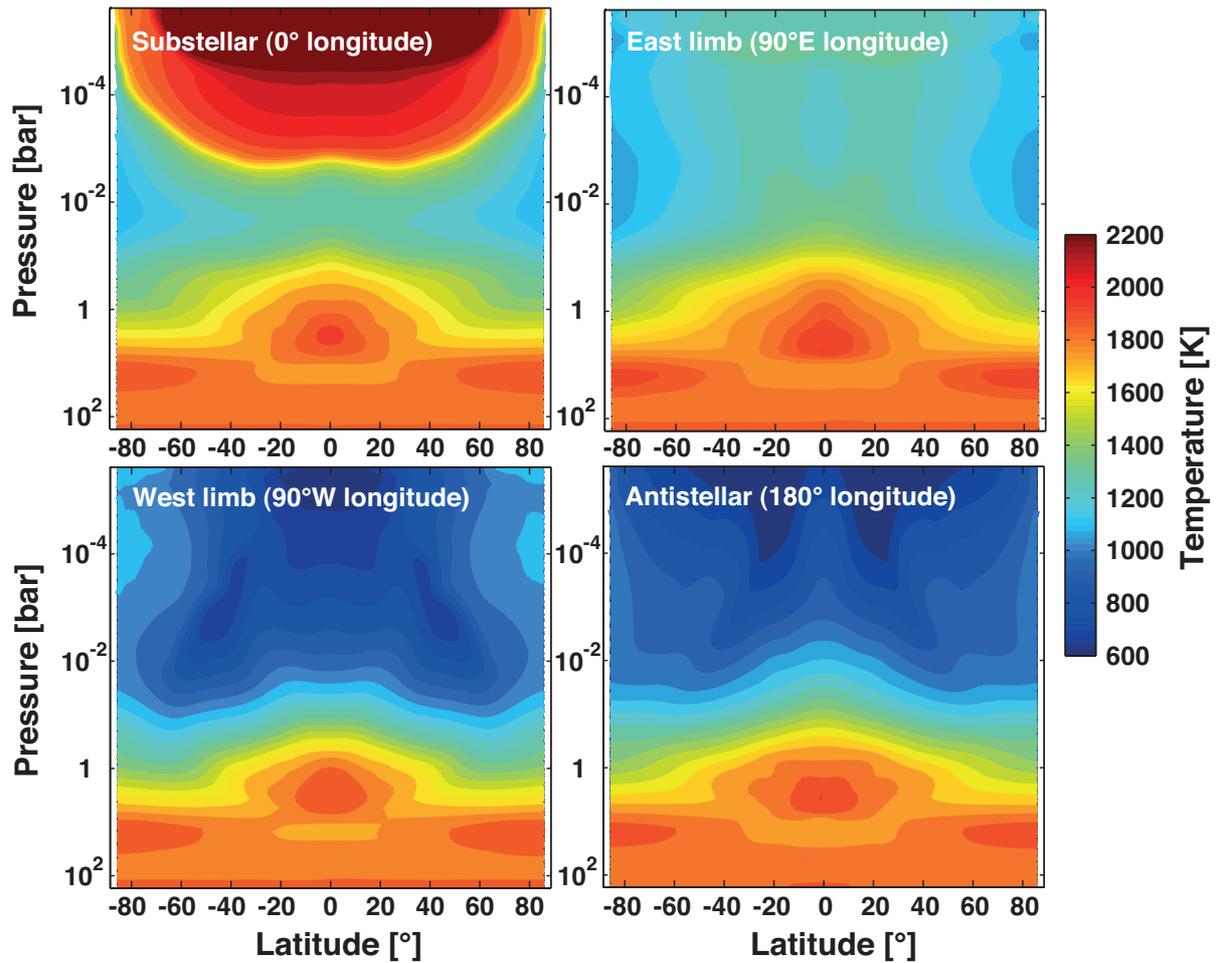}}}
\caption{Simulated temperatures as a function of latitude ($0^\circ$ at the equator, $\pm 90^\circ$ at the poles) and pressure for HD209458\,b obtained from a global 3D global circulation model. The four panels correspond to cuts from pole to pole at different longitudes (clockwise from the upper left): $0^\circ$ (crossing the substellar point), $90^\circ$E (along the east limb), $180^\circ$E (crossing the antistellar point) and $270^\circ$E (along the west limb). A superrotating equatorial jet ($0^\circ$ latitude) is present and characterized by warmer temperatures than its surrounding at pressures of a few bars. The globally west to east circulation is responsible for a pronounced asymmetry between the warmer east limb and colder west limb. [Figure based on \citet{Parmentier+2013}. Courtesy of V. Parmentier.]}
\label{fig:parmentier}
\end{center}
\end{figure}

Figure~\ref{fig:parmentier} illustrates the expected structure for the atmosphere of HD209458b from a modern, tri-dimensional global circulation model coupled with a one-dimensional radiative transfer algorithm \citep{Parmentier+2013}. All the caveats concerning these overforced simulations discussed in section~\ref{sec:dynamics} of course also apply and add to the uncertainties stemming from the poorly known chemical composition. For example, the particular simulation of fig.~\ref{fig:parmentier} assumes the presence of TiO in the atmosphere, which yields very high temperatures at low pressures on the day side of the planet. It is not clear that this molecule is present or has condensed at deeper levels \citep[see also][]{Spiegel+2009}. Aside from that, the eastward equatorial circulation and the strong equator to pole gradient now appear to be a robust feature of these simulations \citep[e.g.,][]{SG02,RauscherMenou2013,Parmentier+2013}. As seen in fig.~\ref{fig:parmentier}, the equatorial jet redistributes heat between the day side and the night side relatively efficiently at large pressures and on the equator, but this is not the case at the poles, and at low pressures, in line with the observational constraints (see section~\ref{sec:exoplanets}).

These strong temperature variations must influence at some point the cooling and contraction histories of hot Jupiters. When opacities variations are not included, they result in more loss of the intrinsic heat and a faster contraction than when assuming that the stellar irradiation is homogeneously redistributed across the planetary surface \citep{GS02,Guillot2010,Budaj+2012,SpiegelBurrows2013}. However, given other uncertainties (e.g., on the chemical composition and opacities to be used), this has been neglected in planetary evolution models thus far. 


\subsubsection{Thermal evolution and inferred compositions}

We have seen in fig.~\ref{fig:mass_rad} that the measured masses and
radii of transiting planets can be globally explained in the framework
of an evolution model including the strong stellar irradiation and the
presence of a variable mass of heavy elements, either in the form of a
central core, or spread in the planet interior. However, when
analyzing the situation for each planet, it appears that several
planets are too large to be reproduced by standard models, i.e., models
using the most up-to-date equations of state, opacities, atmospheric
boundary conditions and assuming that the planetary luminosity
governing its cooling is taken solely from the lost gravitational
potential energy (see section~\ref{sec:virial}).

\begin{figure}[htbp]
\centerline{\resizebox{12cm}{!}{\includegraphics[angle=0]{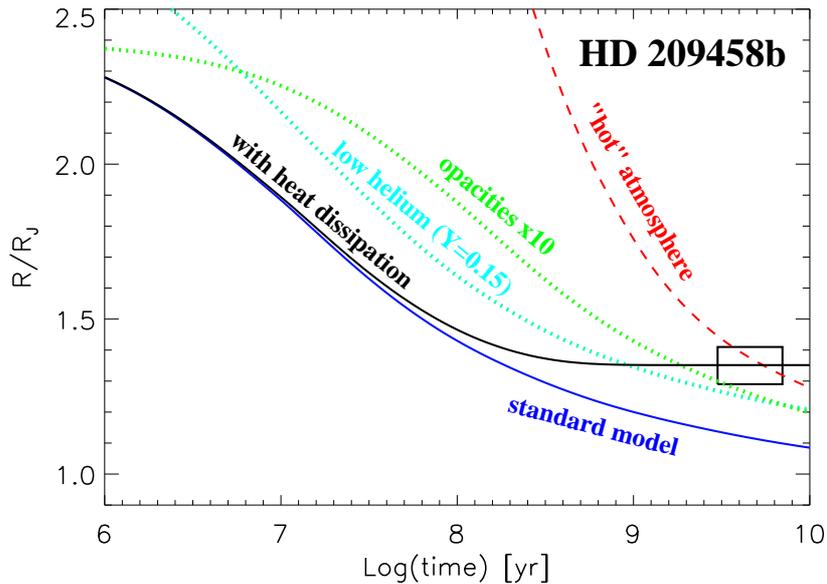}}}
\caption{The contraction of HD209458b as a function of time can
be compared to its measured radius and inferred age shown by the black
box. Standard models (blue curve) for the evolution of that $0.69\,\rm
M_{\rm J}$ planet generally yield a radius that is too small compared
to the observations, even for a solar composition and no central core
(a larger core and -in most cases- larger amounts of heavy elements in
the planet imply an even smaller size for a given
age). Unrealistically low helium abundances or high opacities models
lead to evolution tracks that barely cross the observational box. A
possibility is that heat is dissipated into the deep interior by
stellar tides, either related to a non-zero orbital eccentricity
forced by an unseen companion, or because of a constant transfer of
angular momentum from the heated atmosphere to the interior (black
curve). Alternatively, the atmosphere may be hotter than predicted due
to heating by strong zonal winds and shear instabilities (red curve).}
\label{fig:ev-hd209458b}
\end{figure}

Figure~\ref{fig:ev-hd209458b} illustrates the situation for the
particular case of HD209458b: unless using an unrealistically hot
atmosphere, or arbitrarily increasing the internal opacity, or
decreasing the helium content, one cannot reproduce the observed
radius which is 10 to 20\% larger than calculated using standard models \citep{BLM01, BLL03,
  GS02, Baraffe+03}. The
fact that the measured radius corresponds to a low-pressure
($\sim$mbar) level while the calculated radius corresponds to a level
near 1\,bar is not negligible \citep{BSH03} but too small to
account for the difference. This is problematic because while it is
easy to invoke the presence of a massive core to explain the small
size of a planet, a large size such as that of HD209458b requires
an additional energy source, or significant modifications in the
data/physics involved.

The discovery of many transiting hot Jupiters has shown that this phenomenon is widespread, with at least a third of them being oversized compared to predictions from the standard evolution of a solar-composition planet with no core \citep{Guillot+06,Guillot08,Laughlin+11}. Numerous explanations have been put forward to explain this large size. The first ones, invoking tidal dissipation of eccentricity \citep{BLM01} or inclination \citep{WinnHolman2005} imply that orbital energy is transfered to the planet. These are generally too short-lived \citep[e.g.,][]{Leconte+10} or of a low probability of occurrence  \citep{Levrard+07}. The second ones posit that part of the irradiation energy is transferred into kinetic energy and is then dissipated deeper into the planet. This is the case of weather-noise \citep{SG02},  ohmic dissipation \citep{BS10}, thermal tides \citep{ArrasSocrates10} and turbulent burial \citep{YoudinMitchell10} models. These mechanisms appear quite promising as they are long-lived and generally require only a small fraction of order 1\% or less of the irradiation luminosity to be transported and dissipated at deeper levels to explain the observed planets \citep{GS02}. Finally, a third class of models is based on a reduced cooling, either through an ad hoc increase of opacities \citep{Burrows+07} or inefficient heat transport due to semi-convection \citep{CB07}. Validating these models is becoming possible thanks to a large number of planets allowing statistical tests \citep[see][]{Laughlin+11}, but will require further work.

\begin{figure}
\centerline{\resizebox{10cm}{!}{\includegraphics{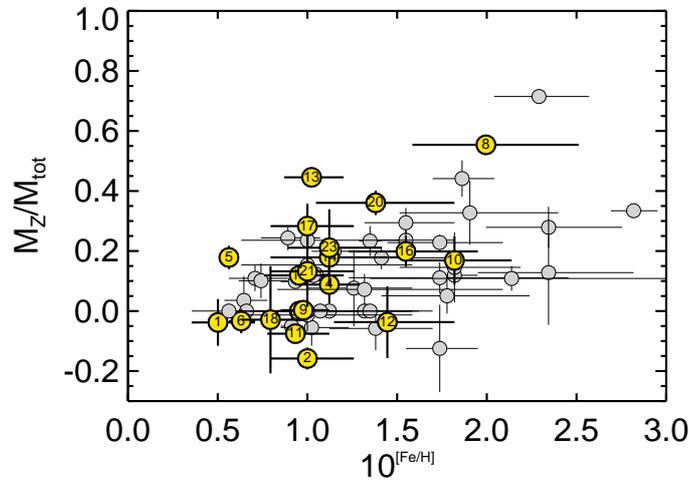}}}
\caption{Mass fraction of heavy elements in the planets as a function of the metal content of their parent star expressed in solar units (e.g., $10^{\rm [Fe/H]} = 3$ implies that the star contains three times more iron than our Sun). The evolution model assumes that 0.25\% of the incoming irradiation flux is dissipated at the planet's center. Circles which are labeled 1 to 23 correspond to the CoRoT giant planets. Gray symbols correspond to a subset of known transiting systems \citep{Guillot08, Laughlin+11}. Unphysical negative values for $M_Z$ correspond to insufficient heat sources leading to a radius that is larger than observed. [From \cite{Moutou+2013}.]}
\label{fig:correlation}
\end{figure}

In any case, the fact that a large number of planets are oversized lends weight to a mechanism that would apply to each planet. Masses of
heavy elements can then be derived by imposing that all planets should be
fitted by the same model with the same hypotheses. This can be done by
inverting the results of \rfig{mass_rad}, as proposed by
\citet{Guillot+06}. On the basis of this hypothesis, figure~\ref{fig:correlation} shows that some of the hot Jupiters contain a large fraction of heavy elements in their interior and that this fraction is correlated with the metallicity of the parent star. The large fraction of heavy elements is inferred both for relatively small giant planets (i.e., Saturn mass) and for planets which are several times the mass of Jupiter. A typical archetype of the first is HD149026\,b which must contain around $70\mea$ of heavy elements, a conclusion that is hard to escape because of the low total mass and high irradiation of the planet \citep[see][]{Ikoma+06, Fortney+06}. For the latter, CoRoT-20b appears to push the models to the limit, with predicted masses of heavy elements in excess of $400\mea$ \citep{Deleuil+12}. 

The correlation between mass of heavy elements in hot Jupiters and stellar metallicity first obtained by \cite{Guillot+06} appears to stand the trial of time \citep{Burrows+07,Guillot08,Laughlin+11,Moutou+2013}, and remains valid when applied to planets with low irradiation levels which do not require additional physics to explain their large sizes \citep{MF11}. It is important to realize that simply accreting slightly more metal-rich gas with the composition of the parent star would lead to a much smaller increase of a few percent at most. This correlation requires an efficient mechanism to collect solids in the protoplanetary disk and bring them into the hot Jupiters, something that is just beginning to be included into planet formation models \citep{Mordasini+12}. 

Another intriguing possibility concerning hot Jupiters is that of a
sustained mass loss due to the high irradiation dose that the planets
receive. Indeed, this effect was predicted \citep{BL95, Guillot+96,
Lammer+03} and detected \citep{VidalMadjar+03,Fossati+10,Bourrier+13}. While its magnitude is still uncertain, it appears to have sculpted the population of planets in very close orbits around their star \citep[e.g.,][]{Lopez+12}. 

The harvest of the Kepler and CoRoT missions opens the possibility to extend these studies to smaller planets. These objects are especially interesting but pose difficult
problems in terms of structure because depending on their formation
history, precise composition and location, they may be fluid, solid,
or they may even possess a global liquid ocean \citep[see][]{Kuchner03,
  Leger+04}.

\section{Implications for planetary formation models}

The giant planets in our Solar System have in common a
large mass of hydrogen and helium, but they are obviously quite
different in their appearances, compositions and internal structures. Although
studies cannot be conducted with the same level of details, we can
safely conclude that extrasolar planets show a greater variety
of compositions and structures, and imagine that their appearances differ even more significantly. 

A parallel study of the structures of our giant planets and of giant
planets orbiting around other stars should provide us with key
information regarding planet formation in the next decade or so. But,
already, some conclusions, some of them robust, others still
tentative, can be drawn:

{\it Giant planets formed in circumstellar disks, before these were
  completely dissipated:}\\
This is a relatively obvious consequence of the fact that giant
  planets are mostly made of hydrogen and helium: these elements had
  to be acquired when they were still present in the disk. Because the
  observed lifetime of gaseous circumstellar disks is of the order of
  a few million years, this implies that these planets formed
  (i.e., acquired most of their final masses) in a few million years
  also, quite faster than terrestrial planets in the Solar System. 

{\it Giant planets migrated:}\\
The
observed orbital distribution of extrasolar planets and the presence of planets extremely close to their star is generally taken a strong evidence for an inward
migration of planets, and various mechanisms have been proposed for
that \citep[see][...etc.]{IL04a, AMBW05, MA05}. Separately, it was
shown that several properties of our 
Solar System can be explained if Jupiter, Saturn, Uranus and Neptune
ended the early formation phase in the presence of a disk with
quasi-circular orbit, and with Saturn, Uranus and Neptune
significantly closer to the Sun than they are now, and that these
three planets subsequently migrated outward \citep{TGML05}.

{\it Accretion played a key role for giant planet formation:}\\
Although formation by direct gas instability still remains a possibility \citep[e.g.,][]{HelledBodenheimer2011,Boley+2011}, 
several indications point towards a formation of giant planets that is
dominated by accretion of heavy elements: First, Jupiter, Saturn,
Uranus and Neptune are all significantly enriched in heavy elements
compared to the Sun. This feature can be reproduced by core-accretion
models, for Jupiter and Saturn at least \citep{AMBW05}. Second,
the probability to find a giant planet around a solar-type star (with
stellar type F, G or K) is a strongly rising function of stellar
metallicity \citep{Gonzalez98, SIM04, FV05}, a property
that is also well-reproduced by standard core accretion models
\citep{IL04b, AMBW05}. Third, the large masses of heavy
elements inferred in some transiting extrasolar planets as well as the
apparent correlation between mass of heavy elements in the planet and
stellar metallicity \citep{Guillot+06,Burrows+07,Guillot08,Laughlin+11} is a strong indication that accretion was
possible and that it was furthermore efficient.

{\it Giant planets were enriched in heavy elements by core accretion,
  planetesimal delivery and/or formation in an enriched protoplanetary
  disk:}\\
The giant planets in our Solar System are unambigously enriched in
  heavy elements compared to the Sun, both globally, and when
  considering their atmosphere. This may also be the case of
  extrasolar planets, although the evidence is still tenuous.
  The accretion of a central core can explain part of the global
  enrichment, but not that of the atmosphere. The accretion of
  planetesimals may be a possible solution but in the case of Jupiter
  at least the rapid drop in accretion efficiency as the planet
  reaches appreciable masses ($\sim 100\mea$ or so) implies that such
  an enrichment would have originally concerned only very deep layers,
  and would require a relatively efficient upper mixing of these
  elements, and possibly an erosion of the central core \citep{GSHS04,WilsonMilitzer2012}. 

Although not unambiguously explained, the fact that Jupiter is also
enriched in noble gases compared to the Sun is a key observation to
understand some of the processes occuring in the early Solar
System. Indeed, noble gases are trapped into solids only at very low
temperatures, and this tells us either that most of the solids that
formed Jupiter were formed at very low temperature to be able to trap
gases such as argon, probably as clathrates \citep{GHML01, HGL04,Mousis+2012}, or
that the planet formed in an enriched disk as 
it was being evaporated \citep{GH06}.  The fact that in Jupiter, argon, krypton and xenon have a comparable enrichment over the solar value within the error bars \citep[see][and table~\ref{tab:comp}]{Lodders2008} slightly favors the latter explanation.

\section{Future prospects}

We have shown that the compositions and structures of giant planets
remain very uncertain. This is an important problem when attempting
to understand and constrain the formation of planets, and the origins
of the Solar System. However, the parallel study of giant planets in
our Solar System by space missions such as Galileo and Cassini, and
of extrasolar planets by both ground based and space programs has led
to rapid improvements in the field, with in particular a precise
determination of the composition of Jupiter's troposphere, and
constraints on the compositions of a dozen of extrasolar planets.

Improvements on our knowledge of the giant planets requires a variety
of efforts. Fortunately, nearly all of these are addressed at least
partially by adequate projects in the next few years. The efforts that
are necessary thus include (but are not limited to):
\begin{itemize}
\item Continue progresses on EOSs in order to obtain reliable results that can be used for a wide range of temperatures and pressures in the astrophysical context. This should be done with the help of laboratory experiments, for instance by using powerful lasers such as the NIF in the USA and
  the M\'egaJoule laser in France. Extensive numerical calculations should be performed as well, in particular with mixtures of elements. 
\item Calculate phase diagrams for a variety of mixtures, in particular involving superionic water, rocks, iron (and hydrogen). The hydrogen-helium phase diagram should also be refined because it is critical to understand the evolution and structure of Jupiter and Saturn. 
\item Have a better yardstick to measure solar and protosolar
  compositions. This has not been fully addressed by the Genesis
  mission and may require another mission and/or progresses in modeling the Sun's composition. 
\item Improve the measurement of Jupiter's gravity and magnetic fields, and
  determine the abundance of water in the deep atmosphere. This will
  be done by the Juno mission \citep{Bolton2010} which is to arrive at Jupiter in 2016, and whose polar orbit skimming a mere 5000\,km above the cloud tops should enable exquisite measurements of these quantities.  
\item Measure with high precision Saturn's gravity field. Saturn's gravitational moments have already been improved, but an important increase in accuracy can be obtained as part of the Cassini Solstice mission \citep{Spilker2012}, while the spacecraft plunges onto the planet. This should lead to
  better constraints, and possibly a determination of whether the
  interior of Saturn rotates as a solid body. 
\item Pursue the discovery of transiting extrasolar planets including some with longer orbital periods and around bright stars. A large number of these objects will enable detailed statistical studies which will be key in understanding this population of objects. 
\item Develop consistent models for the formation, evolution, present structure and magnetic field of Uranus and Neptune, in order to understand ice giants as a class of planets. 
\item It would be highly desirable to send a probe similar to the
  Galileo probe into Saturn's atmosphere \citep[e.g.,][]{Marty+2009}. The comparison of the
  abundance of noble gases would discriminate between different models
  of the enrichment of the giant planets, and the additional
  measurement of key isotopic ratio would provide further tests to
  understand our origins.
\item In the long term, a mission to the ice giants Uranus or Neptune would bring new views of these fascinating planets and help to complete our knowledge of the outer solar system. 
\end{itemize}

Clearly, there is a lot of work on the road, but the prospects for a
much improved knowledge of giant planets and their formation are
bright.

\section*{Acknowledgements}

The manuscript improved significantly thanks to the insightful reviews of Nadine Nettelmann and another reviewer. The authors also wish to thank Vivien Parmentier, Leigh Fletcher, Didier Saumon, Emmanuel Lellouch, Paul Loubeyre, Ravit Helled, Bill Hubbard, Erich Karkoschka, Imke De Pater and Miguel Morales for very useful comments and suggestions.

\bibliographystyle{elsarticle-harv.bst}
\bibliography{geophys_guillot.bib}

\begin{thebibliography}{265}
\expandafter\ifx\csname natexlab\endcsname\relax\def\natexlab#1{#1}\fi
\expandafter\ifx\csname url\endcsname\relax
  \def\url#1{\texttt{#1}}\fi
\expandafter\ifx\csname urlprefix\endcsname\relax\def\urlprefix{URL }\fi

\bibitem[{{Acuna} et~al.(1983){Acuna}, {Connerney}, and
  {Ness}}]{1983JGR....88.8771A}
{Acuna}, M.~H., {Connerney}, J.~E.~P., {Ness}, N.~F., Nov. 1983. {The Z3 zonal
  harmonic model of Saturn's magnetic field Analyses and implications}. \jgr
  88, 8771--8778.

\bibitem[{{Alibert} et~al.(2005){Alibert}, {Mordasini}, {Benz}, and
  {Winisdoerffer}}]{AMBW05}
{Alibert}, Y., {Mordasini}, C., {Benz}, W., {Winisdoerffer}, C., Apr. 2005.
  {Models of giant planet formation with migration and disc evolution}. \aap
  434, 343--353.

\bibitem[{{Anderson} et~al.(1987){Anderson}, {Campbell}, {Jacobson},
  {Sweetnam}, and {Taylor}}]{1987JGR....9214877A}
{Anderson}, J.~D., {Campbell}, J.~K., {Jacobson}, R.~A., {Sweetnam}, D.~N.,
  {Taylor}, A.~H., Dec. 1987. {Radio science with Voyager 2 at Uranus - Results
  on masses and densities of the planet and five principal satellites}. \jgr
  92, 14877--14883.

\bibitem[{{Anderson} and {Schubert}(2007)}]{AndersonSchubert2007}
{Anderson}, J.~D., {Schubert}, G., Sep. 2007. {Saturn's Gravitational Field,
  Internal Rotation, and Interior Structure}. Science 317, 1384--.

\bibitem[{{Arras} and {Socrates}(2010)}]{ArrasSocrates10}
{Arras}, P., {Socrates}, A., May 2010. {Thermal Tides in Fluid Extrasolar
  Planets}. \apj 714, 1--12.

\bibitem[{{Atreya} et~al.(2003){Atreya}, {Mahaffy}, {Niemann}, {Wong}, and
  {Owen}}]{Atreya+2003}
{Atreya}, S.~K., {Mahaffy}, P.~R., {Niemann}, H.~B., {Wong}, M.~H., {Owen},
  T.~C., Feb. 2003. {Composition and origin of the atmosphere of Jupiter - an
  update, and implications for the extrasolar giant planets}. \planss 51,
  105--112.

\bibitem[{{Bahcall} et~al.(1995){Bahcall}, {Pinsonneault}, and
  {Wasserburg}}]{1995RvMP...67..781B}
{Bahcall}, J.~N., {Pinsonneault}, M.~H., {Wasserburg}, G.~J., Oct. 1995. {Solar
  models with helium and heavy-element diffusion}. Reviews of Modern Physics
  67, 781--808.

\bibitem[{{Baines} and {Smith}(1990)}]{BainesSmith1990}
{Baines}, K.~H., {Smith}, H.~W., May 1990. {The atmospheric structure and
  dynamical properties of Neptune derived from ground-based and IUE
  spectrophotometry}. \icarus 85, 65--108.

\bibitem[{{Bakos} et~al.(2009){Bakos}, {Howard}, {Noyes}, {Hartman}, {Torres},
  {Kov{\'a}cs}, {Fischer}, {Latham}, {Johnson}, {Marcy}, {Sasselov},
  {Stefanik}, {Sip{\H o}cz}, {Kov{\'a}cs}, {Esquerdo}, {P{\'a}l},
  {L{\'a}z{\'a}r}, {Papp}, and {S{\'a}ri}}]{Bakos+2009}
{Bakos}, G.~{\'A}., {Howard}, A.~W., {Noyes}, R.~W., {Hartman}, J., {Torres},
  G., {Kov{\'a}cs}, G., {Fischer}, D.~A., {Latham}, D.~W., {Johnson}, J.~A.,
  {Marcy}, G.~W., {Sasselov}, D.~D., {Stefanik}, R.~P., {Sip{\H o}cz}, B.,
  {Kov{\'a}cs}, G., {Esquerdo}, G.~A., {P{\'a}l}, A., {L{\'a}z{\'a}r}, J.,
  {Papp}, I., {S{\'a}ri}, P., Dec. 2009. {HAT-P-13b,c: A Transiting Hot Jupiter
  with a Massive Outer Companion on an Eccentric Orbit}. \apj 707, 446--456.

\bibitem[{{Baraffe} et~al.(2003){Baraffe}, {Chabrier}, {Barman}, {Allard}, and
  {Hauschildt}}]{Baraffe+03}
{Baraffe}, I., {Chabrier}, G., {Barman}, T.~S., {Allard}, F., {Hauschildt},
  P.~H., May 2003. {Evolutionary models for cool brown dwarfs and extrasolar
  giant planets. The case of HD 209458}. \aap 402, 701--712.

\bibitem[{{Baraffe} et~al.(2005){Baraffe}, {Chabrier}, {Barman}, {Selsis},
  {Allard}, and {Hauschildt}}]{Baraffe+05}
{Baraffe}, I., {Chabrier}, G., {Barman}, T.~S., {Selsis}, F., {Allard}, F.,
  {Hauschildt}, P.~H., Jun. 2005. {Hot-Jupiters and hot-Neptunes: A common
  origin?} \aap 436, L47--L51.

\bibitem[{{Barman} et~al.(2001){Barman}, {Hauschildt}, and {Allard}}]{BHA01}
{Barman}, T.~S., {Hauschildt}, P.~H., {Allard}, F., Aug. 2001. {Irradiated
  Planets}. \apj 556, 885--895.

\bibitem[{{Barman} et~al.(2005){Barman}, {Hauschildt}, and {Allard}}]{BHA05}
{Barman}, T.~S., {Hauschildt}, P.~H., {Allard}, F., Oct. 2005. {Phase-Dependent
  Properties of Extrasolar Planet Atmospheres}. \apj 632, 1132--1139.

\bibitem[{{Batygin} et~al.(2009){Batygin}, {Bodenheimer}, and
  {Laughlin}}]{Batygin+2009b}
{Batygin}, K., {Bodenheimer}, P., {Laughlin}, G., Oct. 2009. {Determination of
  the Interior Structure of Transiting Planets in Multiple-Planet Systems}.
  \apjl 704, L49--L53.

\bibitem[{{Batygin} and {Stevenson}(2010)}]{BS10}
{Batygin}, K., {Stevenson}, D.~J., May 2010. {Inflating Hot Jupiters with Ohmic
  Dissipation}. \apjl 714, L238--L243.

\bibitem[{{Bercovici} and {Schubert}(1987)}]{BS87}
{Bercovici}, D., {Schubert}, G., Mar. 1987. Jovian seismology. \icarus 69,
  557--565.

\bibitem[{{Bethkenhagen} et~al.(2013){Bethkenhagen}, {French}, and
  {Redmer}}]{Bethkenhagen+2013}
{Bethkenhagen}, M., {French}, M., {Redmer}, R., Jun. 2013. {Equation of state
  and phase diagram of ammonia at high pressures from ab initio simulations}.
  \jcp 138~(23), 234504.

\bibitem[{{Bodenheimer} et~al.(2003){Bodenheimer}, {Laughlin}, and
  {Lin}}]{BLL03}
{Bodenheimer}, P., {Laughlin}, G., {Lin}, D.~N.~C., Jul. 2003. {On the Radii of
  Extrasolar Giant Planets}. \apj 592, 555--563.

\bibitem[{{Bodenheimer} et~al.(2001){Bodenheimer}, {Lin}, and
  {Mardling}}]{BLM01}
{Bodenheimer}, P., {Lin}, D.~N.~C., {Mardling}, R.~A., Feb. 2001. {On the Tidal
  Inflation of Short-Period Extrasolar Planets}. \apj 548, 466--472.

\bibitem[{{Boley} et~al.(2011){Boley}, {Helled}, and {Payne}}]{Boley+2011}
{Boley}, A.~C., {Helled}, R., {Payne}, M.~J., Jul. 2011. {The Heavy-element
  Composition of Disk Instability Planets Can Range from Sub- to
  Super-nebular}. \apj 735, 30.

\bibitem[{{Bolton}(2010)}]{Bolton2010}
{Bolton}, S.~J., Jan. 2010. {The Juno Mission}. In: {Barbieri}, C.,
  {Chakrabarti}, S., {Coradini}, M., {Lazzarin}, M. (Eds.), IAU Symposium. Vol.
  269 of IAU Symposium. pp. 92--100.

\bibitem[{{Bonev} et~al.(2004){Bonev}, {Militzer}, and {Galli}}]{BMG04}
{Bonev}, S.~A., {Militzer}, B., {Galli}, G., Jan. 2004. {Ab initio simulations
  of dense liquid deuterium: Comparison with gas-gun shock-wave experiments}.
  \prb 69~(1), 014101--+.

\bibitem[{{Boriskov} et~al.(2005){Boriskov}, {Bykov}, {Il'Kaev}, {Selemir},
  {Simakov}, {Trunin}, {Urlin}, {Shuikin}, and {Nellis}}]{Boriskov+2005}
{Boriskov}, G.~V., {Bykov}, A.~I., {Il'Kaev}, R.~I., {Selemir}, V.~D.,
  {Simakov}, G.~V., {Trunin}, R.~F., {Urlin}, V.~D., {Shuikin}, A.~N.,
  {Nellis}, W.~J., Mar. 2005. {Shock compression of liquid deuterium up to 109
  GPa}. \prb 71~(9), 092104.

\bibitem[{{Borysow} et~al.(1997){Borysow}, {Jorgensen}, and
  {Zheng}}]{1997AA...324..185B}
{Borysow}, A., {Jorgensen}, U.~G., {Zheng}, C., Aug. 1997. {Model atmospheres
  of cool, low-metallicity stars: the importance of collision-induced
  absorption.} \aap 324, 185--195.

\bibitem[{{Bouchy} et~al.(2011){Bouchy}, {Deleuil}, {Guillot}, {Aigrain},
  {Carone}, {Cochran}, {Almenara}, {Alonso}, {Auvergne}, {Baglin}, {Barge},
  {Bonomo}, {Bord{\'e}}, {Csizmadia}, {de Bondt}, {Deeg}, {D{\'{\i}}az},
  {Dvorak}, {Endl}, {Erikson}, {Ferraz-Mello}, {Fridlund}, {Gandolfi},
  {Gazzano}, {Gibson}, {Gillon}, {Guenther}, {Hatzes}, {Havel}, {H{\'e}brard},
  {Jorda}, {L{\'e}ger}, {Lovis}, {Llebaria}, {Lammer}, {MacQueen}, {Mazeh},
  {Moutou}, {Ofir}, {Ollivier}, {Parviainen}, {P{\"a}tzold}, {Queloz}, {Rauer},
  {Rouan}, {Santerne}, {Schneider}, {Tingley}, and {Wuchterl}}]{Bouchy+2011}
{Bouchy}, F., {Deleuil}, M., {Guillot}, T., {Aigrain}, S., {Carone}, L.,
  {Cochran}, W.~D., {Almenara}, J.~M., {Alonso}, R., {Auvergne}, M., {Baglin},
  A., {Barge}, P., {Bonomo}, A.~S., {Bord{\'e}}, P., {Csizmadia}, S., {de
  Bondt}, K., {Deeg}, H.~J., {D{\'{\i}}az}, R.~F., {Dvorak}, R., {Endl}, M.,
  {Erikson}, A., {Ferraz-Mello}, S., {Fridlund}, M., {Gandolfi}, D., {Gazzano},
  J.~C., {Gibson}, N., {Gillon}, M., {Guenther}, E., {Hatzes}, A., {Havel}, M.,
  {H{\'e}brard}, G., {Jorda}, L., {L{\'e}ger}, A., {Lovis}, C., {Llebaria}, A.,
  {Lammer}, H., {MacQueen}, P.~J., {Mazeh}, T., {Moutou}, C., {Ofir}, A.,
  {Ollivier}, M., {Parviainen}, H., {P{\"a}tzold}, M., {Queloz}, D., {Rauer},
  H., {Rouan}, D., {Santerne}, A., {Schneider}, J., {Tingley}, B., {Wuchterl},
  G., Jan. 2011. {Transiting exoplanets from the CoRoT space mission. XV.
  CoRoT-15b: a brown-dwarf transiting companion}. \aap 525, A68.

\bibitem[{{Bourrier} et~al.(2013){Bourrier}, {Lecavelier des Etangs}, {Dupuy},
  {Ehrenreich}, {Vidal-Madjar}, {H{\'e}brard}, {Ballester}, {D{\'e}sert},
  {Ferlet}, {Sing}, and {Wheatley}}]{Bourrier+13}
{Bourrier}, V., {Lecavelier des Etangs}, A., {Dupuy}, H., {Ehrenreich}, D.,
  {Vidal-Madjar}, A., {H{\'e}brard}, G., {Ballester}, G.~E., {D{\'e}sert},
  J.-M., {Ferlet}, R., {Sing}, D.~K., {Wheatley}, P.~J., Mar. 2013.
  {Atmospheric escape from HD 189733b observed in H I Lyman-{$\alpha$}:
  detailed analysis of HST/STIS September 2011 observations}. \aap 551, A63.

\bibitem[{{Briggs} and {Sackett}(1989)}]{BS89}
{Briggs}, F.~H., {Sackett}, P.~D., Jul. 1989. {Radio observations of Saturn as
  a probe of its atmosphere and cloud structure}. Icarus 80, 77--103.

\bibitem[{{Budaj} et~al.(2012){Budaj}, {Hubeny}, and {Burrows}}]{Budaj+2012}
{Budaj}, J., {Hubeny}, I., {Burrows}, A., Jan. 2012. {Day and night side core
  cooling of a strongly irradiated giant planet}. \aap 537, A115.

\bibitem[{{Burrows}(2013)}]{Burrows+2014}
{Burrows}, A., Dec. 2013. {Spectra as Windows into Exoplanet Atmospheres}.
  ArXiv e-prints.

\bibitem[{{Burrows} et~al.(2000){Burrows}, {Guillot}, {Hubbard}, {Marley},
  {Saumon}, {Lunine}, and {Sudarsky}}]{Burrows+00}
{Burrows}, A., {Guillot}, T., {Hubbard}, W.~B., {Marley}, M.~S., {Saumon}, D.,
  {Lunine}, J.~I., {Sudarsky}, D., May 2000. {On the Radii of Close-in Giant
  Planets}. \apjl 534, L97--L100.

\bibitem[{{Burrows} et~al.(2007){Burrows}, {Hubeny}, {Budaj}, and
  {Hubbard}}]{Burrows+07}
{Burrows}, A., {Hubeny}, I., {Budaj}, J., {Hubbard}, W.~B., May 2007. {Possible
  Solutions to the Radius Anomalies of Transiting Giant Planets}. \apj 661,
  502--514.

\bibitem[{{Burrows} and {Lunine}(1995)}]{BL95}
{Burrows}, A., {Lunine}, J., Nov. 1995. {Extrasolar Planets - Astronomical
  Questions of Origin and Survival}. \nat 378, 333--+.

\bibitem[{{Burrows} et~al.(1997){Burrows}, {Marley}, {Hubbard}, {Lunine},
  {Guillot}, {Saumon}, {Freedman}, {Sudarsky}, and {Sharp}}]{Burrows+97}
{Burrows}, A., {Marley}, M., {Hubbard}, W.~B., {Lunine}, J.~I., {Guillot}, T.,
  {Saumon}, D., {Freedman}, R., {Sudarsky}, D., {Sharp}, C., Dec. 1997. {A
  Nongray Theory of Extrasolar Giant Planets and Brown Dwarfs}. \apj 491,
  856--+.

\bibitem[{{Burrows} et~al.(2003){Burrows}, {Sudarsky}, and {Hubbard}}]{BSH03}
{Burrows}, A., {Sudarsky}, D., {Hubbard}, W.~B., Sep. 2003. {A Theory for the
  Radius of the Transiting Giant Planet HD 209458b}. \apj 594, 545--551.

\bibitem[{{Busse}(1978)}]{Busse78}
{Busse}, F.~H., 1978. {Magnetohydrodynamics of the Earth's Dynamo}. Annual
  Review of Fluid Mechanics 10, 435--462.

\bibitem[{{Caillabet} et~al.(2011){Caillabet}, {Mazevet}, and
  {Loubeyre}}]{Caillabet+2011}
{Caillabet}, L., {Mazevet}, S., {Loubeyre}, P., Mar. 2011. {Multiphase equation
  of state of hydrogen from ab initio calculations in the range 0.2 to 5 g/cc
  up to 10 eV}. \prb 83~(9), 094101.

\bibitem[{{Campbell} and {Synnott}(1985)}]{1985AJ.....90..364C}
{Campbell}, J.~K., {Synnott}, S.~P., Feb. 1985. {Gravity field of the Jovian
  system from Pioneer and Voyager tracking data}. \aj 90, 364--372.

\bibitem[{{Cavazzoni} et~al.(1999){Cavazzoni}, {Chiarotti}, {Scandolo},
  {Tosatti}, {Bernasconi}, and {Parrinello}}]{Cavazzoni+1999}
{Cavazzoni}, C., {Chiarotti}, G.~L., {Scandolo}, S., {Tosatti}, E.,
  {Bernasconi}, M., {Parrinello}, M., Jan. 1999. {Superionic and Metallic
  States of Water and Ammonia at Giant Planet Conditions}. Science 283, 44.

\bibitem[{{Cecconi} and {Zarka}(2005)}]{2005JGRA..11012203C}
{Cecconi}, B., {Zarka}, P., Dec. 2005. {Model of a variable radio period for
  Saturn}. Journal of Geophysical Research (Space Physics) 110, 12203--+.

\bibitem[{{Chabrier} and {Ashcroft}(1990)}]{ChabrierAshcroft1990}
{Chabrier}, G., {Ashcroft}, N.~W., Aug. 1990. {Linear mixing rule in screened
  binary ionic mixtures}. \pra 42, 2284--2291.

\bibitem[{{Chabrier} and {Baraffe}(2007)}]{CB07}
{Chabrier}, G., {Baraffe}, I., May 2007. {Heat Transport in Giant (Exo)planets:
  A New Perspective}. \apjl 661, L81--L84.

\bibitem[{{Chabrier} et~al.(2007){Chabrier}, {Saumon}, and
  {Winisdoerffer}}]{Chabrier+2007}
{Chabrier}, G., {Saumon}, D., {Winisdoerffer}, C., Jan. 2007. {Hydrogen and
  Helium at High Density and Astrophysical Implications}. \apss 307, 263--267.

\bibitem[{{Chandrasekhar}(1939)}]{Chandrasekhar39}
{Chandrasekhar}, S., 1939. {An introduction to the study of stellar structure}.
  Chicago, Ill., The University of Chicago press [1939].

\bibitem[{{Charbonneau} et~al.(2002){Charbonneau}, {Brown}, {Noyes}, and
  {Gilliland}}]{CBNG02}
{Charbonneau}, D., {Brown}, T.~M., {Noyes}, R.~W., {Gilliland}, R.~L., Mar.
  2002. {Detection of an Extrasolar Planet Atmosphere}. \apj 568, 377--384.

\bibitem[{{Christensen-Dalsgaard} et~al.(1996){Christensen-Dalsgaard},
  {Dappen}, {Ajukov}, {Anderson}, {Antia}, {Basu}, {Baturin}, {Berthomieu},
  {Chaboyer}, {Chitre}, {Cox}, {Demarque}, {Donatowicz}, {Dziembowski},
  {Gabriel}, {Gough}, {Guenther}, {Guzik}, {Harvey}, {Hill}, {Houdek},
  {Iglesias}, {Kosovichev}, {Leibacher}, {Morel}, {Proffitt}, {Provost},
  {Reiter}, {Rhodes}, {Rogers}, {Roxburgh}, {Thompson}, and
  {Ulrich}}]{Christensen-Dalsgaard+1996}
{Christensen-Dalsgaard}, J., {Dappen}, W., {Ajukov}, S.~V., {Anderson}, E.~R.,
  {Antia}, H.~M., {Basu}, S., {Baturin}, V.~A., {Berthomieu}, G., {Chaboyer},
  B., {Chitre}, S.~M., {Cox}, A.~N., {Demarque}, P., {Donatowicz}, J.,
  {Dziembowski}, W.~A., {Gabriel}, M., {Gough}, D.~O., {Guenther}, D.~B.,
  {Guzik}, J.~A., {Harvey}, J.~W., {Hill}, F., {Houdek}, G., {Iglesias}, C.~A.,
  {Kosovichev}, A.~G., {Leibacher}, J.~W., {Morel}, P., {Proffitt}, C.~R.,
  {Provost}, J., {Reiter}, J., {Rhodes}, Jr., E.~J., {Rogers}, F.~J.,
  {Roxburgh}, I.~W., {Thompson}, M.~J., {Ulrich}, R.~K., May 1996. {The Current
  State of Solar Modeling}. Science 272, 1286--1292.

\bibitem[{{Cohen} and {Taylor}(1987)}]{1987RvMP...59.1121C}
{Cohen}, E.~R., {Taylor}, B.~N., 1987. {The 1986 adjustment of the fundamental
  physical constants}. Reviews of Modern Physics 59, 1121--1148.

\bibitem[{{Collins} et~al.(1998){Collins}, {da Silva}, {Celliers}, {Gold},
  {Foord}, {Wallace}, {Ng}, {Weber}, {Budil}, and {Cauble}}]{Collins+98}
{Collins}, G.~W., {da Silva}, L.~B., {Celliers}, P., {Gold}, D.~M., {Foord},
  M.~E., {Wallace}, R.~J., {Ng}, A., {Weber}, S.~V., {Budil}, K.~S., {Cauble},
  R., Aug. 1998. {Measurements of the equation of state of deuterium at the
  fluid insulator-metal transition}. Science 281, 1178--1181.

\bibitem[{{Connerney} et~al.(1982){Connerney}, {Ness}, and
  {Acuna}}]{1982Natur.298...44C}
{Connerney}, J.~E.~P., {Ness}, N.~F., {Acuna}, M.~H., Jul. 1982. {Zonal
  harmonic model of Saturn's magnetic field from Voyager 1 and 2 observations}.
  \nat 298, 44--46.

\bibitem[{{Conrath} et~al.(1987){Conrath}, {Hanel}, {Gautier}, {Marten}, and
  {Lindal}}]{Conrath+1987}
{Conrath}, B., {Hanel}, R., {Gautier}, D., {Marten}, A., {Lindal}, G., Dec.
  1987. {The helium abundance of Uranus from Voyager measurements}. \jgr 92,
  15003--15010.

\bibitem[{{Conrath} and {Gautier}(2000)}]{ConrathGautier2000}
{Conrath}, B.~J., {Gautier}, D., Mar. 2000. {Saturn Helium Abundance: A
  Reanalysis of Voyager Measurements}. \icarus 144, 124--134.

\bibitem[{{Conrath} et~al.(1991){Conrath}, {Gautier}, {Lindal}, {Samuelson},
  and {Shaffer}}]{Conrath+1991}
{Conrath}, B.~J., {Gautier}, D., {Lindal}, G.~F., {Samuelson}, R.~E.,
  {Shaffer}, W.~A., Oct. 1991. {The helium abundance of Neptune from Voyager
  measurements}. \jgr 96, 18907.

\bibitem[{{Cooper} and {Showman}(2005)}]{CS05}
{Cooper}, C.~S., {Showman}, A.~P., Aug. 2005. {Dynamic Meteorology at the
  Photosphere of HD 209458b}. \apjl 629, L45--L48.

\bibitem[{{Crossfield} et~al.(2012){Crossfield}, {Barman}, {Hansen}, {Tanaka},
  and {Kodama}}]{Crossfield+2012}
{Crossfield}, I.~J.~M., {Barman}, T., {Hansen}, B.~M.~S., {Tanaka}, I.,
  {Kodama}, T., Dec. 2012. {Re-evaluating WASP-12b: Strong Emission at 2.315
  {$\mu$}m, Deeper Occultations, and an Isothermal Atmosphere}. \apj 760, 140.

\bibitem[{{Crouzet} et~al.(2012){Crouzet}, {McCullough}, {Burke}, and
  {Long}}]{Crouzet+2012}
{Crouzet}, N., {McCullough}, P.~R., {Burke}, C., {Long}, D., Dec. 2012.
  {Transmission Spectroscopy of Exoplanet XO-2b Observed with Hubble Space
  Telescope NICMOS}. \apj 761, 7.

\bibitem[{{da Silva} et~al.(1997){da Silva}, {Celliers}, {Collins}, {Budil},
  {Holmes}, {Barbee}, {Hammel}, {Kilkenny}, {Wallace}, {Ross}, {Cauble}, {Ng},
  and {Chiu}}]{daSilva+97}
{da Silva}, L.~B., {Celliers}, P., {Collins}, G.~W., {Budil}, K.~S., {Holmes},
  N.~C., {Barbee}, Jr., T.~W., {Hammel}, B.~A., {Kilkenny}, J.~D., {Wallace},
  R.~J., {Ross}, M., {Cauble}, R., {Ng}, A., {Chiu}, G., Jan. 1997. {Absolute
  Equation of State Measurements on Shocked Liquid Deuterium up to 200 GPa (2
  Mbar)}. Physical Review Letters 78, 483--486.

\bibitem[{{Davies} et~al.(1986){Davies}, {Abalakin}, {Bursa}, {Lederle}, and
  {Lieske}}]{1986CeMec..39..103D}
{Davies}, M.~E., {Abalakin}, V.~K., {Bursa}, M., {Lederle}, T., {Lieske},
  J.~H., May 1986. {Report of the IAU/IAG/COSPAR working group on cartographic
  coordinates and rotational elements of the planets and satellites - 1985}.
  Celestial Mechanics 39, 103--113.

\bibitem[{{de Pater} et~al.(1991){de Pater}, {Romani}, and
  {Atreya}}]{dePater+1991}
{de Pater}, I., {Romani}, P.~N., {Atreya}, S.~K., Jun. 1991. {Possible
  microwave absorption by H2S gas in Uranus' and Neptune's atmospheres}.
  \icarus 91, 220--233.

\bibitem[{{de Pater} et~al.(2011){de Pater}, {Sromovsky}, {Hammel}, {Fry},
  {LeBeau}, {Rages}, {Showalter}, and {Matthews}}]{dePater+2011}
{de Pater}, I., {Sromovsky}, L.~A., {Hammel}, H.~B., {Fry}, P.~M., {LeBeau},
  R.~P., {Rages}, K., {Showalter}, M., {Matthews}, K., Sep. 2011. {Post-equinox
  observations of Uranus: Berg's evolution, vertical structure, and track
  towards the equator}. \icarus 215, 332--345.

\bibitem[{{Deleuil} et~al.(2012){Deleuil}, {Bonomo}, {Ferraz-Mello}, {Erikson},
  {Bouchy}, {Havel}, {Aigrain}, {Almenara}, {Alonso}, {Auvergne}, {Baglin},
  {Barge}, {Bord{\'e}}, {Bruntt}, {Cabrera}, {Carpano}, {Cavarroc},
  {Csizmadia}, {Damiani}, {Deeg}, {Dvorak}, {Fridlund}, {H{\'e}brard},
  {Gandolfi}, {Gillon}, {Guenther}, {Guillot}, {Hatzes}, {Jorda}, {L{\'e}ger},
  {Lammer}, {Mazeh}, {Moutou}, {Ollivier}, {Ofir}, {Parviainen}, {Queloz},
  {Rauer}, {Rodr{\'{\i}}guez}, {Rouan}, {Santerne}, {Schneider}, {Tal-Or},
  {Tingley}, {Weingrill}, and {Wuchterl}}]{Deleuil+12}
{Deleuil}, M., {Bonomo}, A.~S., {Ferraz-Mello}, S., {Erikson}, A., {Bouchy},
  F., {Havel}, M., {Aigrain}, S., {Almenara}, J.-M., {Alonso}, R., {Auvergne},
  M., {Baglin}, A., {Barge}, P., {Bord{\'e}}, P., {Bruntt}, H., {Cabrera}, J.,
  {Carpano}, S., {Cavarroc}, C., {Csizmadia}, S., {Damiani}, C., {Deeg}, H.~J.,
  {Dvorak}, R., {Fridlund}, M., {H{\'e}brard}, G., {Gandolfi}, D., {Gillon},
  M., {Guenther}, E., {Guillot}, T., {Hatzes}, A., {Jorda}, L., {L{\'e}ger},
  A., {Lammer}, H., {Mazeh}, T., {Moutou}, C., {Ollivier}, M., {Ofir}, A.,
  {Parviainen}, H., {Queloz}, D., {Rauer}, H., {Rodr{\'{\i}}guez}, A., {Rouan},
  D., {Santerne}, A., {Schneider}, J., {Tal-Or}, L., {Tingley}, B.,
  {Weingrill}, J., {Wuchterl}, G., Feb. 2012. {Transiting exoplanets from the
  CoRoT space mission. XX. CoRoT-20b: A very high density, high eccentricity
  transiting giant planet}. \aap 538, A145.

\bibitem[{{Deming} et~al.(2013){Deming}, {Wilkins}, {McCullough}, {Burrows},
  {Fortney}, {Agol}, {Dobbs-Dixon}, {Madhusudhan}, {Crouzet}, {Desert},
  {Gilliland}, {Haynes}, {Knutson}, {Line}, {Magic}, {Mandell}, {Ranjan},
  {Charbonneau}, {Clampin}, {Seager}, and {Showman}}]{Deming+2013}
{Deming}, D., {Wilkins}, A., {McCullough}, P., {Burrows}, A., {Fortney}, J.~J.,
  {Agol}, E., {Dobbs-Dixon}, I., {Madhusudhan}, N., {Crouzet}, N., {Desert},
  J.-M., {Gilliland}, R.~L., {Haynes}, K., {Knutson}, H.~A., {Line}, M.,
  {Magic}, Z., {Mandell}, A.~M., {Ranjan}, S., {Charbonneau}, D., {Clampin},
  M., {Seager}, S., {Showman}, A.~P., Sep. 2013. {Infrared Transmission
  Spectroscopy of the Exoplanets HD 209458b and XO-1b Using the Wide Field
  Camera-3 on the Hubble Space Telescope}. \apj 774, 95.

\bibitem[{{D{\'e}sert} et~al.(2009){D{\'e}sert}, {Lecavelier des Etangs},
  {H{\'e}brard}, {Sing}, {Ehrenreich}, {Ferlet}, and
  {Vidal-Madjar}}]{Desert+2009}
{D{\'e}sert}, J.-M., {Lecavelier des Etangs}, A., {H{\'e}brard}, G., {Sing},
  D.~K., {Ehrenreich}, D., {Ferlet}, R., {Vidal-Madjar}, A., Jul. 2009. {Search
  for Carbon Monoxide in the Atmosphere of the Transiting Exoplanet HD
  189733b}. \apj 699, 478--485.

\bibitem[{{Desjarlais}(2003)}]{Desjarlais03}
{Desjarlais}, M.~P., Aug. 2003. {Density-functional calculations of the liquid
  deuterium Hugoniot, reshock, and reverberation timing}. \prb 68~(6),
  064204--+.

\bibitem[{{Fegley} and {Lodders}(1994)}]{FL94}
{Fegley}, B.~J., {Lodders}, K., Jul. 1994. {Chemical models of the deep
  atmospheres of Jupiter and Saturn}. Icarus 110, 117--154.

\bibitem[{{Feuchtgruber} et~al.(2013){Feuchtgruber}, {Lellouch}, {Orton}, {de
  Graauw}, {Vandenbussche}, {Swinyard}, {Moreno}, {Jarchow}, {Billebaud},
  {Cavali{\'e}}, {Sidher}, and {Hartogh}}]{Feuchtgruber+2013}
{Feuchtgruber}, H., {Lellouch}, E., {Orton}, G., {de Graauw}, T.,
  {Vandenbussche}, B., {Swinyard}, B., {Moreno}, R., {Jarchow}, C.,
  {Billebaud}, F., {Cavali{\'e}}, T., {Sidher}, S., {Hartogh}, P., Mar. 2013.
  {The D/H ratio in the atmospheres of Uranus and Neptune from Herschel-PACS
  observations}. \aap 551, A126.

\bibitem[{{Fischer} and {Valenti}(2005)}]{FV05}
{Fischer}, D.~A., {Valenti}, J., Apr. 2005. {The Planet-Metallicity
  Correlation}. \apj 622, 1102--1117.

\bibitem[{{Fletcher} et~al.(2011){Fletcher}, {Baines}, {Momary}, {Showman},
  {Irwin}, {Orton}, {Roos-Serote}, and {Merlet}}]{Fletcher+2011}
{Fletcher}, L.~N., {Baines}, K.~H., {Momary}, T.~W., {Showman}, A.~P., {Irwin},
  P.~G.~J., {Orton}, G.~S., {Roos-Serote}, M., {Merlet}, C., Aug. 2011.
  {Saturn's tropospheric composition and clouds from Cassini/VIMS 4.6-5.1
  {$\mu$}m nightside spectroscopy}. \icarus 214, 510--533.

\bibitem[{{Fletcher} et~al.(2009{\natexlab{a}}){Fletcher}, {Orton}, {Teanby},
  and {Irwin}}]{Fletcher+2009b}
{Fletcher}, L.~N., {Orton}, G.~S., {Teanby}, N.~A., {Irwin}, P.~G.~J., Aug.
  2009{\natexlab{a}}. {Phosphine on Jupiter and Saturn from Cassini/CIRS}.
  \icarus 202, 543--564.

\bibitem[{{Fletcher} et~al.(2009{\natexlab{b}}){Fletcher}, {Orton}, {Teanby},
  {Irwin}, and {Bjoraker}}]{Fletcher+2009a}
{Fletcher}, L.~N., {Orton}, G.~S., {Teanby}, N.~A., {Irwin}, P.~G.~J.,
  {Bjoraker}, G.~L., Feb. 2009{\natexlab{b}}. {Methane and its isotopologues on
  Saturn from Cassini/CIRS observations}. \icarus 199, 351--367.

\bibitem[{{Fortney} and {Hubbard}(2003)}]{FH03}
{Fortney}, J.~J., {Hubbard}, W.~B., Jul. 2003. {Phase separation in giant
  planets: inhomogeneous evolution of Saturn}. Icarus 164, 228--243.

\bibitem[{{Fortney} et~al.(2011){Fortney}, {Ikoma}, {Nettelmann}, {Guillot},
  and {Marley}}]{Fortney+2011}
{Fortney}, J.~J., {Ikoma}, M., {Nettelmann}, N., {Guillot}, T., {Marley},
  M.~S., Mar. 2011. {Self-consistent Model Atmospheres and the Cooling of the
  Solar System's Giant Planets}. \apj 729, 32.

\bibitem[{{Fortney} and {Nettelmann}(2010)}]{FortneyNettelmann2010}
{Fortney}, J.~J., {Nettelmann}, N., May 2010. {The Interior Structure,
  Composition, and Evolution of Giant Planets}. \ssr 152, 423--447.

\bibitem[{{Fortney} et~al.(2006){Fortney}, {Saumon}, {Marley}, {Lodders}, and
  {Freedman}}]{Fortney+06}
{Fortney}, J.~J., {Saumon}, D., {Marley}, M.~S., {Lodders}, K., {Freedman},
  R.~S., May 2006. {Atmosphere, Interior, and Evolution of the Metal-rich
  Transiting Planet HD 149026b}. \apj 642, 495--504.

\bibitem[{{Fortov} et~al.(2007){Fortov}, {Ilkaev}, {Arinin}, {Burtzev},
  {Golubev}, {Iosilevskiy}, {Khrustalev}, {Mikhailov}, {Mochalov}, {Ternovoi},
  and {Zhernokletov}}]{Fortov+2007}
{Fortov}, V.~E., {Ilkaev}, R.~I., {Arinin}, V.~A., {Burtzev}, V.~V., {Golubev},
  V.~A., {Iosilevskiy}, I.~L., {Khrustalev}, V.~V., {Mikhailov}, A.~L.,
  {Mochalov}, M.~A., {Ternovoi}, V.~Y., {Zhernokletov}, M.~V., Nov. 2007.
  {Phase Transition in a Strongly Nonideal Deuterium Plasma Generated by
  Quasi-Isentropical Compression at Megabar Pressures}. Physical Review Letters
  99~(18), 185001.

\bibitem[{{Fossati} et~al.(2010){Fossati}, {Haswell}, {Froning}, {Hebb},
  {Holmes}, {Kolb}, {Helling}, {Carter}, {Wheatley}, {Collier Cameron},
  {Loeillet}, {Pollacco}, {Street}, {Stempels}, {Simpson}, {Udry}, {Joshi},
  {West}, {Skillen}, and {Wilson}}]{Fossati+10}
{Fossati}, L., {Haswell}, C.~A., {Froning}, C.~S., {Hebb}, L., {Holmes}, S.,
  {Kolb}, U., {Helling}, C., {Carter}, A., {Wheatley}, P., {Collier Cameron},
  A., {Loeillet}, B., {Pollacco}, D., {Street}, R., {Stempels}, H.~C.,
  {Simpson}, E., {Udry}, S., {Joshi}, Y.~C., {West}, R.~G., {Skillen}, I.,
  {Wilson}, D., May 2010. {Metals in the Exosphere of the Highly Irradiated
  Planet WASP-12b}. \apjl 714, L222--L227.

\bibitem[{{Freedman} et~al.(2008){Freedman}, {Marley}, and
  {Lodders}}]{Freedman+2008}
{Freedman}, R.~S., {Marley}, M.~S., {Lodders}, K., Feb. 2008. {Line and Mean
  Opacities for Ultracool Dwarfs and Extrasolar Planets}. \apjs 174, 504--513.

\bibitem[{{French} et~al.(2012){French}, {Becker}, {Lorenzen}, {Nettelmann},
  {Bethkenhagen}, {Wicht}, and {Redmer}}]{French+2012}
{French}, M., {Becker}, A., {Lorenzen}, W., {Nettelmann}, N., {Bethkenhagen},
  M., {Wicht}, J., {Redmer}, R., Sep. 2012. {Ab Initio Simulations for Material
  Properties along the Jupiter Adiabat}. \apjs 202, 5.

\bibitem[{{French} et~al.(2009){French}, {Mattsson}, {Nettelmann}, and
  {Redmer}}]{French+2009}
{French}, M., {Mattsson}, T.~R., {Nettelmann}, N., {Redmer}, R., Feb. 2009.
  {Equation of state and phase diagram of water at ultrahigh pressures as in
  planetary interiors}. \prb 79~(5), 054107.

\bibitem[{{Gastine} et~al.(2013){Gastine}, {Wicht}, and
  {Aurnou}}]{Gastine+2013}
{Gastine}, T., {Wicht}, J., {Aurnou}, J.~M., Jul. 2013. {Zonal flow regimes in
  rotating anelastic spherical shells: An application to giant planets}.
  \icarus 225, 156--172.

\bibitem[{{Gaulme} et~al.(2011){Gaulme}, {Schmider}, {Gay}, {Guillot}, and
  {Jacob}}]{Gaulme+11}
{Gaulme}, P., {Schmider}, F.-X., {Gay}, J., {Guillot}, T., {Jacob}, C., Jul.
  2011. {Detection of Jovian seismic waves: a new probe of its interior
  structure}. \aap 531, A104.

\bibitem[{{Gautier} et~al.(1995){Gautier}, {Conrath}, {Owen}, {De Pater}, and
  {Atreya}}]{Gautier+95}
{Gautier}, D., {Conrath}, B.~J., {Owen}, T., {De Pater}, I., {Atreya}, S.~K.,
  1995. {The Troposphere of Neptune}. Neptune and Triton, UofA Press, pp.
  547--611.

\bibitem[{{Gautier} et~al.(2001){Gautier}, {Hersant}, {Mousis}, and
  {Lunine}}]{GHML01}
{Gautier}, D., {Hersant}, F., {Mousis}, O., {Lunine}, J.~I., Oct. 2001.
  {Erratum: Enrichments in Volatiles in Jupiter: A New Interpretation of the
  Galileo Measurements}. \apjl 559, L183--L183.

\bibitem[{{Giampieri} et~al.(2006){Giampieri}, {Dougherty}, {Smith}, and
  {Russell}}]{2006Natur.441...62G}
{Giampieri}, G., {Dougherty}, M.~K., {Smith}, E.~J., {Russell}, C.~T., May
  2006. {A regular period for Saturn's magnetic field that may track its
  internal rotation}. \nat 441, 62--64.

\bibitem[{{Gibson} et~al.(2011){Gibson}, {Pont}, and {Aigrain}}]{Gibson+2011}
{Gibson}, N.~P., {Pont}, F., {Aigrain}, S., Mar. 2011. {A new look at NICMOS
  transmission spectroscopy of HD 189733, GJ-436 and XO-1: no conclusive
  evidence for molecular features}. \mnras 411, 2199--2213.

\bibitem[{{Gierasch} et~al.(2000){Gierasch}, {Ingersoll}, {Banfield}, {Ewald},
  {Helfenstein}, {Simon-Miller}, {Vasavada}, {Breneman}, {Senske}, and {A4
  Galileo Imaging Team}}]{Gierasch+00}
{Gierasch}, P.~J., {Ingersoll}, A.~P., {Banfield}, D., {Ewald}, S.~P.,
  {Helfenstein}, P., {Simon-Miller}, A., {Vasavada}, A., {Breneman}, H.~H.,
  {Senske}, D.~A., {A4 Galileo Imaging Team}, Feb. 2000. {Observation of moist
  convection in Jupiter's atmosphere}. \nat 403, 628--630.

\bibitem[{{Glatzmaier} et~al.(2009){Glatzmaier}, {Evonuk}, and
  {Rogers}}]{Glatzmaier+2009}
{Glatzmaier}, G., {Evonuk}, M., {Rogers}, T., Feb. 2009. {Differential rotation
  in giant planets maintained by density-stratified turbulent convection}.
  Geophysical and Astrophysical Fluid Dynamics 103, 31--51.

\bibitem[{{Gonzalez}(1998)}]{Gonzalez98}
{Gonzalez}, G., Jun. 1998. {Spectroscopic analyses of the parent stars of
  extrasolar planetary system candidates}. \aap 334, 221--238.

\bibitem[{{Goukenleuque} et~al.(2000){Goukenleuque}, {B{\'e}zard}, {Joguet},
  {Lellouch}, and {Freedman}}]{Goukenleuque+00}
{Goukenleuque}, C., {B{\'e}zard}, B., {Joguet}, B., {Lellouch}, E., {Freedman},
  R., Feb. 2000. {A Radiative Equilibrium Model of 51 Peg b}. Icarus 143,
  308--323.

\bibitem[{{Guervilly} et~al.(2012){Guervilly}, {Cardin}, and
  {Schaeffer}}]{Guervilly+2012}
{Guervilly}, C., {Cardin}, P., {Schaeffer}, N., Mar. 2012. {A dynamo driven by
  zonal jets at the upper surface: Applications to giant planets}. \icarus 218,
  100--114.

\bibitem[{{Guillot}(1999{\natexlab{a}})}]{Guillot99a}
{Guillot}, T., Oct. 1999{\natexlab{a}}. {A comparison of the interiors of
  Jupiter and Saturn}. \planss 47, 1183--1200.

\bibitem[{{Guillot}(1999{\natexlab{b}})}]{Guillot99b}
{Guillot}, T., Oct. 1999{\natexlab{b}}. {Interior of Giant Planets Inside and
  Outside the Solar System}. Science 286, 72--77.

\bibitem[{{Guillot}(2005)}]{Guillot05}
{Guillot}, T., Jan. 2005. {THE INTERIORS OF GIANT PLANETS: Models and
  Outstanding Questions}. Annual Review of Earth and Planetary Sciences 33,
  493--530.

\bibitem[{{Guillot}(2008)}]{Guillot08}
{Guillot}, T., Aug. 2008. {The composition of transiting giant extrasolar
  planets}. Physica Scripta Volume T 130~(1), 014023.

\bibitem[{{Guillot}(2010)}]{Guillot2010}
{Guillot}, T., Sep. 2010. {On the radiative equilibrium of irradiated planetary
  atmospheres}. \aap 520, A27.

\bibitem[{{Guillot} et~al.(1996){Guillot}, {Burrows}, {Hubbard}, {Lunine}, and
  {Saumon}}]{Guillot+96}
{Guillot}, T., {Burrows}, A., {Hubbard}, W.~B., {Lunine}, J.~I., {Saumon}, D.,
  Mar. 1996. {Giant Planets at Small Orbital Distances}. \apjl 459, L35--L39.

\bibitem[{{Guillot} and {Hueso}(2006)}]{GH06}
{Guillot}, T., {Hueso}, R., Mar. 2006. {The composition of Jupiter: sign of a
  (relatively) late formation in a chemically evolved protosolar disc}. \mnras
  367, L47--L51.

\bibitem[{{Guillot} et~al.(2006){Guillot}, {Santos}, {Pont}, {Iro}, {Melo}, and
  {Ribas}}]{Guillot+06}
{Guillot}, T., {Santos}, N.~C., {Pont}, F., {Iro}, N., {Melo}, C., {Ribas}, I.,
  Jul. 2006. {A correlation between the heavy element content of transiting
  extrasolar planets and the metallicity of their parent stars}. \aap 453,
  L21--L24.

\bibitem[{{Guillot} and {Showman}(2002)}]{GS02}
{Guillot}, T., {Showman}, A.~P., Apr. 2002. Evolution of ``51 pegasus b-like''
  planets. \aap 385, 156--165.

\bibitem[{{Guillot} et~al.(2004){Guillot}, {Stevenson}, {Hubbard}, and
  {Saumon}}]{GSHS04}
{Guillot}, T., {Stevenson}, D.~J., {Hubbard}, W.~B., {Saumon}, D., 2004. {The
  interior of Jupiter}. Jupiter.~The Planet, Satellites and Magnetosphere, pp.
  35--57.

\bibitem[{{Gulkis} et~al.(1978){Gulkis}, {Janssen}, and {Olsen}}]{GJO78}
{Gulkis}, S., {Janssen}, M.~A., {Olsen}, E.~T., Apr. 1978. {Evidence for the
  depletion of ammonia in the Uranus atmosphere}. Icarus 34, 10--19.

\bibitem[{{Gurnett} et~al.(2005){Gurnett}, {Kurth}, {Hospodarsky}, {Persoon},
  {Averkamp}, {Cecconi}, {Lecacheux}, {Zarka}, {Canu}, {Cornilleau-Wehrlin},
  {Galopeau}, {Roux}, {Harvey}, {Louarn}, {Bostrom}, {Gustafsson}, {Wahlund},
  {Desch}, {Farrell}, {Kaiser}, {Goetz}, {Kellogg}, {Fischer}, {Ladreiter},
  {Rucker}, {Alleyne}, and {Pedersen}}]{2005Sci...307.1255G}
{Gurnett}, D.~A., {Kurth}, W.~S., {Hospodarsky}, G.~B., {Persoon}, A.~M.,
  {Averkamp}, T.~F., {Cecconi}, B., {Lecacheux}, A., {Zarka}, P., {Canu}, P.,
  {Cornilleau-Wehrlin}, N., {Galopeau}, P., {Roux}, A., {Harvey}, C., {Louarn},
  P., {Bostrom}, R., {Gustafsson}, G., {Wahlund}, J.-E., {Desch}, M.~D.,
  {Farrell}, W.~M., {Kaiser}, M.~L., {Goetz}, K., {Kellogg}, P.~J., {Fischer},
  G., {Ladreiter}, H.-P., {Rucker}, H., {Alleyne}, H., {Pedersen}, A., Feb.
  2005. {Radio and Plasma Wave Observations at Saturn from Cassini's Approach
  and First Orbit}. Science 307, 1255--1259.

\bibitem[{{Hammel} et~al.(2005){Hammel}, {de Pater}, {Gibbard}, {Lockwood}, and
  {Rages}}]{Hammel+05}
{Hammel}, H.~B., {de Pater}, I., {Gibbard}, S.~G., {Lockwood}, G.~W., {Rages},
  K., May 2005. {New cloud activity on Uranus in 2004: First detection of a
  southern feature at 2.2 {$\mu$}m}. Icarus 175, 284--288.

\bibitem[{{Hansen}(2008)}]{Hansen2008}
{Hansen}, B.~M.~S., Dec. 2008. {On the Absorption and Redistribution of Energy
  in Irradiated Planets}. \apjs 179, 484--508.

\bibitem[{{Harrington} et~al.(2006){Harrington}, {Hansen}, {Luszcz}, {Seager},
  {Deming}, {Menou}, {Cho}, and {Richardson}}]{Harrington+06}
{Harrington}, J., {Hansen}, B.~M., {Luszcz}, S.~H., {Seager}, S., {Deming}, D.,
  {Menou}, K., {Cho}, J.~Y.-K., {Richardson}, L.~J., Oct. 2006. {The
  Phase-Dependent Infrared Brightness of the Extrasolar Planet {$\upsilon$}
  Andromedae b}. Science 314, 623--626.

\bibitem[{{Hedman} and {Nicholson}(2013)}]{HedmanNicholson2013}
{Hedman}, M.~M., {Nicholson}, P.~D., Jul. 2013. {Kronoseismology: Using Density
  Waves in Saturn's C Ring to Probe the Planet's Interior}. \aj 146, 12.

\bibitem[{{Heimpel} and {G{\'o}mez P{\'e}rez}(2011)}]{HeimpelGomezPerez2011}
{Heimpel}, M., {G{\'o}mez P{\'e}rez}, N., Jul. 2011. {On the relationship
  between zonal jets and dynamo action in giant planets}. \grl 38, 14201.

\bibitem[{{Helled}(2011)}]{Helled2011}
{Helled}, R., Jul. 2011. {Constraining Saturn's Core Properties by a
  Measurement of Its Moment of Inertia -- Implications to the Cassini Solstice
  Mission}. \apjl 735, L16.

\bibitem[{{Helled} et~al.(2011{\natexlab{a}}){Helled}, {Anderson}, {Podolak},
  and {Schubert}}]{Helled+2011}
{Helled}, R., {Anderson}, J.~D., {Podolak}, M., {Schubert}, G., Jan.
  2011{\natexlab{a}}. {Interior Models of Uranus and Neptune}. \apj 726, 15.

\bibitem[{{Helled} et~al.(2010){Helled}, {Anderson}, and
  {Schubert}}]{Helled+2010}
{Helled}, R., {Anderson}, J.~D., {Schubert}, G., Nov. 2010. {Uranus and
  Neptune: Shape and rotation}. \icarus 210, 446--454.

\bibitem[{{Helled} et~al.(2011{\natexlab{b}}){Helled}, {Anderson}, {Schubert},
  and {Stevenson}}]{Helled+2011b}
{Helled}, R., {Anderson}, J.~D., {Schubert}, G., {Stevenson}, D.~J., Dec.
  2011{\natexlab{b}}. {Jupiter's moment of inertia: A possible determination by
  Juno}. \icarus 216, 440--448.

\bibitem[{{Helled} and {Bodenheimer}(2011)}]{HelledBodenheimer2011}
{Helled}, R., {Bodenheimer}, P., Feb. 2011. {The effects of metallicity and
  grain growth and settling on the early evolution of gaseous protoplanets}.
  \icarus 211, 939--947.

\bibitem[{{Helled} and {Guillot}(2013)}]{HelledGuillot2013}
{Helled}, R., {Guillot}, T., Apr. 2013. {Interior Models of Saturn: Including
  the Uncertainties in Shape and Rotation}. \apj 767, 113.

\bibitem[{{Hersant} et~al.(2004){Hersant}, {Gautier}, and {Lunine}}]{HGL04}
{Hersant}, F., {Gautier}, D., {Lunine}, J.~I., Jun. 2004. {Enrichment in
  volatiles in the giant planets of the Solar System}. \planss 52, 623--641.

\bibitem[{{Hicks} et~al.(2009){Hicks}, {Boehly}, {Celliers}, {Eggert}, {Moon},
  {Meyerhofer}, and {Collins}}]{Hicks+2009}
{Hicks}, D.~G., {Boehly}, T.~R., {Celliers}, P.~M., {Eggert}, J.~H., {Moon},
  S.~J., {Meyerhofer}, D.~D., {Collins}, G.~W., Jan. 2009. {Laser-driven single
  shock compression of fluid deuterium from 45 to 220 GPa}. \prb 79~(1),
  014112.

\bibitem[{{Hubbard}(1968)}]{Hubbard68}
{Hubbard}, W.~B., Jun. 1968. {Thermal structure of Jupiter}. \apj 152,
  745--754.

\bibitem[{{Hubbard}(1977)}]{Hubbard77}
{Hubbard}, W.~B., Feb. 1977. {The Jovian surface condition and cooling rate}.
  Icarus 30, 305--310.

\bibitem[{{Hubbard}(1989)}]{1989oeps.book..539H}
{Hubbard}, W.~B., 1989. {Structure and composition of giant planet interiors}.
  Origin and Evolution of Planetary and Satellite Atmospheres, pp. 539--563.

\bibitem[{{Hubbard}(1999)}]{Hubbard1999}
{Hubbard}, W.~B., Feb. 1999. {NOTE: Gravitational Signature of Jupiter's Deep
  Zonal Flows}. \icarus 137, 357--359.

\bibitem[{{Hubbard}(2013)}]{Hubbard2013}
{Hubbard}, W.~B., May 2013. {Concentric Maclaurin Spheroid Models of Rotating
  Liquid Planets}. \apj 768, 43.

\bibitem[{{Hubbard} et~al.(1999){Hubbard}, {Guillot}, {Marley}, {Burrows},
  {Lunine}, and {Saumon}}]{Hubbard+99}
{Hubbard}, W.~B., {Guillot}, T., {Marley}, M.~S., {Burrows}, A., {Lunine},
  J.~I., {Saumon}, D.~S., Oct. 1999. {Comparative evolution of Jupiter and
  Saturn}. \planss 47, 1175--1182.

\bibitem[{{Hubbard} et~al.(1995){Hubbard}, {Pearl}, {Podolak}, and
  {Stevenson}}]{Hubbard+95}
{Hubbard}, W.~B., {Pearl}, J.~C., {Podolak}, M., {Stevenson}, D.~J., 1995. {The
  Interior of Neptune}. Neptune and Triton, UofA Press, pp. 109--138.

\bibitem[{{Hueso} et~al.(2002){Hueso}, {S{\'a}nchez-Lavega}, and
  {Guillot}}]{HSG02}
{Hueso}, R., {S{\'a}nchez-Lavega}, A., {Guillot}, T., Oct. 2002. {A model for
  large-scale convective storms in Jupiter}. Journal of Geophysical Research
  (Planets) 107, 5--1.

\bibitem[{{Ida} and {Lin}(2004{\natexlab{a}})}]{IL04a}
{Ida}, S., {Lin}, D.~N.~C., Mar. 2004{\natexlab{a}}. {Toward a Deterministic
  Model of Planetary Formation. I. A Desert in the Mass and Semimajor Axis
  Distributions of Extrasolar Planets}. \apj 604, 388--413.

\bibitem[{{Ida} and {Lin}(2004{\natexlab{b}})}]{IL04b}
{Ida}, S., {Lin}, D.~N.~C., Nov. 2004{\natexlab{b}}. {Toward a Deterministic
  Model of Planetary Formation. II. The Formation and Retention of Gas Giant
  Planets around Stars with a Range of Metallicities}. \apj 616, 567--572.

\bibitem[{{Ikoma} et~al.(2006){Ikoma}, {Guillot}, {Genda}, {Tanigawa}, and
  {Ida}}]{Ikoma+06}
{Ikoma}, M., {Guillot}, T., {Genda}, H., {Tanigawa}, T., {Ida}, S., Oct. 2006.
  {On the Origin of HD 149026b}. \apj 650, 1150--1159.

\bibitem[{{Ingersoll} et~al.(1995){Ingersoll}, {Barnet}, {Beebe}, {Flasar},
  {Hinson}, {Limaye}, {Sromovsky}, and {Suomi}}]{Ingersoll+95}
{Ingersoll}, A.~P., {Barnet}, C.~D., {Beebe}, R.~F., {Flasar}, F.~M., {Hinson},
  D.~P., {Limaye}, S.~S., {Sromovsky}, L.~A., {Suomi}, V.~E., 1995. {Dynamic
  Meteorology of Neptune}. Neptune and Triton, UofA Press, pp. 613--682.

\bibitem[{{Iorio}(2010)}]{Iorio2010}
{Iorio}, L., Aug. 2010. {Juno, the angular momentum of Jupiter and the
  Lense-Thirring effect}. New Astronomy 15, 554--560.

\bibitem[{{Iro} et~al.(2005){Iro}, {B{\'e}zard}, and {Guillot}}]{IBG05}
{Iro}, N., {B{\'e}zard}, B., {Guillot}, T., Jun. 2005. {A time-dependent
  radiative model of HD 209458b}. \aap 436, 719--727.

\bibitem[{{Irwin} et~al.(2014){Irwin}, {Lellouch}, {de Bergh}, {Courtin},
  {B{\'e}zard}, {Fletcher}, {Orton}, {Teanby}, {Calcutt}, {Tice}, {Hurley}, and
  {Davis}}]{Irwin+2014}
{Irwin}, P.~G.~J., {Lellouch}, E., {de Bergh}, C., {Courtin}, R., {B{\'e}zard},
  B., {Fletcher}, L.~N., {Orton}, G.~S., {Teanby}, N.~A., {Calcutt}, S.~B.,
  {Tice}, D., {Hurley}, J., {Davis}, G.~R., Jan. 2014. {Line-by-line analysis
  of Neptune's near-IR spectrum observed with Gemini/NIFS and VLT/CRIRES}.
  \icarus 227, 37--48.

\bibitem[{{Jackiewicz} et~al.(2012){Jackiewicz}, {Nettelmann}, {Marley}, and
  {Fortney}}]{Jackiewicz+2012}
{Jackiewicz}, J., {Nettelmann}, N., {Marley}, M., {Fortney}, J., Aug. 2012.
  {Forward and inverse modeling for jovian seismology}. \icarus 220, 844--854.

\bibitem[{{Jacobson}(2009)}]{Jacobson2009}
{Jacobson}, R.~A., May 2009. {The Orbits of the Neptunian Satellites and the
  Orientation of the Pole of Neptune}. \aj 137, 4322--4329.

\bibitem[{{Jacobson} et~al.(2006){Jacobson}, {Antreasian}, {Bordi}, {Criddle},
  {Ionasescu}, {Jones}, {Mackenzie}, {Meek}, {Parcher}, {Pelletier}, {Owen},
  {Roth}, {Roundhill}, and {Stauch}}]{2006AJ....132.2520J}
{Jacobson}, R.~A., {Antreasian}, P.~G., {Bordi}, J.~J., {Criddle}, K.~E.,
  {Ionasescu}, R., {Jones}, J.~B., {Mackenzie}, R.~A., {Meek}, M.~C.,
  {Parcher}, D., {Pelletier}, F.~J., {Owen}, Jr., W.~M., {Roth}, D.~C.,
  {Roundhill}, I.~M., {Stauch}, J.~R., Dec. 2006. {The Gravity Field of the
  Saturnian System from Satellite Observations and Spacecraft Tracking Data}.
  \aj 132, 2520--2526.

\bibitem[{{Karkoschka} and {Tomasko}(2011)}]{KarkoschkaTomasko2011}
{Karkoschka}, E., {Tomasko}, M.~G., Jan. 2011. {The haze and methane
  distributions on Neptune from HST-STIS spectroscopy}. \icarus 211, 780--797.

\bibitem[{{Kaspi} et~al.(2013){Kaspi}, {Showman}, {Hubbard}, {Aharonson}, and
  {Helled}}]{Kaspi+2013}
{Kaspi}, Y., {Showman}, A.~P., {Hubbard}, W.~B., {Aharonson}, O., {Helled}, R.,
  May 2013. {Atmospheric confinement of jet streams on Uranus and Neptune}.
  \nat 497, 344--347.

\bibitem[{{Kippenhahn} and {Weigert}(1994)}]{KW94}
{Kippenhahn}, R., {Weigert}, A., 1994. {Stellar Structure and Evolution}.
  Stellar Structure and Evolution, XVI, 468 pp.~192 figs..~ Springer-Verlag
  Berlin Heidelberg New York.~Also Astronomy and Astrophysics Library.

\bibitem[{{Knudson} and {Desjarlais}(2009)}]{KnudsonDesjarlais2009}
{Knudson}, M.~D., {Desjarlais}, M.~P., Nov. 2009. {Shock Compression of Quartz
  to 1.6 TPa: Redefining a Pressure Standard}. Physical Review Letters
  103~(22), 225501.

\bibitem[{{Knudson} et~al.(2012){Knudson}, {Desjarlais}, {Lemke}, {Mattsson},
  {French}, {Nettelmann}, and {Redmer}}]{Knudson+2012}
{Knudson}, M.~D., {Desjarlais}, M.~P., {Lemke}, R.~W., {Mattsson}, T.~R.,
  {French}, M., {Nettelmann}, N., {Redmer}, R., Mar. 2012. {Probing the
  Interiors of the Ice Giants: Shock Compression of Water to 700 GPa and
  3.8g/cm$^{3}$}. Physical Review Letters 108~(9), 091102.

\bibitem[{{Knudson} et~al.(2004){Knudson}, {Hanson}, {Bailey}, {Hall}, {Asay},
  and {Deeney}}]{Knudson+04}
{Knudson}, M.~D., {Hanson}, D.~L., {Bailey}, J.~E., {Hall}, C.~A., {Asay},
  J.~R., {Deeney}, C., Apr. 2004. {Principal Hugoniot, reverberating wave, and
  mechanical reshock measurements of liquid deuterium to 400 GPa using plate
  impact techniques}. \prb 69~(14), 144209--+.

\bibitem[{{Knutson} et~al.(2007){Knutson}, {Charbonneau}, {Allen}, {Fortney},
  {Agol}, {Cowan}, {Showman}, {Cooper}, and {Megeath}}]{Knutson+07}
{Knutson}, H.~A., {Charbonneau}, D., {Allen}, L.~E., {Fortney}, J.~J., {Agol},
  E., {Cowan}, N.~B., {Showman}, A.~P., {Cooper}, C.~S., {Megeath}, S.~T., May
  2007. {A map of the day-night contrast of the extrasolar planet HD 189733b}.
  \nat 447, 183--186.

\bibitem[{{Kramm} et~al.(2012){Kramm}, {Nettelmann}, {Fortney},
  {Neuh{\"a}user}, and {Redmer}}]{Kramm+2012}
{Kramm}, U., {Nettelmann}, N., {Fortney}, J.~J., {Neuh{\"a}user}, R., {Redmer},
  R., Feb. 2012. {Constraining the interior of extrasolar giant planets with
  the tidal Love number k$_{2}$ using the example of HAT-P-13b}. \aap 538,
  A146.

\bibitem[{{Kuchner}(2003)}]{Kuchner03}
{Kuchner}, M.~J., Oct. 2003. {Volatile-rich Earth-Mass Planets in the Habitable
  Zone}. \apjl 596, L105--L108.

\bibitem[{{Kunde} et~al.(1982){Kunde}, {Hanel}, {Maguire}, {Gautier},
  {Baluteau}, {Marten}, {Chedin}, {Husson}, and {Scott}}]{Kunde+1982}
{Kunde}, V., {Hanel}, R., {Maguire}, W., {Gautier}, D., {Baluteau}, J.~P.,
  {Marten}, A., {Chedin}, A., {Husson}, N., {Scott}, N., Dec. 1982. {The
  tropospheric gas composition of Jupiter's north equatorial belt /NH3, PH3,
  CH3D, GeH4, H2O/ and the Jovian D/H isotopic ratio}. \apj 263, 443--467.

\bibitem[{{Lammer} et~al.(2003){Lammer}, {Selsis}, {Ribas}, {Guinan}, {Bauer},
  and {Weiss}}]{Lammer+03}
{Lammer}, H., {Selsis}, F., {Ribas}, I., {Guinan}, E.~F., {Bauer}, S.~J.,
  {Weiss}, W.~W., Dec. 2003. {Atmospheric Loss of Exoplanets Resulting from
  Stellar X-Ray and Extreme-Ultraviolet Heating}. \apjl 598, L121--L124.

\bibitem[{{Laughlin} et~al.(2011){Laughlin}, {Crismani}, and
  {Adams}}]{Laughlin+11}
{Laughlin}, G., {Crismani}, M., {Adams}, F.~C., Mar. 2011. {On the Anomalous
  Radii of the Transiting Extrasolar Planets}. \apjl 729, L7.

\bibitem[{{Leconte} and {Chabrier}(2012)}]{LeconteChabrier2012}
{Leconte}, J., {Chabrier}, G., Apr. 2012. {A new vision of giant planet
  interiors: Impact of double diffusive convection}. \aap 540, A20.

\bibitem[{{Leconte} et~al.(2010){Leconte}, {Chabrier}, {Baraffe}, and
  {Levrard}}]{Leconte+10}
{Leconte}, J., {Chabrier}, G., {Baraffe}, I., {Levrard}, B., Jun. 2010. {Is
  tidal heating sufficient to explain bloated exoplanets? Consistent
  calculations accounting for finite initial eccentricity}. \aap 516, A64.

\bibitem[{{L{\'e}ger} et~al.(2004){L{\'e}ger}, {Selsis}, {Sotin}, {Guillot},
  {Despois}, {Mawet}, {Ollivier}, {Lab{\`e}que}, {Valette}, {Brachet},
  {Chazelas}, and {Lammer}}]{Leger+04}
{L{\'e}ger}, A., {Selsis}, F., {Sotin}, C., {Guillot}, T., {Despois}, D.,
  {Mawet}, D., {Ollivier}, M., {Lab{\`e}que}, A., {Valette}, C., {Brachet}, F.,
  {Chazelas}, B., {Lammer}, H., Jun. 2004. {A new family of planets?
  ``Ocean-Planets''}. Icarus 169, 499--504.

\bibitem[{{Lellouch} et~al.(2001){Lellouch}, {B{\'e}zard}, {Fouchet},
  {Feuchtgruber}, {Encrenaz}, and {de Graauw}}]{Lellouch+2001}
{Lellouch}, E., {B{\'e}zard}, B., {Fouchet}, T., {Feuchtgruber}, H.,
  {Encrenaz}, T., {de Graauw}, T., May 2001. {The deuterium abundance in
  Jupiter and Saturn from ISO-SWS observations}. \aap 370, 610--622.

\bibitem[{{Levrard} et~al.(2007){Levrard}, {Correia}, {Chabrier}, {Baraffe},
  {Selsis}, and {Laskar}}]{Levrard+07}
{Levrard}, B., {Correia}, A.~C.~M., {Chabrier}, G., {Baraffe}, I., {Selsis},
  F., {Laskar}, J., Jan. 2007. {Tidal dissipation within hot Jupiters: a new
  appraisal}. \aap 462, L5--L8.

\bibitem[{{Lian} and {Showman}(2010)}]{LianShowman2010}
{Lian}, Y., {Showman}, A.~P., May 2010. {Generation of equatorial jets by
  large-scale latent heating on the giant planets}. \icarus 207, 373--393.

\bibitem[{{Lindal}(1992{\natexlab{a}})}]{1992AJ....103..967L}
{Lindal}, G.~F., Mar. 1992{\natexlab{a}}. {The atmosphere of Neptune - an
  analysis of radio occultation data acquired with Voyager 2}. \aj 103,
  967--982.

\bibitem[{{Lindal}(1992{\natexlab{b}})}]{Lindal92}
{Lindal}, G.~F., Mar. 1992{\natexlab{b}}. {The atmosphere of Neptune - an
  analysis of radio occultation data acquired with Voyager 2}. \aj 103,
  967--982.

\bibitem[{{Lindal} et~al.(1985){Lindal}, {Sweetnam}, and
  {Eshleman}}]{1985AJ.....90.1136L}
{Lindal}, G.~F., {Sweetnam}, D.~N., {Eshleman}, V.~R., Jun. 1985. {The
  atmosphere of Saturn - an analysis of the Voyager radio occultation
  measurements}. \aj 90, 1136--1146.

\bibitem[{{Lindal} et~al.(1981){Lindal}, {Wood}, {Levy}, {Anderson},
  {Sweetnam}, {Hotz}, {Buckles}, {Holmes}, {Doms}, {Eshleman}, {Tyler}, and
  {Croft}}]{1981JGR....86.8721L}
{Lindal}, G.~F., {Wood}, G.~E., {Levy}, G.~S., {Anderson}, J.~D., {Sweetnam},
  D.~N., {Hotz}, H.~B., {Buckles}, B.~J., {Holmes}, D.~P., {Doms}, P.~E.,
  {Eshleman}, V.~R., {Tyler}, G.~L., {Croft}, T.~A., Sep. 1981. {The atmosphere
  of Jupiter - an analysis of the Voyager radio occultation measurements}. \jgr
  86, 8721--8727.

\bibitem[{{Little} et~al.(1999){Little}, {Anger}, {Ingersoll}, {Vasavada},
  {Senske}, {Breneman}, {Borucki}, and {The Galileo SSI Team}}]{Little+99}
{Little}, B., {Anger}, C.~D., {Ingersoll}, A.~P., {Vasavada}, A.~R., {Senske},
  D.~A., {Breneman}, H.~H., {Borucki}, W.~J., {The Galileo SSI Team}, Dec.
  1999. {Galileo Images of Lightning on Jupiter}. Icarus 142, 306--323.

\bibitem[{{Liu} and {Schneider}(2011)}]{LiuSchneider2011}
{Liu}, J., {Schneider}, T., Nov. 2011. {Convective Generation of Equatorial
  Superrotation in Planetary Atmospheres}. Journal of Atmospheric Sciences 68,
  2742--2756.

\bibitem[{{Liu} et~al.(2013){Liu}, {Schneider}, and {Kaspi}}]{Liu+2013}
{Liu}, J., {Schneider}, T., {Kaspi}, Y., May 2013. {Predictions of thermal and
  gravitational signals of Jupiter's deep zonal winds}. \icarus 224, 114--125.

\bibitem[{{Lodders}(2008)}]{Lodders2008}
{Lodders}, K., Feb. 2008. {The Solar Argon Abundance}. \apj 674, 607--611.

\bibitem[{{Lodders} and {Fegley}(1994)}]{LF94}
{Lodders}, K., {Fegley}, Jr., B., Dec. 1994. {The origin of carbon monoxide in
  Neptunes's atmosphere}. Icarus 112, 368--375.

\bibitem[{{Lodders} et~al.(2009){Lodders}, {Palme}, and {Gail}}]{Lodders+2009}
{Lodders}, K., {Palme}, H., {Gail}, H.-P., 2009. {Abundances of the Elements in
  the Solar System}. Landolt B{\"o}rnstein, 44.

\bibitem[{{Lopez} et~al.(2012){Lopez}, {Fortney}, and {Miller}}]{Lopez+12}
{Lopez}, E.~D., {Fortney}, J.~J., {Miller}, N., Dec. 2012. {How Thermal
  Evolution and Mass-loss Sculpt Populations of Super-Earths and Sub-Neptunes:
  Application to the Kepler-11 System and Beyond}. \apj 761, 59.

\bibitem[{{Lorenzen} et~al.(2011){Lorenzen}, {Holst}, and
  {Redmer}}]{Lorenzen+2011}
{Lorenzen}, W., {Holst}, B., {Redmer}, R., Dec. 2011. {Metallization in
  hydrogen-helium mixtures}. \prb 84~(23), 235109.

\bibitem[{{Loubeyre} et~al.(2012){Loubeyre}, {Brygoo}, {Eggert}, {Celliers},
  {Spaulding}, {Rygg}, {Boehly}, {Collins}, and {Jeanloz}}]{Loubeyre+2012}
{Loubeyre}, P., {Brygoo}, S., {Eggert}, J., {Celliers}, P.~M., {Spaulding},
  D.~K., {Rygg}, J.~R., {Boehly}, T.~R., {Collins}, G.~W., {Jeanloz}, R., Oct.
  2012. {Extended data set for the equation of state of warm dense hydrogen
  isotopes}. \prb 86~(14), 144115.

\bibitem[{{Loubeyre} et~al.(1991){Loubeyre}, {Letoullec}, and
  {Pinceaux}}]{Loubeyre+1991}
{Loubeyre}, P., {Letoullec}, R., {Pinceaux}, J.~P., May 1991. {A new
  determination of the binary phase diagram of H$_{2}$-He mixtures at 296 K}.
  Journal of Physics Condensed Matter 3, 3183--3192.

\bibitem[{{Mardling}(2010)}]{Mardling2010}
{Mardling}, R.~A., Sep. 2010. {The determination of planetary structure in
  tidally relaxed inclined systems}. \mnras 407, 1048--1069.

\bibitem[{{Marley} et~al.(2006){Marley}, {Fortney}, {Hubickyj}, {Bodenheimer},
  and {Lissauer}}]{Marley+06}
{Marley}, M.~S., {Fortney}, J.~J., {Hubickyj}, O., {Bodenheimer}, P.,
  {Lissauer}, J.~J., Sep. 2006. {On the Luminosity of Young Jupiters}. ArXiv
  Astrophysics e-prints.

\bibitem[{{Marley} et~al.(1995){Marley}, {G{\'o}mez}, and {Podolak}}]{MGP95}
{Marley}, M.~S., {G{\'o}mez}, P., {Podolak}, M., Nov. 1995. {Monte Carlo
  interior models for Uranus and Neptune}. \jgr 100, 23349--23354.

\bibitem[{{Marley} and {Porco}(1993)}]{MarleyPorco1993}
{Marley}, M.~S., {Porco}, C.~C., Dec. 1993. {Planetary acoustic mode seismology
  - Saturn's rings}. \icarus 106, 508.

\bibitem[{{Marley} et~al.(1996){Marley}, {Saumon}, {Guillot}, {Freedman},
  {Hubbard}, {Burrows}, and {Lunine}}]{Marley+1996}
{Marley}, M.~S., {Saumon}, D., {Guillot}, T., {Freedman}, R.~S., {Hubbard},
  W.~B., {Burrows}, A., {Lunine}, J.~I., Jun. 1996. {Atmospheric, Evolutionary,
  and Spectral Models of the Brown Dwarf Gliese 229 B}. Science 272,
  1919--1921.

\bibitem[{{Marty} et~al.(2011){Marty}, {Chaussidon}, {Wiens}, {Jurewicz}, and
  {Burnett}}]{Marty+2011}
{Marty}, B., {Chaussidon}, M., {Wiens}, R.~C., {Jurewicz}, A.~J.~G., {Burnett},
  D.~S., Jun. 2011. {A $^{15}$N-Poor Isotopic Composition for the Solar System
  As Shown by Genesis Solar Wind Samples}. Science 332, 1533--.

\bibitem[{{Marty} et~al.(2009){Marty}, {Guillot}, {Coustenis}, {Achilleos},
  {Alibert}, {Asmar}, {Atkinson}, {Atreya}, {Babasides}, {Baines}, {Balint},
  {Banfield}, {Barber}, {B{\'e}zard}, {Bjoraker}, {Blanc}, {Bolton},
  {Chanover}, {Charnoz}, {Chassefi{\`e}re}, {Colwell}, {Deangelis},
  {Dougherty}, {Drossart}, {Flasar}, {Fouchet}, {Frampton}, {Franchi},
  {Gautier}, {Gurvits}, {Hueso}, {Kazeminejad}, {Krimigis}, {Jambon}, {Jones},
  {Langevin}, {Leese}, {Lellouch}, {Lunine}, {Milillo}, {Mahaffy}, {Mauk},
  {Morse}, {Moreira}, {Moussas}, {Murray}, {Mueller-Wodarg}, {Owen},
  {Pogrebenko}, {Prang{\'e}}, {Read}, {Sanchez-Lavega}, {Sarda}, {Stam},
  {Tinetti}, {Zarka}, and {Zarnecki}}]{Marty+2009}
{Marty}, B., {Guillot}, T., {Coustenis}, A., {Achilleos}, N., {Alibert}, Y.,
  {Asmar}, S., {Atkinson}, D., {Atreya}, S., {Babasides}, G., {Baines}, K.,
  {Balint}, T., {Banfield}, D., {Barber}, S., {B{\'e}zard}, B., {Bjoraker},
  G.~L., {Blanc}, M., {Bolton}, S., {Chanover}, N., {Charnoz}, S.,
  {Chassefi{\`e}re}, E., {Colwell}, J.~E., {Deangelis}, E., {Dougherty}, M.,
  {Drossart}, P., {Flasar}, F.~M., {Fouchet}, T., {Frampton}, R., {Franchi},
  I., {Gautier}, D., {Gurvits}, L., {Hueso}, R., {Kazeminejad}, B., {Krimigis},
  T., {Jambon}, A., {Jones}, G., {Langevin}, Y., {Leese}, M., {Lellouch}, E.,
  {Lunine}, J., {Milillo}, A., {Mahaffy}, P., {Mauk}, B., {Morse}, A.,
  {Moreira}, M., {Moussas}, X., {Murray}, C., {Mueller-Wodarg}, I., {Owen},
  T.~C., {Pogrebenko}, S., {Prang{\'e}}, R., {Read}, P., {Sanchez-Lavega}, A.,
  {Sarda}, P., {Stam}, D., {Tinetti}, G., {Zarka}, P., {Zarnecki}, J., Mar.
  2009. {Kronos: exploring the depths of Saturn with probes and remote sensing
  through an international mission}. Experimental Astronomy 23, 947--976.

\bibitem[{{Mayor} and {Queloz}(1995)}]{MQ95}
{Mayor}, M., {Queloz}, D., Nov. 1995. {A Jupiter-Mass Companion to a Solar-Type
  Star}. \nat 378, 355--+.

\bibitem[{{McMahon} et~al.(2012){McMahon}, {Morales}, {Pierleoni}, and
  {Ceperley}}]{McMahon+2012}
{McMahon}, J.~M., {Morales}, M.~A., {Pierleoni}, C., {Ceperley}, D.~M., Oct.
  2012. {The properties of hydrogen and helium under extreme conditions}.
  Reviews of Modern Physics 84, 1607--1653.

\bibitem[{{Militzer} et~al.(2001){Militzer}, {Ceperley}, {Kress}, {Johnson},
  {Collins}, and {Mazevet}}]{Militzer+01}
{Militzer}, B., {Ceperley}, D.~M., {Kress}, J.~D., {Johnson}, J.~D., {Collins},
  L.~A., {Mazevet}, S., Dec. 2001. {Calculation of a Deuterium Double Shock
  Hugoniot from Ab Initio Simulations}. Physical Review Letters 87~(26),
  A265502+.

\bibitem[{{Militzer} and {Hubbard}(2013)}]{MilitzerHubbard2013}
{Militzer}, B., {Hubbard}, W.~B., Sep. 2013. {Ab Initio Equation of State for
  Hydrogen-Helium Mixtures with Recalibration of the Giant-planet Mass-Radius
  Relation}. \apj 774, 148.

\bibitem[{{Militzer} et~al.(2008){Militzer}, {Hubbard}, {Vorberger}, {Tamblyn},
  and {Bonev}}]{Militzer+2008}
{Militzer}, B., {Hubbard}, W.~B., {Vorberger}, J., {Tamblyn}, I., {Bonev},
  S.~A., Nov. 2008. {A Massive Core in Jupiter Predicted from First-Principles
  Simulations}. \apjl 688, L45--L48.

\bibitem[{{Miller} and {Fortney}(2011)}]{MF11}
{Miller}, N., {Fortney}, J.~J., Aug. 2011. {The Heavy-element Masses of
  Extrasolar Giant Planets, Revealed}. \apjl 736, L29.

\bibitem[{{Mochalov} et~al.(2012){Mochalov}, {Il'kaev}, {Fortov}, {Mikhailov},
  {Makarov}, {Arinin}, {Blikov}, {Baurin}, {Komrakov}, {Ogorodnikov},
  {Ryzhkov}, {Pronin}, and {Yukhimchuk}}]{Mochalov+2012}
{Mochalov}, M.~A., {Il'kaev}, R.~I., {Fortov}, V.~E., {Mikhailov}, A.~L.,
  {Makarov}, Y.~M., {Arinin}, V.~A., {Blikov}, A.~O., {Baurin}, A.~Y.,
  {Komrakov}, V.~A., {Ogorodnikov}, V.~A., {Ryzhkov}, A.~V., {Pronin}, E.~A.,
  {Yukhimchuk}, A.~A., Oct. 2012. {Measurement of quasi-isentropic
  compressibility of helium and deuterium at pressures of 1500-2000 GPa}.
  Soviet Journal of Experimental and Theoretical Physics 115, 614--625.

\bibitem[{{Moorhead} and {Adams}(2005)}]{MA05}
{Moorhead}, A.~V., {Adams}, F.~C., Nov. 2005. {Giant planet migration through
  the action of disk torques and planet planet scattering}. Icarus 178,
  517--539.

\bibitem[{{Morales} et~al.(2013{\natexlab{a}}){Morales}, {Hamel}, {Caspersen},
  and {Schwegler}}]{Morales+2013b}
{Morales}, M.~A., {Hamel}, S., {Caspersen}, K., {Schwegler}, E., May
  2013{\natexlab{a}}. {Hydrogen-helium demixing from first principles: From
  diamond anvil cells to planetary interiors}. \prb 87~(17), 174105.

\bibitem[{{Morales} et~al.(2013{\natexlab{b}}){Morales}, {McMahon},
  {Pierleoni}, and {Ceperley}}]{Morales+2013a}
{Morales}, M.~A., {McMahon}, J.~M., {Pierleoni}, C., {Ceperley}, D.~M., Feb.
  2013{\natexlab{b}}. {Nuclear Quantum Effects and Nonlocal
  Exchange-Correlation Functionals Applied to Liquid Hydrogen at High
  Pressure}. Physical Review Letters 110~(6), 065702.

\bibitem[{{Morales} et~al.(2010){Morales}, {Pierleoni}, and
  {Ceperley}}]{Morales+2010}
{Morales}, M.~A., {Pierleoni}, C., {Ceperley}, D.~M., Feb. 2010. {Equation of
  state of metallic hydrogen from coupled electron-ion Monte Carlo
  simulations}. \pre 81~(2), 021202.

\bibitem[{{Morales} et~al.(2009){Morales}, {Schwegler}, {Ceperley},
  {Pierleoni}, {Hamel}, and {Caspersen}}]{Morales+2009}
{Morales}, M.~A., {Schwegler}, E., {Ceperley}, D., {Pierleoni}, C., {Hamel},
  S., {Caspersen}, K., Feb. 2009. {Phase separation in hydrogen-helium mixtures
  at Mbar pressures}. Proceedings of the National Academy of Science 106, 1324.

\bibitem[{{Mordasini}(2013)}]{Mordasini2013}
{Mordasini}, C., Oct. 2013. {Luminosity of young Jupiters revisited. Massive
  cores make hot planets}. \aap 558, A113.

\bibitem[{{Mordasini} et~al.(2012){Mordasini}, {Alibert}, {Georgy},
  {Dittkrist}, {Klahr}, and {Henning}}]{Mordasini+12}
{Mordasini}, C., {Alibert}, Y., {Georgy}, C., {Dittkrist}, K.-M., {Klahr}, H.,
  {Henning}, T., Nov. 2012. {Characterization of exoplanets from their
  formation. II. The planetary mass-radius relationship}. \aap 547, A112.

\bibitem[{{Mousis} et~al.(2012){Mousis}, {Lunine}, {Madhusudhan}, and
  {Johnson}}]{Mousis+2012}
{Mousis}, O., {Lunine}, J.~I., {Madhusudhan}, N., {Johnson}, T.~V., May 2012.
  {Nebular Water Depletion as the Cause of Jupiter's Low Oxygen Abundance}.
  \apjl 751, L7.

\bibitem[{{Moutou} et~al.(2013){Moutou}, {Deleuil}, {Guillot}, {Baglin},
  {Bord{\'e}}, {Bouchy}, {Cabrera}, {Csizmadia}, and {Deeg}}]{Moutou+2013}
{Moutou}, C., {Deleuil}, M., {Guillot}, T., {Baglin}, A., {Bord{\'e}}, P.,
  {Bouchy}, F., {Cabrera}, J., {Csizmadia}, S., {Deeg}, H.~J., Nov. 2013.
  {CoRoT: Harvest of the exoplanet program}. \icarus 226, 1625--1634.

\bibitem[{{Ness} et~al.(1986){Ness}, {Acuna}, {Behannon}, {Burlaga},
  {Connerney}, and {Lepping}}]{1986Sci...233...85N}
{Ness}, N.~F., {Acuna}, M.~H., {Behannon}, K.~W., {Burlaga}, L.~F.,
  {Connerney}, J.~E.~P., {Lepping}, R.~P., Jul. 1986. {Magnetic fields at
  Uranus}. Science 233, 85--89.

\bibitem[{{Ness} et~al.(1989){Ness}, {Acuna}, {Burlaga}, {Connerney}, and
  {Lepping}}]{1989Sci...246.1473N}
{Ness}, N.~F., {Acuna}, M.~H., {Burlaga}, L.~F., {Connerney}, J.~E.~P.,
  {Lepping}, R.~P., Dec. 1989. {Magnetic fields at Neptune}. Science 246,
  1473--1478.

\bibitem[{{Nettelmann} et~al.(2012){Nettelmann}, {Becker}, {Holst}, and
  {Redmer}}]{Nettelmann+2012}
{Nettelmann}, N., {Becker}, A., {Holst}, B., {Redmer}, R., May 2012. {Jupiter
  Models with Improved Ab Initio Hydrogen Equation of State (H-REOS.2)}. \apj
  750, 52.

\bibitem[{{Nettelmann} et~al.(2013{\natexlab{a}}){Nettelmann}, {Helled},
  {Fortney}, and {Redmer}}]{Nettelmann+2013a}
{Nettelmann}, N., {Helled}, R., {Fortney}, J.~J., {Redmer}, R., Mar.
  2013{\natexlab{a}}. {New indication for a dichotomy in the interior structure
  of Uranus and Neptune from the application of modified shape and rotation
  data}. \planss 77, 143--151.

\bibitem[{{Nettelmann} et~al.(2013{\natexlab{b}}){Nettelmann}, {P{\"u}stow},
  and {Redmer}}]{Nettelmann+2013b}
{Nettelmann}, N., {P{\"u}stow}, R., {Redmer}, R., Jul. 2013{\natexlab{b}}.
  {Saturn layered structure and homogeneous evolution models with different
  EOSs}. \icarus 225, 548--557.

\bibitem[{{Noll} and {Larson}(1991)}]{NollLarson1991}
{Noll}, K.~S., {Larson}, H.~P., Jan. 1991. {The spectrum of Saturn from 1990 to
  2230/cm - Abundances of AsH3, CH3D, CO, GeH4, NH3, and PH3}. \icarus 89,
  168--189.

\bibitem[{{Noll} et~al.(1990){Noll}, {Larson}, and {Geballe}}]{Noll+1990}
{Noll}, K.~S., {Larson}, H.~P., {Geballe}, T.~R., Feb. 1990. {The abundance of
  AsH3 in Jupiter}. \icarus 83, 494--499.

\bibitem[{{Owen} et~al.(1999){Owen}, {Mahaffy}, {Niemann}, {Atreya}, {Donahue},
  {Bar-Nun}, and {de Pater}}]{Owen+99}
{Owen}, T., {Mahaffy}, P., {Niemann}, H.~B., {Atreya}, S., {Donahue}, T.,
  {Bar-Nun}, A., {de Pater}, I., Nov. 1999. {A low-temperature origin for the
  planetesimals that formed Jupiter}. \nat 402, 269--270.

\bibitem[{{Parmentier} and {Guillot}(2014)}]{ParmentierGuillot2014}
{Parmentier}, V., {Guillot}, T., Feb. 2014. {A non-grey analytical model for
  irradiated atmospheres. I: Derivation}. \aap, in press.

\bibitem[{{Parmentier} et~al.(2013){Parmentier}, {Showman}, and
  {Lian}}]{Parmentier+2013}
{Parmentier}, V., {Showman}, A.~P., {Lian}, Y., Oct. 2013. {3D mixing in hot
  Jupiters atmospheres. I. Application to the day/night cold trap in HD
  209458b}. \aap 558, A91.

\bibitem[{{Pearl} and {Conrath}(1991)}]{PC91}
{Pearl}, J.~C., {Conrath}, B.~J., Oct. 1991. {The albedo, effective
  temperature, and energy balance of Neptune, as determined from Voyager data}.
  \jgr 96, 18921--+.

\bibitem[{{Podolak} et~al.(1991){Podolak}, {Hubbard}, and {Stevenson}}]{PHS91}
{Podolak}, M., {Hubbard}, W.~B., {Stevenson}, D.~J., 1991. {Model of Uranus'
  interior and magnetic field}. Uranus, UofA Press, pp. 29--61.

\bibitem[{{Podolak} et~al.(2000){Podolak}, {Podolak}, and {Marley}}]{PPM00}
{Podolak}, M., {Podolak}, J.~I., {Marley}, M.~S., Feb. 2000. {Further
  investigations of random models of Uranus and Neptune}. \planss 48, 143--151.

\bibitem[{{Podolak} et~al.(1995){Podolak}, {Weizman}, and {Marley}}]{PWM95}
{Podolak}, M., {Weizman}, A., {Marley}, M., Dec. 1995. {Comparative models of
  Uranus and Neptune}. \planss 43, 1517--1522.

\bibitem[{{Pont} et~al.(2013){Pont}, {Sing}, {Gibson}, {Aigrain}, {Henry}, and
  {Husnoo}}]{Pont+13}
{Pont}, F., {Sing}, D.~K., {Gibson}, N.~P., {Aigrain}, S., {Henry}, G.,
  {Husnoo}, N., Jul. 2013. {The prevalence of dust on the exoplanet HD 189733b
  from Hubble and Spitzer observations}. \mnras 432, 2917--2944.

\bibitem[{{Rages} et~al.(2002){Rages}, {Hammel}, and {Lockwood}}]{Rages+02}
{Rages}, K., {Hammel}, H.~B., {Lockwood}, G.~W., Sep. 2002. {A Prominent
  Apparition of Neptune's South Polar Feature}. Icarus 159, 262--265.

\bibitem[{{Rauscher} and {Menou}(2013)}]{RauscherMenou2013}
{Rauscher}, E., {Menou}, K., Feb. 2013. {Three-dimensional Atmospheric
  Circulation Models of HD 189733b and HD 209458b with Consistent Magnetic Drag
  and Ohmic Dissipation}. \apj 764, 103.

\bibitem[{{Read} et~al.(2009){Read}, {Dowling}, and {Schubert}}]{Read+2009}
{Read}, P.~L., {Dowling}, T.~E., {Schubert}, G., Jul. 2009. {Saturn's rotation
  period from its atmospheric planetary-wave configuration}. \nat 460,
  608--610.

\bibitem[{{Redmer} et~al.(2011){Redmer}, {Mattsson}, {Nettelmann}, and
  {French}}]{Redmer+2011}
{Redmer}, R., {Mattsson}, T.~R., {Nettelmann}, N., {French}, M., Jan. 2011.
  {The phase diagram of water and the magnetic fields of Uranus and Neptune}.
  \icarus 211, 798--803.

\bibitem[{{Ribas}(2006)}]{Ribas06}
{Ribas}, I., Aug. 2006. {Masses and Radii of Low-Mass Stars: Theory Versus
  Observations}. \apss 304, 89--92.

\bibitem[{{Robinson} and {Catling}(2014)}]{RobinsonCatling2014}
{Robinson}, T.~D., {Catling}, D.~C., Jan. 2014. {Common 0.1bar tropopause in
  thick atmospheres set by pressure-dependent infrared transparency}. Nature
  Geoscience 7, 12--15.

\bibitem[{{Rosenblum} et~al.(2011){Rosenblum}, {Garaud}, {Traxler}, and
  {Stellmach}}]{Rosenblum+2011}
{Rosenblum}, E., {Garaud}, P., {Traxler}, A., {Stellmach}, S., Apr. 2011.
  {Turbulent Mixing and Layer Formation in Double-diffusive Convection:
  Three-dimensional Numerical Simulations and Theory}. \apj 731, 66.

\bibitem[{{Rossow}(1978)}]{Rossow1978}
{Rossow}, W.~B., Oct. 1978. {Cloud microphysics - Analysis of the clouds of
  Earth, Venus, Mars, and Jupiter}. \icarus 36, 1--50.

\bibitem[{{Roulston} and {Stevenson}(1995)}]{RS95}
{Roulston}, M.~S., {Stevenson}, D.~J., 1995. {Prediction of neon depletion in
  Jupiter's atmosphere}. In: EOS. Vol.~76. p. 343.

\bibitem[{{Russell} and {Dougherty}(2010)}]{RussellDougherty2010}
{Russell}, C.~T., {Dougherty}, M.~K., May 2010. {Magnetic Fields of the Outer
  Planets}. \ssr 152, 251--269.

\bibitem[{{Salpeter}(1973)}]{Salpeter73}
{Salpeter}, E.~E., Apr. 1973. {On Convection and Gravitational Layering in
  Jupiter and in Stars of Low Mass}. \apjl 181, L83+.

\bibitem[{{Sanchez-Lavega} et~al.(1996){Sanchez-Lavega}, {Lecacheux}, {Gomez},
  {Colas}, {Laques}, {Noll}, {Gilmore}, {Miyazaki}, and
  {Parker}}]{SanchezLavega+96}
{Sanchez-Lavega}, A., {Lecacheux}, J., {Gomez}, J.~M., {Colas}, F., {Laques},
  P., {Noll}, K., {Gilmore}, D., {Miyazaki}, I., {Parker}, D., Feb. 1996.
  {Large-scale storms in Saturn's atmosphere during 1994}. Science 271,
  631--634.

\bibitem[{{Sano} et~al.(2011){Sano}, {Ozaki}, {Sakaiya}, {Shigemori}, {Ikoma},
  {Kimura}, {Miyanishi}, {Endo}, {Shiroshita}, {Takahashi}, {Jitsui}, {Hori},
  {Hironaka}, {Iwamoto}, {Kadono}, {Nakai}, {Okuchi}, {Otani}, {Shimizu},
  {Kondo}, {Kodama}, and {Mima}}]{Sano+2011}
{Sano}, T., {Ozaki}, N., {Sakaiya}, T., {Shigemori}, K., {Ikoma}, M., {Kimura},
  T., {Miyanishi}, K., {Endo}, T., {Shiroshita}, A., {Takahashi}, H., {Jitsui},
  T., {Hori}, Y., {Hironaka}, Y., {Iwamoto}, A., {Kadono}, T., {Nakai}, M.,
  {Okuchi}, T., {Otani}, K., {Shimizu}, K., {Kondo}, T., {Kodama}, R., {Mima},
  K., Feb. 2011. {Laser-shock compression and Hugoniot measurements of liquid
  hydrogen to 55 GPa}. \prb 83~(5), 054117.

\bibitem[{{Santos} et~al.(2004){Santos}, {Israelian}, and {Mayor}}]{SIM04}
{Santos}, N.~C., {Israelian}, G., {Mayor}, M., Mar. 2004. {Spectroscopic [Fe/H]
  for 98 extra-solar planet-host stars. Exploring the probability of planet
  formation}. \aap 415, 1153--1166.

\bibitem[{{Sato} et~al.(2005){Sato}, {Fischer}, {Henry}, {Laughlin}, {Butler},
  {Marcy}, {Vogt}, {Bodenheimer}, {Ida}, {Toyota}, {Wolf}, {Valenti}, {Boyd},
  {Johnson}, {Wright}, {Ammons}, {Robinson}, {Strader}, {McCarthy}, {Tah}, and
  {Minniti}}]{Sato+05}
{Sato}, B., {Fischer}, D.~A., {Henry}, G.~W., {Laughlin}, G., {Butler}, R.~P.,
  {Marcy}, G.~W., {Vogt}, S.~S., {Bodenheimer}, P., {Ida}, S., {Toyota}, E.,
  {Wolf}, A., {Valenti}, J.~A., {Boyd}, L.~J., {Johnson}, J.~A., {Wright},
  J.~T., {Ammons}, M., {Robinson}, S., {Strader}, J., {McCarthy}, C., {Tah},
  K.~L., {Minniti}, D., Nov. 2005. {The N2K Consortium. II. A Transiting Hot
  Saturn around HD 149026 with a Large Dense Core}. \apj 633, 465--473.

\bibitem[{{Saumon} et~al.(1995){Saumon}, {Chabrier}, and {van Horn}}]{SCvH95}
{Saumon}, D., {Chabrier}, G., {van Horn}, H.~M., Aug. 1995. {An Equation of
  State for Low-Mass Stars and Giant Planets}. \apjs 99, 713--+.

\bibitem[{{Saumon} and {Guillot}(2004)}]{SG04}
{Saumon}, D., {Guillot}, T., Jul. 2004. {Shock Compression of Deuterium and the
  Interiors of Jupiter and Saturn}. \apj 609, 1170--1180.

\bibitem[{{Saumon} et~al.(1996){Saumon}, {Hubbard}, {Burrows}, {Guillot},
  {Lunine}, and {Chabrier}}]{Saumon+96}
{Saumon}, D., {Hubbard}, W.~B., {Burrows}, A., {Guillot}, T., {Lunine}, J.~I.,
  {Chabrier}, G., Apr. 1996. {A Theory of Extrasolar Giant Planets}. \apj 460,
  993--+.

\bibitem[{{Schneider} et~al.(2011){Schneider}, {Dedieu}, {Le Sidaner},
  {Savalle}, and {Zolotukhin}}]{Schneider+11}
{Schneider}, J., {Dedieu}, C., {Le Sidaner}, P., {Savalle}, R., {Zolotukhin},
  I., Aug. 2011. {Defining and cataloging exoplanets: the exoplanet.eu
  database}. \aap 532, A79.

\bibitem[{{Schouten} et~al.(1991){Schouten}, {de Kuijper}, and
  {Michels}}]{Schouten+1991}
{Schouten}, J.~A., {de Kuijper}, A., {Michels}, J.~P.~J., Oct. 1991. {Critical
  line of He-H$_{2}$ up to 2500 K and the influence of attraction on
  fluid-fluid separation}. \prb 44, 6630--6634.

\bibitem[{{Seager} and {Deming}(2010)}]{SD10}
{Seager}, S., {Deming}, D., Sep. 2010. {Exoplanet Atmospheres}. \araa 48,
  631--672.

\bibitem[{{Seager} and {Sasselov}(2000)}]{SS00}
{Seager}, S., {Sasselov}, D.~D., Jul. 2000. {Theoretical Transmission Spectra
  during Extrasolar Giant Planet Transits}. \apj 537, 916--921.

\bibitem[{{Seiff} et~al.(1998){Seiff}, {Kirk}, {Knight}, {Young}, {Mihalov},
  {Young}, {Milos}, {Schubert}, {Blanchard}, and {Atkinson}}]{Seiff+98}
{Seiff}, A., {Kirk}, D.~B., {Knight}, T.~C.~D., {Young}, R.~E., {Mihalov},
  J.~D., {Young}, L.~A., {Milos}, F.~S., {Schubert}, G., {Blanchard}, R.~C.,
  {Atkinson}, D., Sep. 1998. {Thermal structure of Jupiter's atmosphere near
  the edge of a 5-{$\mu$}m hot spot in the north equatorial belt}. \jgr 103,
  22857--22890.

\bibitem[{{Showman} and {Guillot}(2002)}]{SG02}
{Showman}, A.~P., {Guillot}, T., Apr. 2002. {Atmospheric circulation and tides
  of ``51 Pegasus b-like'' planets}. \aap 385, 166--180.

\bibitem[{{Showman} and {Ingersoll}(1998)}]{SI98}
{Showman}, A.~P., {Ingersoll}, A.~P., Apr. 1998. {Interpretation of Galileo
  Probe Data and Implications for Jupiter's Dry Downdrafts}. Icarus 132,
  205--220.

\bibitem[{{Showman} et~al.(2011){Showman}, {Kaspi}, and
  {Flierl}}]{Showman+2011}
{Showman}, A.~P., {Kaspi}, Y., {Flierl}, G.~R., Feb. 2011. {Scaling laws for
  convection and jet speeds in the giant planets}. \icarus 211, 1258--1273.

\bibitem[{{Simon-Miller} et~al.(2002){Simon-Miller}, {Gierasch}, {Beebe},
  {Conrath}, {Flasar}, {Achterberg}, and {the Cassini CIRS
  Team}}]{SimonMiller+02}
{Simon-Miller}, A.~A., {Gierasch}, P.~J., {Beebe}, R.~F., {Conrath}, B.,
  {Flasar}, F.~M., {Achterberg}, R.~K., {the Cassini CIRS Team}, Jul. 2002.
  {New Observational Results Concerning Jupiter's Great Red Spot}. Icarus 158,
  249--266.

\bibitem[{{Sing} et~al.(2009){Sing}, {D{\'e}sert}, {Lecavelier Des Etangs},
  {Ballester}, {Vidal-Madjar}, {Parmentier}, {Hebrard}, and
  {Henry}}]{Sing+2009}
{Sing}, D.~K., {D{\'e}sert}, J.-M., {Lecavelier Des Etangs}, A., {Ballester},
  G.~E., {Vidal-Madjar}, A., {Parmentier}, V., {Hebrard}, G., {Henry}, G.~W.,
  Oct. 2009. {Transit spectrophotometry of the exoplanet HD 189733b. I.
  Searching for water but finding haze with HST NICMOS}. \aap 505, 891--899.

\bibitem[{{Snellen} et~al.(2010){Snellen}, {de Kok}, {de Mooij}, and
  {Albrecht}}]{Snellen+10}
{Snellen}, I.~A.~G., {de Kok}, R.~J., {de Mooij}, E.~J.~W., {Albrecht}, S.,
  Jun. 2010. {The orbital motion, absolute mass and high-altitude winds of
  exoplanet HD209458b}. \nat 465, 1049--1051.

\bibitem[{{Soderlund} et~al.(2013){Soderlund}, {Heimpel}, {King}, and
  {Aurnou}}]{Soderlund+2013}
{Soderlund}, K.~M., {Heimpel}, M.~H., {King}, E.~M., {Aurnou}, J.~M., May 2013.
  {Turbulent models of ice giant internal dynamics: Dynamos, heat transfer, and
  zonal flows}. \icarus 224, 97--113.

\bibitem[{{Spiegel} and {Burrows}(2013)}]{SpiegelBurrows2013}
{Spiegel}, D.~S., {Burrows}, A., Jul. 2013. {Thermal Processes Governing
  Hot-Jupiter Radii}. \apj 772, 76.

\bibitem[{{Spiegel} et~al.(2009){Spiegel}, {Silverio}, and
  {Burrows}}]{Spiegel+2009}
{Spiegel}, D.~S., {Silverio}, K., {Burrows}, A., Jul. 2009. {Can TiO Explain
  Thermal Inversions in the Upper Atmospheres of Irradiated Giant Planets?}
  \apj 699, 1487--1500.

\bibitem[{{Spilker}(2012)}]{Spilker2012}
{Spilker}, L.~J., Mar. 2012. {Cassini: Science Highlights from the Equinox and
  Solstice Missions}. In: Lunar and Planetary Institute Science Conference
  Abstracts. Vol.~43 of Lunar and Planetary Institute Science Conference
  Abstracts. p. 1358.

\bibitem[{{Sromovsky} et~al.(2011){Sromovsky}, {Fry}, and
  {Kim}}]{Sromovsky+2011}
{Sromovsky}, L.~A., {Fry}, P.~M., {Kim}, J.~H., Sep. 2011. {Methane on Uranus:
  The case for a compact CH $_{4}$ cloud layer at low latitudes and a severe CH
  $_{4}$ depletion at high-latitudes based on re-analysis of Voyager
  occultation measurements and STIS spectroscopy}. \icarus 215, 292--312.

\bibitem[{{Stanley} and {Bloxham}(2004)}]{StanleyBloxham2004}
{Stanley}, S., {Bloxham}, J., Mar. 2004. {Convective-region geometry as the
  cause of Uranus' and Neptune's unusual magnetic fields}. \nat 428, 151--153.

\bibitem[{{Stanley} and {Glatzmaier}(2010)}]{StanleyGlatzmaier2010}
{Stanley}, S., {Glatzmaier}, G.~A., May 2010. {Dynamo Models for Planets Other
  Than Earth}. \ssr 152, 617--649.

\bibitem[{{Stevenson}(1982)}]{Stevenson82}
{Stevenson}, D.~J., 1982. {Interiors of the giant planets}. Annual Review of
  Earth and Planetary Sciences 10, 257--295.

\bibitem[{{Stevenson}(1983)}]{1983RPPh...46..555S}
{Stevenson}, D.~J., May 1983. {Planetary magnetic fields}. Reports of Progress
  in Physics 46, 555--557.

\bibitem[{{Stevenson}(1985{\natexlab{a}})}]{1985Icar...62....4S}
{Stevenson}, D.~J., Apr. 1985{\natexlab{a}}. {Cosmochemistry and structure of
  the giant planets and their satellites}. Icarus 62, 4--15.

\bibitem[{{Stevenson}(1985{\natexlab{b}})}]{Stevenson1985}
{Stevenson}, D.~J., Apr. 1985{\natexlab{b}}. {Cosmochemistry and structure of
  the giant planets and their satellites}. \icarus 62, 4--15.

\bibitem[{{Stevenson} and {Salpeter}(1977{\natexlab{a}})}]{SS77b}
{Stevenson}, D.~J., {Salpeter}, E.~E., Oct. 1977{\natexlab{a}}. {The dynamics
  and helium distribution in hydrogen-helium fluid planets}. \apjs 35,
  239--261.

\bibitem[{{Stevenson} and {Salpeter}(1977{\natexlab{b}})}]{SS77a}
{Stevenson}, D.~J., {Salpeter}, E.~E., Oct. 1977{\natexlab{b}}. {The phase
  diagram and transport properties for hydrogen-helium fluid planets}. \apjs
  35, 221--237.

\bibitem[{{Sudarsky} et~al.(2003){Sudarsky}, {Burrows}, and {Hubeny}}]{SBH03}
{Sudarsky}, D., {Burrows}, A., {Hubeny}, I., May 2003. {Theoretical Spectra and
  Atmospheres of Extrasolar Giant Planets}. \apj 588, 1121--1148.

\bibitem[{{Trafton}(1967)}]{Trafton67}
{Trafton}, L.~M., Feb. 1967. {Model atmospheres of the major planets}. \apj
  147, 765--781.

\bibitem[{{Tsiganis} et~al.(2005){Tsiganis}, {Gomes}, {Morbidelli}, and
  {Levison}}]{TGML05}
{Tsiganis}, K., {Gomes}, R., {Morbidelli}, A., {Levison}, H.~F., May 2005.
  {Origin of the orbital architecture of the giant planets of the Solar
  System}. \nat 435, 459--461.

\bibitem[{{Vidal-Madjar} et~al.(2011){Vidal-Madjar}, {Huitson}, {Lecavelier Des
  Etangs}, {Sing}, {Ferlet}, {D{\'e}sert}, {H{\'e}brard}, {Boisse},
  {Ehrenreich}, and {Moutou}}]{Vidal-Madjar+2011}
{Vidal-Madjar}, A., {Huitson}, C.~M., {Lecavelier Des Etangs}, A., {Sing},
  D.~K., {Ferlet}, R., {D{\'e}sert}, J.-M., {H{\'e}brard}, G., {Boisse}, I.,
  {Ehrenreich}, D., {Moutou}, C., Sep. 2011. {The upper atmosphere of the
  exoplanet HD209458b revealed by the sodium D lines: temperature-pressure
  profile, ionization layer and thermosphere (Corrigendum)}. \aap 533, C4.

\bibitem[{{Vidal-Madjar} et~al.(2003){Vidal-Madjar}, {Lecavelier des Etangs},
  {D{\'e}sert}, {Ballester}, {Ferlet}, {H{\'e}brard}, and
  {Mayor}}]{VidalMadjar+03}
{Vidal-Madjar}, A., {Lecavelier des Etangs}, A., {D{\'e}sert}, J.-M.,
  {Ballester}, G.~E., {Ferlet}, R., {H{\'e}brard}, G., {Mayor}, M., Mar. 2003.
  {An extended upper atmosphere around the extrasolar planet HD209458b}. \nat
  422, 143--146.

\bibitem[{{Visscher} and {Moses}(2011)}]{VisscherMoses2011}
{Visscher}, C., {Moses}, J.~I., Sep. 2011. {Quenching of Carbon Monoxide and
  Methane in the Atmospheres of Cool Brown Dwarfs and Hot Jupiters}. \apj 738,
  72.

\bibitem[{{von Zahn} et~al.(1998{\natexlab{a}}){von Zahn}, {Hunten}, and
  {Lehmacher}}]{1998JGR...10322815V}
{von Zahn}, U., {Hunten}, D.~M., {Lehmacher}, G., Sep. 1998{\natexlab{a}}.
  {Helium in Jupiter's atmosphere: Results from the Galileo probe helium
  interferometer experiment}. \jgr 103, 22815--22830.

\bibitem[{{von Zahn} et~al.(1998{\natexlab{b}}){von Zahn}, {Hunten}, and
  {Lehmacher}}]{VonZahn+1998}
{von Zahn}, U., {Hunten}, D.~M., {Lehmacher}, G., Sep. 1998{\natexlab{b}}.
  {Helium in Jupiter's atmosphere: Results from the Galileo probe helium
  interferometer experiment}. \jgr 103, 22815--22830.

\bibitem[{{Vorberger} et~al.(2007){Vorberger}, {Tamblyn}, {Militzer}, and
  {Bonev}}]{Vorberger+2007}
{Vorberger}, J., {Tamblyn}, I., {Militzer}, B., {Bonev}, S.~A., Jan. 2007.
  {Hydrogen-helium mixtures in the interiors of giant planets}. \prb 75~(2),
  024206.

\bibitem[{{Vorontsov} et~al.(1976){Vorontsov}, {Zharkov}, and
  {Lubimov}}]{VZL76}
{Vorontsov}, S.~V., {Zharkov}, V.~N., {Lubimov}, V.~M., Jan. 1976. The free
  oscillations of jupiter and saturn. \icarus 27, 109--118.

\bibitem[{{Wahl} et~al.(2013){Wahl}, {Wilson}, and {Militzer}}]{Wahl+2013}
{Wahl}, S.~M., {Wilson}, H.~F., {Militzer}, B., Aug. 2013. {Solubility of Iron
  in Metallic Hydrogen and Stability of Dense Cores in Giant Planets}. \apj
  773, 95.

\bibitem[{{Warwick} et~al.(1989){Warwick}, {Evans}, {Peltzer}, {Peltzer},
  {Romig}, {Sawyer}, {Riddle}, {Schweitzer}, {Desch}, and
  {Kaiser}}]{1989Sci...246.1498W}
{Warwick}, J.~W., {Evans}, D.~R., {Peltzer}, G.~R., {Peltzer}, R.~G., {Romig},
  J.~H., {Sawyer}, C.~B., {Riddle}, A.~C., {Schweitzer}, A.~E., {Desch}, M.~D.,
  {Kaiser}, M.~L., Dec. 1989. {Voyager planetary radio astronomy at Neptune}.
  Science 246, 1498--1501.

\bibitem[{{Warwick} et~al.(1986){Warwick}, {Evans}, {Romig}, {Sawyer}, {Desch},
  {Kaiser}, {Alexander}, {Gulkis}, and {Poynter}}]{1986Sci...233..102W}
{Warwick}, J.~W., {Evans}, D.~R., {Romig}, J.~H., {Sawyer}, C.~B., {Desch},
  M.~D., {Kaiser}, M.~L., {Alexander}, J.~K., {Gulkis}, S., {Poynter}, R.~L.,
  Jul. 1986. {Voyager 2 radio observations of Uranus}. Science 233, 102--106.

\bibitem[{{Weir} et~al.(1996){Weir}, {Mitchell}, and {Nellis}}]{WMN96}
{Weir}, S.~T., {Mitchell}, A.~C., {Nellis}, W.~J., Mar. 1996. {Metallization of
  Fluid Molecular Hydrogen at 140 GPa (1.4 Mbar)}. Physical Review Letters 76,
  1860--1863.

\bibitem[{{Wilson} and {Militzer}(2010)}]{WilsonMilitzer2010}
{Wilson}, H.~F., {Militzer}, B., Mar. 2010. {Sequestration of Noble Gases in
  Giant Planet Interiors}. Physical Review Letters 104~(12), 121101.

\bibitem[{{Wilson} and {Militzer}(2012)}]{WilsonMilitzer2012}
{Wilson}, H.~F., {Militzer}, B., Jan. 2012. {Solubility of Water Ice in
  Metallic Hydrogen: Consequences for Core Erosion in Gas Giant Planets}. \apj
  745, 54.

\bibitem[{{Winn} and {Holman}(2005)}]{WinnHolman2005}
{Winn}, J.~N., {Holman}, M.~J., Aug. 2005. {Obliquity Tides on Hot Jupiters}.
  \apjl 628, L159--L162.

\bibitem[{{Wong} et~al.(2004){Wong}, {Mahaffy}, {Atreya}, {Niemann}, and
  {Owen}}]{Wong+04}
{Wong}, M.~H., {Mahaffy}, P.~R., {Atreya}, S.~K., {Niemann}, H.~B., {Owen},
  T.~C., Sep. 2004. {Updated Galileo probe mass spectrometer measurements of
  carbon, oxygen, nitrogen, and sulfur on Jupiter}. Icarus 171, 153--170.

\bibitem[{{Wright} et~al.(2011){Wright}, {Fakhouri}, {Marcy}, {Han}, {Feng},
  {Johnson}, {Howard}, {Fischer}, {Valenti}, {Anderson}, and
  {Piskunov}}]{Wright+11}
{Wright}, J.~T., {Fakhouri}, O., {Marcy}, G.~W., {Han}, E., {Feng}, Y.,
  {Johnson}, J.~A., {Howard}, A.~W., {Fischer}, D.~A., {Valenti}, J.~A.,
  {Anderson}, J., {Piskunov}, N., Apr. 2011. The exoplanet orbit database.
  \pasp 123, 412--422.

\bibitem[{{Youdin} and {Mitchell}(2010)}]{YoudinMitchell10}
{Youdin}, A.~N., {Mitchell}, J.~L., Oct. 2010. {The Mechanical Greenhouse:
  Burial of Heat by Turbulence in Hot Jupiter Atmospheres}. \apj 721,
  1113--1126.

\bibitem[{{Zharkov} and {Trubitsyn}(1974)}]{ZT74}
{Zharkov}, V.~N., {Trubitsyn}, V.~P., Feb. 1974. {Internal constitution and the
  figures of the giant planets}. Physics of the Earth and Planetary Interiors
  8, 105--107.

\bibitem[{{Zharkov} and {Trubitsyn}(1978)}]{ZT78}
{Zharkov}, V.~N., {Trubitsyn}, V.~P., 1978. {Physics of planetary interiors}.
  Astronomy and Astrophysics Series, Tucson: Pachart, 1978.

\end{thebibliography}

\end{document}